\documentclass[11pt,a4paper]{article}

\usepackage[english]{babel}
\usepackage[utf8]{inputenc}

\usepackage{epsf}
\usepackage{cite}
\usepackage{graphicx}
\usepackage{subcaption} 

\usepackage{amsmath}
\usepackage{amssymb}
\usepackage{mathrsfs}
\usepackage[normalem]{ulem}
\usepackage{mathbbol}
\usepackage{multirow}
\usepackage{slashed}
\usepackage{amsfonts}
\usepackage{bbding}
\usepackage{array}
\usepackage[
colorlinks=true
,urlcolor=black
,anchorcolor=black
,citecolor=black
,filecolor=black
,linkcolor=black
,menucolor=black
,linktocpage=true
,pdfproducer=medialab
,pdfa=true
]{hyperref}

\usepackage{color} 
\usepackage[left=2cm,right=2cm,top=2cm,bottom=2cm]{geometry}
\usepackage{tikz}
\usetikzlibrary{shapes,arrows,positioning,automata,backgrounds,calc,er,patterns}
\usepackage[compat=1.1.0]{tikz-feynman}
\usepackage{contour}
\usepackage{placeins}
\allowdisplaybreaks
\begin{document}
\begin{center}
\vspace*{1mm}
\vspace{1.3cm}
{\Large\bf
Exploring asymmetries in three-body cLFV lepton
decays: \vspace*{3mm}\\
probing CP violation in HNL extensions of the SM
}

\vspace*{1.2cm}
{\bf Adrian Darricau $^\text{a}$, 
Jonathan~Kriewald $^\text{b}$,  Ana~M.~Teixeira $^\text{a}$ }

\vspace*{.5cm}
$^\text{a}$ Laboratoire de Physique de Clermont Auvergne (UMR 6533), CNRS/IN2P3,\\
Univ. Clermont Auvergne, 4 Av. Blaise Pascal, 63178 Aubi\`ere Cedex,
France
 
\vspace*{.2cm}
$^\text{b}$ {Jožef Stefan Institut, Jamova Cesta 39, P. O. Box 3000, 1001 Ljubljana, Slovenia}
\end{center}

\vspace*{5mm}
\begin{abstract}
\noindent
In the context of Standard Model extensions via Majorana sterile fermions, the presence of additional CP violating phases (Dirac and Majorana) has been shown to be at source of important effects in charged lepton flavour violating (cLFV) transitions and decays. 
Here we will consider further angular observables that can be studied for polarised $\tau$ and $\mu$ cLFV decays. These include, among others, parity asymmetries and time-reversal asymmetries for generic cLFV 3-body decays,  $\ell_\alpha^+ \to \ell_\beta^+ \ell_\gamma^+ \ell_\delta^-$. We address relevant correlations between the different classes of observables, and show that one can have sizeable asymmetries, which can be used to further probe this interesting class of SM extensions.
Our study leads to the prediction of particular patterns of angular observables, which would allow to potentially falsify the model, should a cLFV signal be observed.

\end{abstract}

\section{Introduction}
The discovery of neutrino oscillations signals a clear departure from the Standard Model (SM). The violation of flavour in the neutral lepton sector opens a wide door for charged lepton flavour violation; likewise, the now present leptonic mixings further allow new sources of CP violation.
Numerous mechanisms have been proposed to explain light neutrino masses, calling upon the addition of new states to the SM. Among the most attractive and yet simple ones are SM extensions via (Majorana) sterile fermions. 

Depending on the actual mechanism of neutrino mass generation - and of the scale at which it is realised - the expected phenomenological implications strongly vary. Appealing low-scale models are expected to give rise to extensive contributions to numerous observables, be them charged lepton flavour violation (cLFV), lepton number violation (LNV) and/or CP violation (CPV). The presence of new CP violating phases is of particular interest, as these could play a relevant role in the explanation of the baryon asymmetry of the Universe via leptogenesis. 
Models of New Physics (NP) featuring heavy sterile fermions, with a typical scale around the TeV, and which have non-negligible mixings to the light (active) neutrinos, thus become particularly interesting - in view of the associated rich phenomenological implications~\cite{Riemann:1982rq,Illana:1999ww,Mann:1983dv,Illana:2000ic,Alonso:2012ji,Ilakovac:1994kj,Ma:1979px,Gronau:1984ct,Deppisch:2004fa,Deppisch:2005zm,Dinh:2012bp,Hambye:2013jsa,Abada:2014kba,Abada:2015oba,Abada:2015zea,Abada:2016vzu,Calibbi:2017uvl,Abada:2018nio,Arganda:2014dta,Marcano:2019rmk,Calderon:2022alb,Blondel:2014bra,Antusch:2016ejd,Cai:2017mow,Blondel:2022qqo,Abdullahi:2022jlv,Abada:2022wvh,Giffin:2022rei,Drewes:2022rsk,Ovchynnikov:2023wgg,Ajmal:2024kwi,Antusch:2024otj,Chakraborty:2022pcc,Mekala:2023diu,Kwok:2023dck,Li:2023tbx,Mikulenko:2023ezx,Urquia-Calderon:2023dkf,deLima:2024ohf,Marcos:2024yfm}. 

Searches for cLFV transitions and decays are at the core of a rich experimental programme which includes colliders and low-energy dedicated facilities. The experimental prospects for observing a cLFV signal are outstanding, especially concerning pure leptonic transitions, such as radiative and 3-body decays, or muon-electron conversion in nuclei. 
In view of this, cLFV observables have been intensively explored, especially in view of the insight they might offer regarding the nature of the underlying NP model at their origin; moreover, the interplay between several cLFV observables has been shown to be a very useful probe to discriminate between the nature of the (dominating) NP contributions, and possibly their scale. 

The prospects for cLFV in the context of SM extensions via heavy neutral leptons (HNL) have been the object of numerous dedicated studies in recent years~\cite{Calibbi:2017uvl,Hambye:2013jsa,Riemann:1982rq,Illana:1999ww,Mann:1983dv,Illana:2000ic,Alonso:2012ji,Ilakovac:1994kj,Ma:1979px,Gronau:1984ct,Deppisch:2004fa,Deppisch:2005zm,Dinh:2012bp,Abada:2014kba,Abada:2015oba,Abada:2015zea,Abada:2016vzu,Abada:2018nio,Arganda:2014dta,Kriewald:2024cnt,Crivellin:2022cve,Abada:2022asx,Granelli:2022eru,Urquia-Calderon:2022ufc,Urquia-Calderon:2025wjx}, especially since the heavy states can be at the origin of sizeable contributions to a wide array of observables, potentially within future experimental reach. 
Particularly in the simpler case in which no new phases are present in association with the heavy fermions, several interesting correlations between observables were identified, thus offering the possibility to probe operator dominance in certain regimes, and falsify these constructions in the advent of (multiple) experimental measurements. 

However, new leptonic CPV phases (Dirac and Majorana) are in general present 
in SM extensions via HNL (and are in fact needed should one wish to address the baryon asymmetry of the Universe via leptogenesis). 
As extensively argued in~\cite{Abada:2021zcm,Abada:2022asx,Kriewald:2024cnt}, 
the presence of CPV phases can have a strong impact on cLFV transitions, affecting their total rates, and  
disrupting the cLFV correlation patterns which would be expected in the CP conserving limit (in association with the dominance of a given topology). 

Clearly, the set of cLFV observables ($\ell_\alpha \to \ell_\beta \gamma$, $\ell_\alpha \to \ell_\beta \ell_\gamma \ell_\delta$, neutrinoless $\mu-e$ conversion, ...) 
does not suffice to disentangle the distinct operator contributions in the most general CP violating case. 
One must thus extend the number of observables under scrutiny in order 
to shed light on cancellation effects due to CP violation.
A promising possibility is to consider angular structures, of particular relevance in the case of cLFV 3-body decays. 
Especially in the case of a discovery, measurements of the helicity of the initial/final state particles, and the final state angular distributions can play a critical role. 
For the case of polarised decaying leptons, and noticing that the decay amplitudes can always be 
decomposed into spin-independent and spin-dependent parts, one can construct $P$ and $T$-odd triple products from momenta and spins~\cite{Kuno:1999jp,Kitano:2000fg,Goto:2010sn}. The latter are sensitive to interference terms in the helicity amplitudes, and to the relative physical phases between operators. The seminal works of~\cite{Kuno:1999jp,Kitano:2000fg} thus proposed to further consider three asymmetries (in the angular distributions), sensitive to the violation of parity and time-reversal in cLFV $\mu^+\to e^+ e^+ e^-$ decays: 
\begin{equation}
\mathcal{A}^{\mu^+\to e^+ e^+ e^-}_P\,, \quad 
\mathcal{A}^{\mu^+\to e^+ e^+ e^-}_{P^\prime}\,, \quad 
\mathcal{A}^{\mu^+\to e^+ e^+ e^-}_T\,.
\end{equation}
Recently, and especially in view of the high degree
of polarisation of muon beams at PSI (Mu3e experiment)~\cite{Mu3e:2020gyw}, the role of the asymmetries has been revisited, with the goal of discriminating between the operators contributing to polarised $\mu^+\to e^+ e^+ e^-$ decays~\cite{Bolton:2022lrg}. 
Relying on an effective field theory (EFT) approach, the obtained results evidenced the discriminating power of the $P$- and $T$-odd asymmetries: 
the observation of a non-zero $T$-odd asymmetry would signal the contribution to the amplitude 
of an interference term between dipole and vector operators, which is $T$-violating and (assuming CPT conservation) CP-odd as well.  
Since this process is purely leptonic and strong phases are thus absent, any non-zero measurement of the $T$=violating asymmetry clearly points towards new CP-violating contributions in the short-distance operators.
Likewise, and in association with the also very dynamic searches for the electric dipole moments (EDM) of charged leptons $d_\ell$, it was suggested that asymmetries may offer a first discovery of leptonic CP violation (other than the Dirac phase of the Pontecorvo-Maki-Nakagawa-Sakata (PMNS) matrix), even before a measurement of $d_e$~\cite{Redigolo:2024ztw}.  

\medskip
In this study, driven by the potential discriminating power of this class of angular observables, we explore the prospects for a comprehensive set of asymmetries (and angular distributions): we study all possible 3-body leptonic decays, $\ell_\alpha^+ \to \ell_\beta^+ \ell_\gamma^+ \ell_\delta^-$ (including both cLFV muon and tau-lepton decays), and consider a larger set of associated observables, including forward-backward asymmetries (also in association with $P$-, $P^\prime$- and $T$-asymmetries), providing a detailed description of the kinematical analysis.  
Motivated by the appeal and phenomenological interest of SM extensions via heavy neutral leptons, we have considered a minimal ad-hoc model in which two heavy Majorana sterile states are added to the SM content. 
This minimalistic approach can be interpreted as a first step leading to the analysis of complete NP models (seesaws, or other frameworks featuring sterile fermions), and the findings can be generalised to larger constructions in which cLFV and leptonic CP violation arise from the mixings of active and (heavy) sterile neutrinos.

As detailed in the manuscript, 
our findings support the fact that the new observables can contribute to restore the discriminating power of cLFV observables (even in the presence of CP violating phases), and shed light on the dominance of certain operators. 
In the presence of leptonic CP violating phases, there is no clear operator dominance for 
$\mu^+ \to e^+ e^+ e^-$ decays, and one requires complementary information (i.e. measurements) from radiative decays $\mu \to e \gamma$, especially to assess the dipole contributions. 
Due the sizeable $Z$-penguin contributions in this class of extensions, 3-body cLFV tau decay rates can be well within future reach; this then allows exploring a large set of observables to better identify distinctive angular patterns. 
Moreover, the comparatively smaller
increase in future sensitivities for cLFV tau decays translates in very strong predictions regarding potential measurements of the asymmetries (should a $\tau$ cLFV signal be observed), which add to the probing power of the latter observables for this NP construction.

This manuscript is organised as follows: in Section~\ref{sec:Asyms}
we carry out a detailed computation of the angular asymmetries, providing the results in the case of a SM extension via HNL, and then conduct a  brief overview of the experimental prospects in Section~\ref{sec:exp-prospects}.
In Section~\ref{sec:results} we collect our findings - this includes a first exploration of the relevant new parameters, as well as studies of angular distributions; subsequently we carry out a comprehensive survey of the parameter space, further highlighting the prospects for future experiments. 
Our conclusions and final remarks are summarised in Section~\ref{sec:concs}. 
Several appendices offer further information regarding the computation of the asymmetries and characterisation of the NP model under study.

\section{Asymmetries in three-body cLFV decays}\label{sec:Asyms}

As mentioned in the Introduction, we will consider distinct types of asymmetries, in association with the nature of the flavour violation in the cLFV 3-body decays. It is convenient to identify three classes of decays:
\begin{itemize}
    \item [(i)] $\ell_\alpha^+ \to 3 \ell_\beta^+$ (same-flavour final state leptons); 
    \item [(ii-a)] $\ell_\alpha^+ \to \ell_\beta^+ \ell_\gamma^+ \ell_\gamma^-$ (final state composed of distinguishable leptons); 
\item [(ii-b)] $\ell_\alpha^+ \to \ell_\beta^+ \ell_\beta^+ \ell_\delta^-$ (transitions in which the lepton flavour changes by two units).
\end{itemize}
For all cases of 3-body decays, one can define a time-reversal asymmetry, $\mathcal{A}^{\ell_\alpha^+ \to \ell_\beta^+ \ell_\gamma^+ \ell_\delta^-}_T$, as well as two parity-reversal asymmetries, which we denote $\mathcal{A}^{\ell_\alpha^+ \to \ell_\beta^+ \ell_\gamma^+ \ell_\delta^-}_P$ and $\mathcal{A}^{\ell_\alpha^+ \to \ell_\beta^+ \ell_\gamma^+ \ell_\delta^-}_{P^\prime}$. 
For certain cases 
one can further introduce forward-backward asymmetries and angular forward-backward asymmetries, respectively denoted 
$\mathcal{A}^{\ell_\alpha^+ \to \ell_\beta^+ \ell_\gamma^+ \ell_\gamma^-}_{FB}$ and 
$\mathcal{A}^{\ell_\alpha^+ \to \ell_\beta^+ \ell_\gamma^+ \ell_\gamma^-}_{FB, X}$ (with $X=T, P, P^\prime$).

\subsection{Three-body decays: kinematics}
\begin{figure}[h!]
    \centering
    \includegraphics[width=0.45 \textwidth] {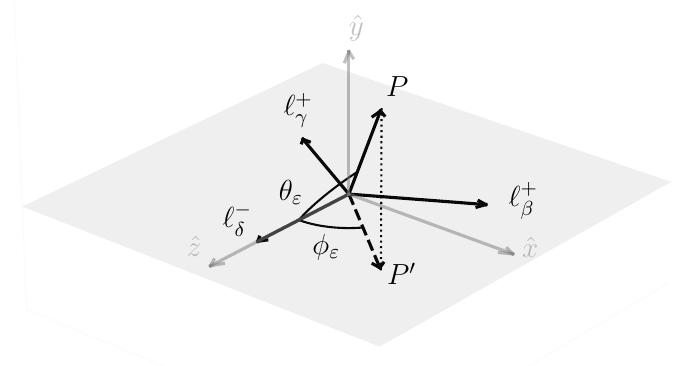}
    \caption{Rest frame of the decaying charged lepton $\ell_\alpha^+$, with $\vec{P}$ denoting its polarisation. }
    \label{fig:ref-frame}
\end{figure}

Leading to the definition of the asymmetries, one must identify a convenient frame, defined by angular and kinematical variables which are relevant for the following  3-body decay 
\begin{equation}\label{eq:genericdecay}
    \ell_\alpha^+ (p) \to 
    \ell_\beta^+ (k_1)\, \ell_\gamma^+ (k_2)\, \ell_\delta^-(k_3)\,.
\end{equation}
As schematically depicted in Fig.~\ref{fig:ref-frame}, and following~\cite{Bolton:2022lrg}, 
in the rest frame of the decaying lepton, one   
defines the angle $\theta_\varepsilon$ to lie between the polarisation of the decaying lepton, and the outgoing negatively charged lepton, and $\phi_\varepsilon$ as the azimuthal angle in the plane perpendicular to the outgoing negatively charged particle. 
This leads to the following conventions for the 4-momenta (as well as polarisation 4-vector of the  decaying particle):
\begin{align}\label{eq:kinematics:convention}
    p^\mu & = (m_\alpha,0,0,0)\,,\nonumber\\ 
    k_1 & = (E_1, k_1^x, 0, k_1^z)\,,\nonumber\\
    k_2 & = (E_2, k_2^x, 0, k_2^z)\,,\nonumber\\
    k_3 & = (E_3, 0, 0, \vert \vec{k_3} \vert)\,,\nonumber\\
    \varepsilon^\mu & = (0,P \sin \theta_\varepsilon \cos \phi_\varepsilon, P \sin \theta_\varepsilon \sin \phi_\varepsilon, P \cos \theta_\varepsilon)\,, 
\end{align}
and we also introduce the usual associated quantities, 
\begin{equation}\label{eq:reducedmomenta}
    s_i = (p - k_i)^2\,, \quad \text{with}\quad
    s_1 + s_2 + s_3  = m_\alpha^2 + m_\beta^2 + m_\gamma^2 + m_\delta^2\,.
\end{equation}
Denoting the polarisation four-vector of the decaying lepton as $(0, \vec{P})$, one has the following expression for the double-differential decay width, decomposed in spin-independent and spin-dependent part
\begin{align}\label{eq:DoubleDiffDec}
    \frac{d \Gamma_{\ell_\alpha^+ \to \ell_\gamma^+ \ell_\beta^+ \ell_\beta^-}}{d \Omega_\epsilon} &= \frac{\Gamma_{\ell_\alpha^+ \to \ell_\gamma^+ \ell_\beta^+ \ell_\beta^-}}{4 \pi} \left[1 + P\left(\mathcal{A}^P_{\ell_\alpha^+ \to \ell_\gamma^+ \ell_\beta^+ \ell_\beta^-}  \cos \theta_\varepsilon \right.\right. \nonumber \\
    &+ \left. \left.\mathcal{A}^{P^\prime}_{\ell_\alpha^+ \to \ell_\gamma^+ \ell_\beta^+ \ell_\beta^-}  \sin \theta_\varepsilon \cos \phi_\varepsilon + \mathcal{A}^T_{\ell_\alpha^+ \to \ell_\gamma^+ \ell_\beta^+ \ell_\beta^-}  \sin \theta_\varepsilon \sin \phi_\varepsilon\right) \right]\,.
\end{align}
While the total decay rate can be obtained by integrating over the entire phase space, the angular coefficients $\mathcal A_{P,P',T}$ can be projected out via asymmetric integrations over the angles as
\begin{align}\label{eq:asym:def}
    \Gamma_{\ell_\alpha^+ \to \ell_\beta^+ \ell_\gamma^+ \ell_\delta^-} & =  \int_0^{2 \pi} d \phi_\varepsilon \int_{-1}^1 d \cos \theta_\varepsilon \int_\Omega d \Omega \,\frac{d \Gamma_{\ell_\alpha^+ \to \ell_\beta^+ \ell_\gamma^+ \ell_\delta^-}}{d \Omega_\varepsilon \,d \Omega}\,,\nonumber\\
    P\,\mathcal{A}_{\ell_\alpha^+ \to \ell_\beta^+ \ell_\gamma^+ \ell_\delta^-}^P & = 
    \frac{1}{\Gamma_{\ell_\alpha^+ \to \ell_\beta^+ \ell_\gamma^+ \ell_\delta^-}} 
    \,\int_0^{2 \pi} d \phi_\varepsilon \left(\int_{0}^1 - \int_{-1}^0 \right) d \cos \theta_\varepsilon \int_\Omega d \Omega\, \frac{d \Gamma_{\ell_\alpha^+ \to \ell_\beta^+ \ell_\gamma^+ \ell_\delta^-}}{d \Omega_\varepsilon \,d \Omega}\,,\nonumber\\
    P\,\mathcal{A}_{\ell_\alpha^+ \to \ell_\beta^+ \ell_\gamma^+ \ell_\delta^-}^{P^\prime} & = \frac{1}{\Gamma_{\ell_\alpha^+ \to \ell_\beta^+ \ell_\gamma^+ \ell_\delta^-}} 
    \,  \left(\int_{-\frac{\pi}{2}}^\frac{\pi}{2} - \int_{\frac{\pi}{2}}^\frac{3 \pi}{2} \right) d \phi_\varepsilon \int_{-1}^1 d \cos \theta_\varepsilon \int_\Omega \,d \Omega \frac{d \Gamma_{\ell_\alpha^+ \to \ell_\beta^+ \ell_\gamma^+ \ell_\delta^-}}{d \Omega_\varepsilon \,d \Omega}\,,\nonumber\\
    P\,\mathcal{A}_{\ell_\alpha^+ \to \ell_\beta^+ \ell_\gamma^+ \ell_\delta^-}^T & = \frac{1}{\Gamma_{\ell_\alpha^+ \to \ell_\beta^+ \ell_\gamma^+ \ell_\delta^-}} 
    \,  \left(\int_{0}^\pi - \int_{- \pi}^0 \right) d \phi_\varepsilon \int_{-1}^1 d \cos \theta_\varepsilon \int_\Omega \,d \Omega \frac{d \Gamma_{\ell_\alpha^+ \to \ell_\beta^+ \ell_\gamma^+ \ell_\delta^-}}{d \Omega_\varepsilon \,d \Omega}\,,
\end{align}
in which $P \equiv \vert \vec{P} \vert$, and 
$\Gamma_{\ell_\alpha^+ \to \ell_\beta^+ \ell_\gamma^+ \ell_\delta^-}$ is the total width of the cLFV 3-body decay.

The details of the computation (as needed to properly parametrise the integration boundaries) are given in Appendix~\ref{app:asym-computation}; 
nevertheless, we highlight here a few important points (particularly relevant for the asymmetries in the decays (i) and (ii-b), in which the final state contains undistinguishable fermions) concerning the integration region in $d\Omega = ds_1 ds_2$. For the integral over $s_1$, one simply has 
\begin{equation}
        s_1^{-}  = (m_\gamma + m_\delta)^2\, \quad \text{and} \quad 
    s_1^{+}  = (m_\alpha - m_\beta)^2\,.
\end{equation}
In the chosen reference frame, one can readily obtain the usual bounds (see, e.g.~\cite{Byckling:1971vca, Workman:2022ynf}),
\begin{align}
    s_2^{\pm} (s_1) & = m_\beta^2 + m_\delta^2 + \frac{(m_\alpha^2 - m_\beta^2 - s_1) \,(s_1 - m_\gamma^2 + m_\delta^2)}{2 \,s_1} \pm \frac{m_\alpha^2}{2}\, \lambda \! \left( \frac{m_\beta^2}{m_\alpha^2}, \frac{s_1}{m_\alpha^2} \right) \,\,\lambda \! \left( \frac{m_\gamma^2}{s_1}, \frac{m_\delta^2}{s_1} \right)\,.
\end{align}
with $\lambda(x,y)$ the reduced Källén function (see Eq.~\eqref{eqn:kallen} in  Appendix~\ref{app:asym-computation}).
We note here, that in the case of same-flavour final states the sign of the azimuthal angle $\phi_\varepsilon$ is not well defined (see, for instance~\cite{Goto:2010sn}).
Without loss of generality (and without loss of phase space), we further restrict the integration region with an energy ordering of the same-charge same-flavour final states as $E_1 \gtrless E_2$ (either ordering being  acceptable), with the energies defined in the rest-frame of the decaying particle as 
\begin{equation}
    E_1 = \frac{m_\alpha^2 + m_\beta^2 - s_1}{2 m_\alpha}\,,
    E_2 = \frac{m_\alpha^2 + m_\gamma^2 - s_2}{2 m_\alpha}\,.
\end{equation}
In the case of opposite-flavour final states, the full integration region must be taken.
However, the energy ordering can be used to further introduce ``forward-backward'' asymmetries by  defining an asymmetric integration region in the Dalitz plane as
\begin{align}\label{eq:Omega:def}
    \int_{\Omega(s_1,s_2)} \to \left(\int_{E_1 > E_2} - \int_{E_1 < E_2}\right)\,,
\end{align}
where the $E_1 \gtrless E_2$ region is defined by Eq.~(\ref{eq:IntegrationRegion}), given in Appendix~\ref{app:asym-computation}.
This allows to introduce new observables which are dubbed $\mathcal A_{FB},\,\mathcal{A}_{FB,P},\,\mathcal A_{FB,P'}$ and $\mathcal A_{FB,T}$.
As mentioned before, a detailed computation of the asymmetries can be found in Appendix~\ref{app:asym-computation}; likewise, a dedicated discussion of the evaluation of the phase space for the distinct cases is also provided. 

\subsection{Angular asymmetries in SM extensions via HNL}
Following the previous discussion, we now present the expressions for the distinct amplitudes considered. As mentioned in the Introduction we work in the framework of a SM extension via 2 heavy sterile fermions. A detailed description of the model (modified interactions, parametrisations and constraints) is given in Appendix~\ref{app:HNL-extensions}.
Likewise, the form factors and loop functions,  which are relevant for cLFV decays in the presence of $n_S$ heavy sterile Majorana states, can be found in Appendix~\ref{app:3body:HNL}. 

For the differently flavoured configurations of the final state charged leptons, the $P$-asymmetries can be cast as
\begin{align}\label{eq:Pasym}
    \mathcal{A}^P_{\ell_\alpha^+ \to \ell_\beta^+ \ell_\beta^+ \ell_\beta^-} & = 
    \frac{\alpha_w^4}{147\, 456 \pi^3} \frac{m_\alpha^5}{M_W^4} \frac{1}{\Gamma_{\ell_\alpha^+ \to \ell_\beta^+ \ell_\beta^+ \ell_\beta^-}} 
    \left[-3 \left| F_{\text{Box}}^{\alpha \beta \beta \beta}\right|^2-40 s_w^4 \left| F_\gamma^{\alpha \beta}\right|^2+\left(-40 s_w^4+48 s_w^2-12\right) \left| F_Z^{\alpha \beta}\right|^2\right. \nonumber \\
    & \left.+48 s_w^4 \left(4 \ln \left(\frac{1}{\hat{\epsilon }^2}\right)-23\right) \left| G_\gamma^{\alpha \beta}\right|^2-48 s_w^2 \text{Re}\left(G_\gamma^{\alpha \beta} \left(F_{\text{Box}}^{\alpha \beta \beta \beta}\right)^*\right)-24 s_w^2 \text{Re}\left(F_\gamma^{\alpha \beta} \left(F_{\text{Box}}^{\alpha \beta \beta \beta}\right)^*\right)\right. \nonumber \\
    & +\left(24 s_w^2-12\right) \text{Re}\left(F_Z^{\alpha \beta} \left(F_{\text{Box}}^{\alpha \beta \beta \beta}\right)^*\right)-96 s_w^4 \text{Re}\left(G_\gamma^{\alpha \beta} \left(F_\gamma^{\alpha \beta}\right)^*\right)\nonumber \\
    & \left. +96 \left(s_w^2-1\right) s_w^2 \text{Re}\left(G_\gamma^{\alpha \beta} \left(F_Z^{\alpha \beta}\right)^*\right) +16 \left(5 s_w^2-3\right) s_w^2 \text{Re}\left(F_\gamma^{\alpha \beta} \left(F_Z^{\alpha \beta}\right)^*\right) \right]\,,\\
    \mathcal{A}^P_{\ell_\alpha^+ \to \ell_\gamma^+ \ell_\beta^+ \ell_\beta^-} & = \frac{\alpha_w^4}{73\, 728 \pi^3} \frac{m_\alpha^5}{M_W^4} \frac{1}{\Gamma_{\ell_\alpha^+ \to \ell_\gamma^+ \ell_\beta^+ \ell_\beta^-}} \left[-3 \left| F_{\text{Box}}^{\alpha \gamma \beta \beta}\right|^2-8 s_w^4 \left| F_\gamma^{\alpha \gamma}\right|^2+\left(-8 s_w^4+12 s_w^2-3\right) \left| F_Z^{\alpha \gamma}\right|^2\right. \nonumber \\
    & \left.+48 s_w^4 \left(2 \ln \left(\frac{1}{\hat{\epsilon }^2}\right)-9\right) \left| G_\gamma^{\alpha \gamma}\right|^2-24 s_w^2 \text{Re}\left(G_\gamma^{\alpha \gamma} \left(F_{\text{Box}}^{\alpha \gamma \beta \beta}\right)^*\right)-12 s_w^2 \text{Re}\left(F_\gamma^{\alpha \gamma} \left(F_{\text{Box}}^{\alpha \gamma \beta \beta}\right)^*\right)\right. \nonumber \\
    & \left.+\left(12 s_w^2-6\right) \text{Re}\left(F_Z^{\alpha \gamma} \left(F_{\text{Box}}^{\alpha \gamma \beta \beta}\right)^*\right)-24 s_w^2 \text{Re}\left(G_\gamma^{\alpha \gamma} \left(F_Z^{\alpha \gamma}\right)^*\right) \right. \nonumber \\
    & \left.
    +4 \left(4 s_w^2-3\right) s_w^2 \text{Re}\left(F_\gamma^{\alpha \gamma} \left(F_Z^{\alpha \gamma}\right)^*\right) \right]\,,\\
    \mathcal{A}^P_{\ell_\alpha^+ \to \ell_\gamma^+ \ell_\gamma^+ \ell_\beta^-} & = - \frac{\alpha_w^4}{49\, 152 \pi^3} \frac{m_\alpha^5}{M_W^4} \frac{1}{\Gamma_{\ell_\alpha^+ \to \ell_\beta^+ \ell_\beta^+ \ell_\beta^-}} \vert F_{\text{Box}}^{\alpha \gamma \gamma \beta} \vert^2 \nonumber \\
    & = -1\,.
\end{align}
Concerning the $P^\prime$-asymmetry, the relevant expressions can be written as 
\begin{align}\label{eq:Pprimeasym}
    \mathcal{A}^{P^\prime}_{\ell_\alpha^+ \to \ell_\gamma^+ \ell_\beta^+ \ell_\beta^-} & = \frac{\alpha_w^4}{26\, 880 \pi^2} \frac{m_\alpha^5}{M_W^4} \frac{1}{\Gamma_{\ell_\alpha^+ \to \ell_\gamma^+ \ell_\beta^+ \ell_\beta^-}}  \left[21 s_w^4 \left| G_\gamma^{\alpha \gamma}\right|^2 - s_w^4 \left| F_\gamma^{\alpha \gamma}\right|^2-s_w^4 \left| F_Z^{\alpha \gamma}\right|^2\right. \nonumber \\
    & \left.+3 s_w^2 \text{Re}\left(G_\gamma^{\alpha \gamma} \left(F_{\text{Box}}^{\alpha \gamma \beta \beta}\right)^*\right)+3 s_w^2 \text{Re}\left(G_\gamma^{\alpha \gamma} \left(F_Z^{\alpha \gamma}\right)^*\right)+2 s_w^4 \text{Re}\left(F_\gamma^{\alpha \gamma} \left(F_Z^{\alpha \gamma}\right)^*\right) \right]\,,\\
    \mathcal{A}^{P^\prime}_{\ell_\alpha^+ \to \ell_\beta^+ \ell_\beta^+ \ell_\beta^-} & = \frac{\alpha_w^4}{40\, 320 \pi^3} \frac{m_\alpha^5}{M_W^4} \frac{1}{\Gamma_{\ell_\alpha^+ \to \ell_\beta^+ \ell_\beta^+ \ell_\beta^-}} \left[- 2 s_w^4 \left| F_\gamma^{\alpha \beta}\right|^2-2 s_w^4 \left| F_Z^{\alpha \beta}\right|^2\right. \nonumber \\
    & \left.+135 \left| G_\gamma^{\alpha \beta}\right|^2+9 s_w^2 \text{Re}\left(G_\gamma^{\alpha \beta} \left(F_{\text{Box}}^{\alpha \beta \beta \beta}\right)^*\right)+12 s_w^4 \text{Re} \left(G_\gamma^{\alpha \beta} \left(F_\gamma^{\alpha \beta}\right)^*\right)\right. \nonumber \\
    & \left.+6 s_w^2 \left( 3 - 2 s_w^2 \right) \text{Re}\left(G_\gamma^{\alpha \beta} \left(F_Z^{\alpha \beta}\right)^*\right)+4 s_w^4 \text{Re}\left(F_\gamma^{\alpha \beta} \left(F_Z^{\alpha \beta}\right)^*\right) \right]\,.
\end{align}
Finally, the $T$-asymmetries are given by
\begin{align}\label{eq:Tasym}
    \mathcal{A}^T_{\ell_\alpha^+ \to \ell_\gamma^+ \ell_\beta^+ \ell_\beta^-} & = - \frac{\alpha_w^4\, s_w^2}{35\, 840 \pi^2} \frac{m_\alpha^5}{M_W^4} \frac{1}{\Gamma_{\ell_\alpha^+ \to \ell_\gamma^+ \ell_\beta^+ \ell_\beta^-}}  \text{Im} \left( \left(F_{\text{Box}}^{\alpha \gamma \beta \beta} + F_Z^{\alpha \gamma}\right) \left(G_\gamma^{\alpha \gamma}\right)^* \right),\\
    \mathcal{A}^T_{\ell_\alpha^+ \to \ell_\beta^+ \ell_\beta^+ \ell_\beta^-} & = - \frac{\alpha_w^4\, s_w^2}{13 \, 440 \pi^3} \frac{m_\alpha^5}{M_W^4} \frac{1}{\Gamma_{\ell_\alpha^+ \to \ell_\beta^+ \ell_\beta^+ \ell_\beta^-}}  \text{Im} \left( \left( 3 F_{\text{Box}}^{\alpha \beta \beta \beta}  + 2 \left(3 - 2 s_w^2 \right) F_Z^{\alpha \beta} + 
    4 s_w^2 F_\gamma^{\alpha \beta}\right) \left(G_\gamma^{\alpha \beta}\right)^* \right)\,.
\end{align}

We notice that these results have been given up to order $\mathcal{O} \left(\hat{\epsilon}^0 \right)$, with $\hat{\epsilon} \equiv m_\delta/m_\alpha$. They offer a satisfactory analytical approximation in the case of electron-to-muon or electron-to-tau mass ratios (typically exhibiting a discrepancy of less than 1\% with respect to the full numerical computation). 
However, in the case of $\hat{\epsilon} = m_\mu/m_\tau$, one must include $\mathcal{O} \left(\hat{\epsilon}^2 \right)$ corrections (or higher) for the forward-backward asymmetries, as the discrepancies 
for the phase space prefactor can be as large as 
$50 \%$ compared to the full numerical calculation (without any approximation). For the sake of brevity, we refrain from presenting here the full result - up to second order in $\hat{\epsilon}$.
We further note here that, in order to avoid such intricacies, our results rely on numerical evaluations of the full expressions without expanding in small mass ratios.

\section{Experimental prospects}\label{sec:exp-prospects}
The strategy of the Mu3e collaboration in their search for $\mu\to 3e$ decays heavily relies on the precise momentum and vertex reconstruction for the final state electrons, in order to suppress backgrounds from the Michel decay ($\mu\to e\nu\bar \nu$) and the rare SM-allowed decay ($\mu\to 3e\nu\bar\nu$).
Furthermore, the polarisation of the initial state muon is precisely known, owing to their production from pion decays.
Since all final state momenta can be accurately reconstructed, Mu3e will be almost trivially able to measure the angular distributions of the decay as well, the only limitation being the amount of recorded events.
If a sufficient number of events is indeed observed, the construction and measurement of the angular observables here proposed and studied is the logical next step.

While the muons in the Mu3e experiment are naturally highly polarised with average polarisations close to $100\%$~\cite{Mu3e:2020gyw}, the situation for $\tau$-leptons is very different.
In particular, the average $\tau$-polarisation measured at the $Z$-pole is only 
$\simeq 15\%$~\cite{ALEPH:2001uca,CMS:2023mgq}, which will strongly limit the experimental observability of the angular asymmetries.
However, at the $Z$-pole, the $\tau$-spins in $Z\to\tau^+\tau^-$ are highly correlated~\cite{ALEPH:1997wux}, such that one can exploit measurements of the $\tau$-spin on the ``tag-hemisphere'' to determine the $\tau$-spin on the ``measurement-hemisphere''.
Specifically, one would expect measurements of cLFV $\tau$-decays always in association with another $\tau$ decaying into a SM-allowed final state such as $\tau^\pm\to\pi^\pm\nu$, $\tau^\pm\to\rho(\to\pi^\pm\pi^0)\nu$, $\tau^\pm\to a_1^\pm(\to\pi^+\pi^-\pi^+)\nu$ and $\tau^\pm\to\ell^\pm\nu\bar\nu$.
All of these decays offer different ``spin-analysing powers'' ranging from $\simeq 30\%$ in the leptonic decays up to $100\%$ in $\tau\to\pi\nu$~\cite{Hagiwara:1989fn,Alemany:1991ki}.
This allows determining the $\tau$-spins on an event-by-event basis (rather than averaging over all events), such that highly-polarised (sub-)samples can be taken, while somewhat sacrificing statistics.
A selection of events with $\tau\to\pi\nu$ at the tag-hemisphere would in principle allow having samples with polarisations reaching close to $\sim100\%$, limited only by the $\tau$-spin correlation of $\sim99\%$, and by the experimental resolution of the $\pi$-momentum.

At lower energies, as at Belle-II~\cite{Belle-II:2018jsg,Banerjee:2022sgf} and at the proposed super-tau-charm-factory (STCF)~\cite{Achasov:2023gey}, similar considerations apply, although further care must be taken concerning the spin-density matrix (with significant $\gamma-Z$ interference~\cite{Banerjee:2022sgf}).
At the STCF, the $\tau$-polarisation could be further enhanced by polarising one or both of the initial electron/positron beams.
While a detailed study of the spin-correlations of the different $\tau$ production and decay modes is beyond the scope of this work, the above considerations underline the technical feasibility of measuring the angular distributions here proposed.

In the end, the biggest limitation to studying  angular distributions will be the actual rates of the cLFV decay, which are impossible to know before a discovery is made.
However, in case of discovery, a rich experimental research programme should be established to thoroughly exploit the additional information accessible via the kinematical structures in the cLFV three-body decays of charged leptons.
The numerical studies carried out in the following section highlight the necessity of full kinematical analyses in order to disentangle the type of NP responsible for such cLFV transitions.

\section{Results and discussion}\label{sec:results}
Before carrying out a comprehensive survey of the prospects for the considered asymmetries in cLFV 
three-body decays, we begin by a first exploratory study concerning the impact 
of the new degrees freedom of the ``3+2'' ad-hoc extensions regarding the observables. Subsequently we will carry out a full phenomenological study, via a comprehensive exploration of the parameter space, taking into account all relevant constraints on these minimal NP models.

\subsection{First exploration of the HNL parameter space}
As detailed in Appendix~\ref{app:HNL-extensions}, 
this minimal extension is characterised by an enlarged spectrum, featuring two additional massive states with masses $m_{4,5}$, and extended leptonic mixings. The unitary $5\times 5$ leptonic mixing matrix can be cast in terms of 10 real mixing angles $\theta_{ij}$ (including the solar, atmospheric and reactor angles), 6 Dirac and 4 Majorana phases (including those possibly present in the would-be PMNS matrix), respectively denoted $\delta_{ij}$  and $\phi_i$.   

In order to assess the role of the new degrees of freedom, we consider some illustrative (benchmark) choices for certain parameters. Relying on previous analysis of cLFV decays in CPV HNL extensions of the SM~\cite{Abada:2021zcm}, we begin by taking the following 
values for the active-sterile mixings:
$s_{14} = -s_{1 5} = 0.0006$, $s_{24} = s_{2 5} = 0.008$, $s_{34} = s_{3 5} = 0.038$ (with 
$s_{ij} \equiv \sin \theta_{ij}$); we also choose 
$\delta_{24} = -\delta_{25} = \pi/6$, setting all other phases to 0. 
In this first step, no phenomenological constraints are taken into account (experimental constraints, etc.); likewise, no requirement is made regarding the experimental sensitivity to the widths of the cLFV decays (i.e. if the latter lie within future experimental reach).

\subsubsection{Distinguishable leptons in the final state - (ii-a)}
\begin{figure}[h!]
\centering
\includegraphics[width=0.48 \textwidth] {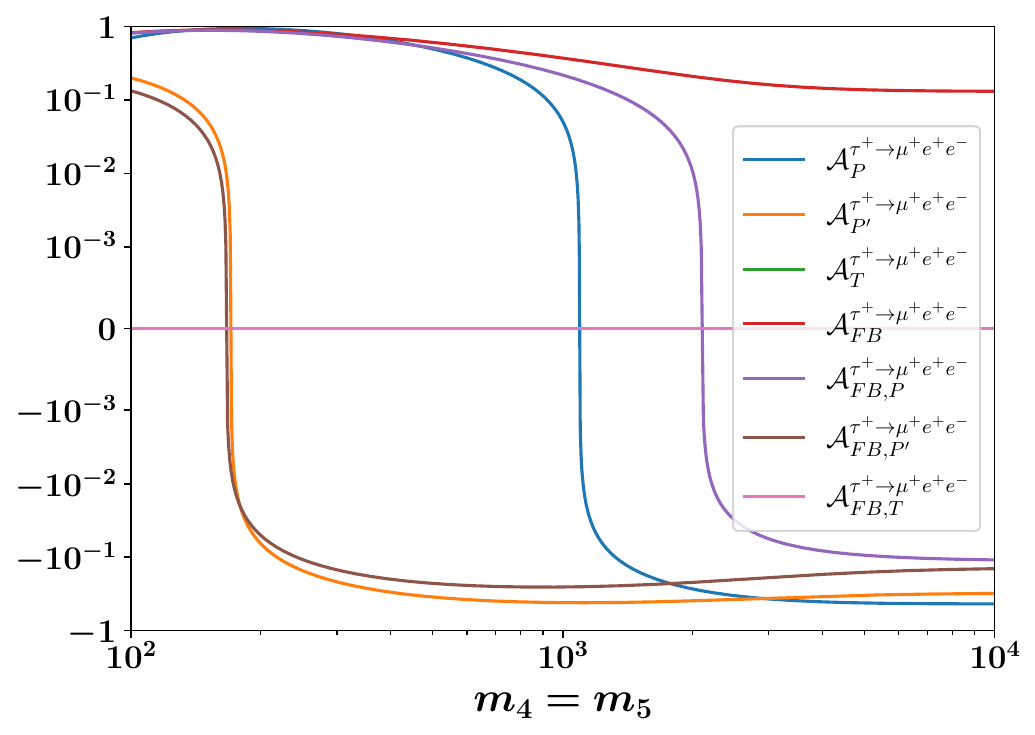}
\caption{Asymmetries for the $\tau^+ \to \mu^+ e^+ e^-$ decay as a function of the degenerate heavy sterile masses $m_{4,5}$ (in GeV). We have chosen 
$s_{14} = s_{1 5} = 0.006$, $s_{24} = -s_{2 5} = -0.02$, $s_{34} = s_{3 5} = 0.036$, with all phases set to 0.
The coloured full lines denote the distinct asymmetries associated with the decay: $P$ (blue), $P^\prime$ (orange), $T$ (green), $FB$ (red),
$FB, P$ (purple), $FB, P^\prime$ (brown) and 
$FB, T$ (pink).
}\label{fig:AsymMass3211:m4m5}
\end{figure}

\begin{figure}[h!]
\centering
\centering
\includegraphics[width=0.48 \textwidth] {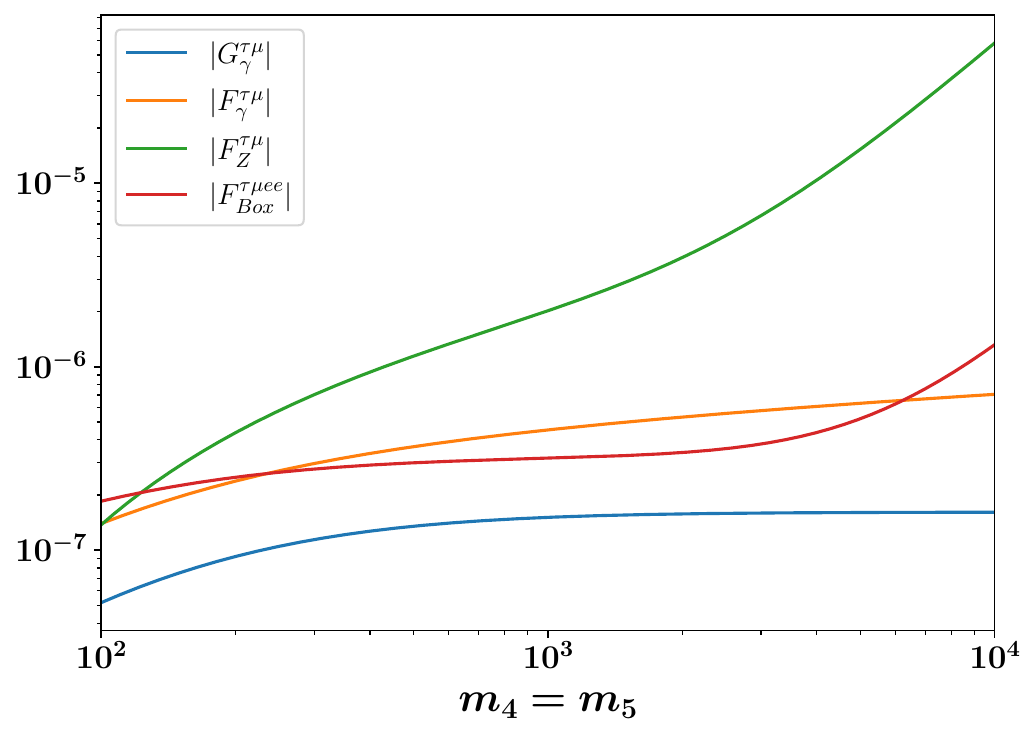}
\caption{Form factors contributing to the 
$\tau^+ \to \mu^+ e^+ e^-$ decay as a function of the degenerate heavy sterile masses $m_{4,5}$ (in GeV): absolute values of the dipole, anapole and $Z$ penguins, as well as the box contributions, respectively associated with  blue, orange, green and red lines.
}\label{fig:FF-AsymMass3211:m4m5}
\end{figure}

We begin by considering the asymmetries associated with $\tau^+ \to \mu^+ e^+ e^-$ decays, and their dependency on the mass of the new heavy mediators 
(for a discussion of the total cLFV decay widths, and in particular the role of phases, see~\cite{Abada:2021zcm}). 
In Fig.~\ref{fig:AsymMass3211:m4m5}, we present the 
dependency of the different asymmetries on the masses of the heavy sterile states which are assumed to be degenerate ($m_4=m_5$). Although the observed behaviours are not straightforward, certain striking 
features can be understood from the expressions of Eqs.~(\ref{eq:Pasym} - \ref{eq:Tasym}), together  
with the dependency of the relevant form factors, which is displayed in Fig.~\ref{fig:FF-AsymMass3211:m4m5}. 

As manifest in Fig.~\ref{fig:AsymMass3211:m4m5},  under these very simplified hypotheses, several interesting points can be raised. 
Firstly, notice that in comparison with the general behaviour of the full decay widths with respect to the heavy fermion mass scale (see, e.g.~\cite{Abada:2021zcm}), the asymmetries exhibit a significantly different dependency. 
The observed behaviour strongly reflects the dominance of different operators (and the interference between them) for the considered mass regimes. 
This can be confirmed from inspection of Fig.~\ref{fig:FF-AsymMass3211:m4m5}, in which we present the absolute values of the contributing form factors, again versus the mass of the (degenerate) heavy states. 
Let us also notice that the $\mathcal{A}_{FB, X}$ closely follow the behaviour of the $\mathcal{A}_{X}$ counterparts, which is a direct consequence of their definition.

\begin{figure}[h!]
\centering
\includegraphics[width=0.48 \textwidth] {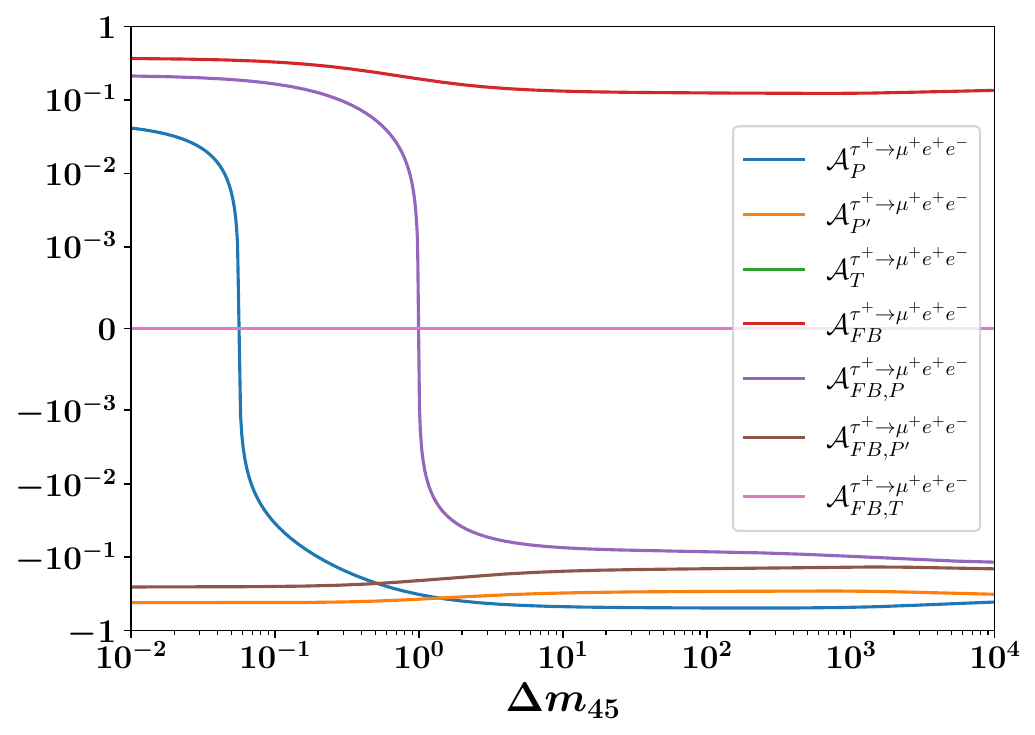}
\caption{Asymmetries for the $\tau^+ \to \mu^+ e^+ e^-$ decay as function of the heavy sterile mass difference $\Delta m_{45} = m_5-m_4$ (in GeV), setting $m_4 = 1$~TeV. 
We have chosen 
$s_{14} = s_{1 5} = 0.006$, $s_{24} = -s_{2 5} = -0.02$, $s_{34} = s_{3 5} = 0.036$, with all phases set to 0.
Line and colour code as in Fig.~\ref{fig:AsymMass3211:m4m5}.
}
\label{fig:l1l2l2AsymMassSplit}
\end{figure}

It is also relevant to consider the role of the changing operator-dominance between the contributions of each of the heavy states. 
In Fig.~\ref{fig:l1l2l2AsymMassSplit}, we display the dependency of the ensemble of the studied asymmetries for the $\tau^+ \to \mu^+ e^+ e^-$ decay, now as a function of the mass difference between the HNL, $\Delta m_{45} = m_5-m_4$.
We consider the same choice for the mixing angles and phases. 
Especially for non-negligible mass splittings (typically above 1~GeV), one can verify that certain asymmetries can significantly change (as is the case of $\mathcal{A}_{P^\prime}$,  $\mathcal{A}_{T}$ (and the $FB$ counterparts); this primarily reflects the evolution from dominant box-contributions to $Z$-penguin dominance (which occurs above 1~GeV).
This is due to the rapidly growing $Z$-penguin contribution in this region ($m_4$ is set to $1\:\mathrm{TeV}$), such that interference patterns between operators in the almost degenerate case are broken and the contribution of the $Z$-penguin dominates.

\begin{figure}[h!]
\centering
\includegraphics[width=0.48 \textwidth] {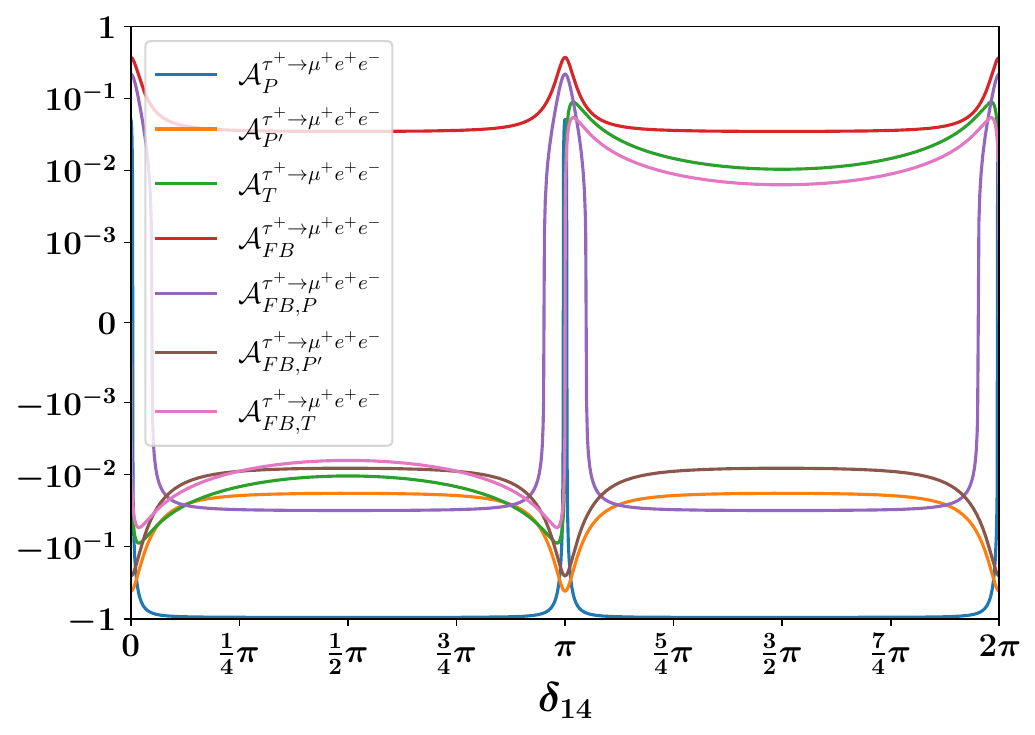}\vspace*{3mm}
\\
\centering
\includegraphics[width=0.48 \textwidth] {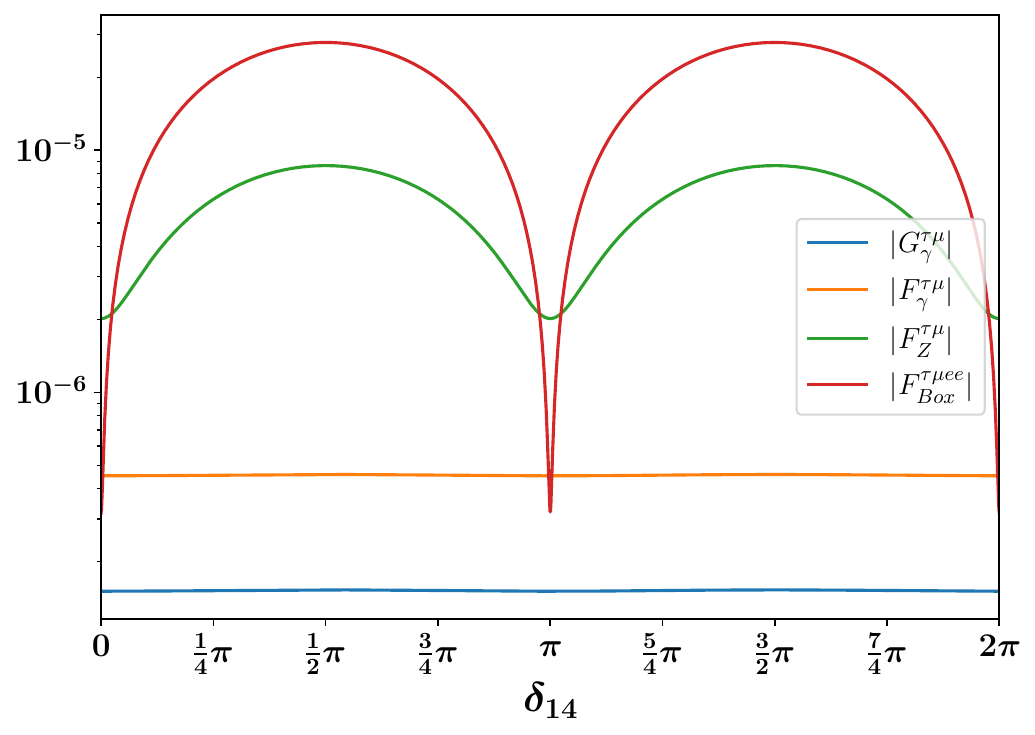}
\hspace*{3mm}
\includegraphics[width=0.48 \textwidth] {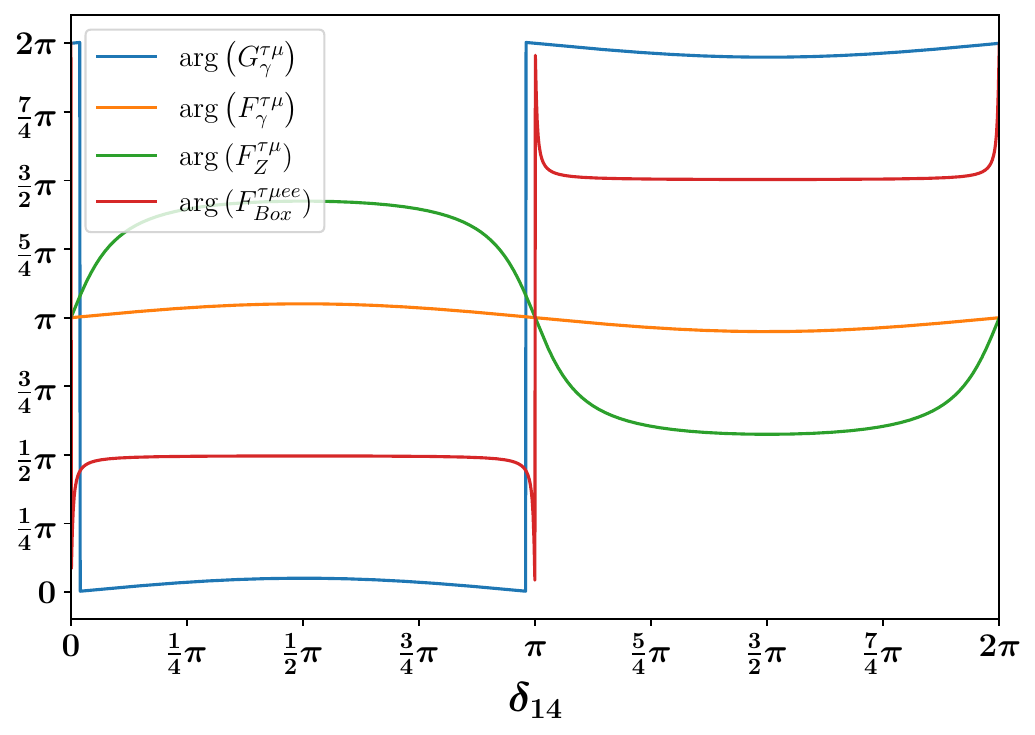}
\caption{
Asymmetries for the $\tau^+ \to \mu^+ e^+ e^-$ decay vs. the Dirac phase, $\delta_{14}$. We choose $s_{14} = s_{1 5} = 0.006$, $s_{24} = -s_{2 5} = -0.02$, $s_{34} = s_{3 5} = 0.036$, with all other phases set to zero, and take $m_4 = m_5 = 1$~TeV.
The top panel displays the distinct asymmetries (with the line and colour code as in Fig.~\ref{fig:AsymMass3211:m4m5}); the lower panels respectively illustrate the dependency of the form factors - absolute value (on the left) and argument (on the right) - on $\delta_{24}$.}
\label{fig:l1l2l2AsymDelta}
\end{figure}

Clearly, the asymmetries are direct probes of new CP violation sources. 
In Fig.~\ref{fig:l1l2l2AsymDelta}, 
we illustrate the dependency of the $\tau^+ \to \mu^+ e^+ e^-$ decay 
asymmetries on a Dirac phase, $\delta_{24}$ which is crucial for the transition (with all other phases, Dirac and Majorana set to zero), again relying on the same choice of mixing angles.  For simplicity we consider degenerate heavy states 
($m_4 = m_5 = 1$~TeV).
While the cyclic pattern of the phases is to be expected (driven by the associated behaviour of the box and $Z$-penguin contributions, see~\cite{Abada:2021zcm}), the most relevant point to be drawn from 
Fig.~\ref{fig:l1l2l2AsymDelta} concerns the impact of a single phase on the considered asymmetries, which exhibit very difference dependencies. 
This simple illustration also allows inferring how complex the overall dependency of the asymmetries will be, once realistic regimes are considered (with all phases generically non-zero).

\begin{figure}[h!]
\centering
\includegraphics[width=0.48 \textwidth] {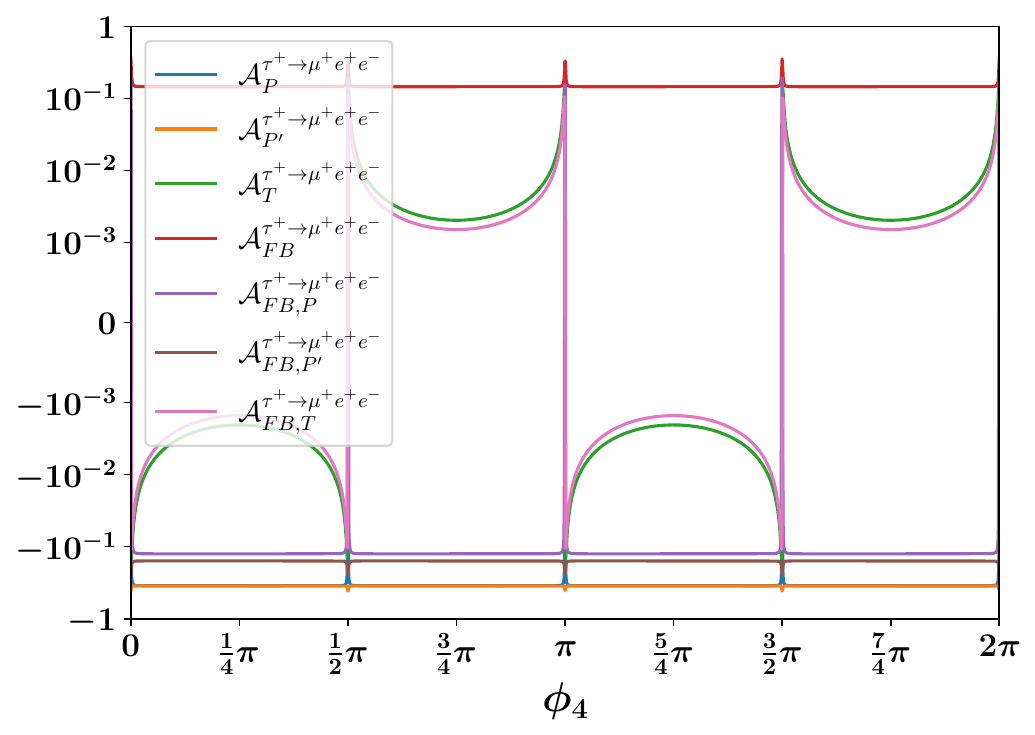}
\vspace*{3mm}\\
\centering
\includegraphics[width=0.48 \textwidth] {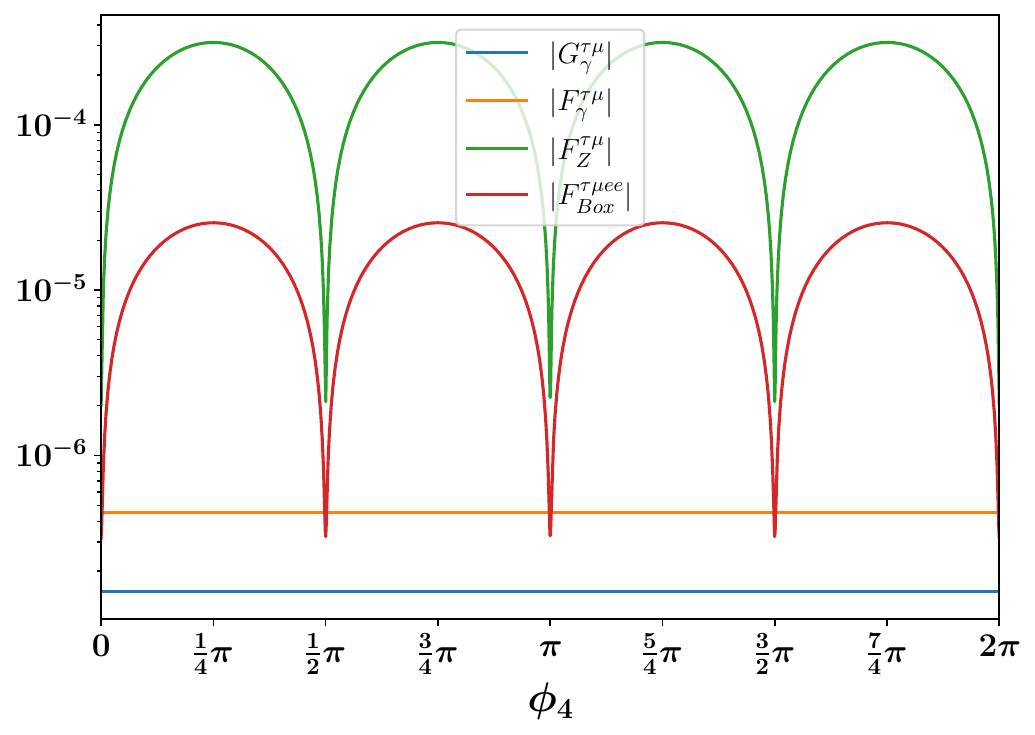}
\hspace*{3mm}
\includegraphics[width=0.48 \textwidth] {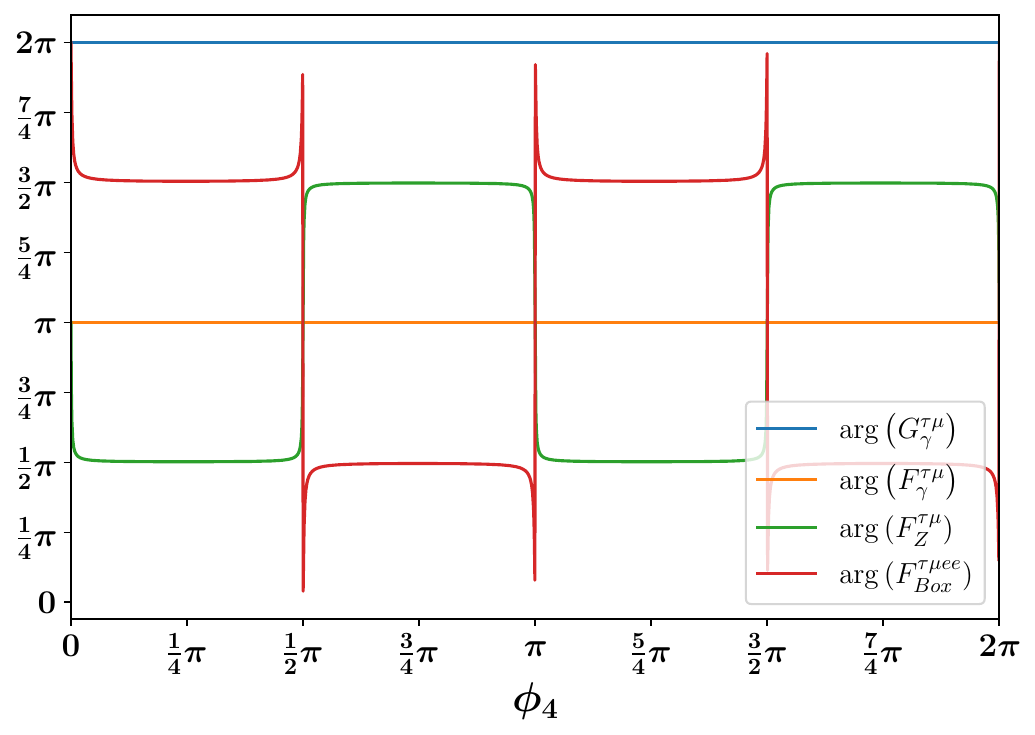}
\caption{
Asymmetries for the $\tau^+ \to \mu^+ e^+ e^-$ decay vs. the Majorana phase, $\varphi_4$. 
We take $s_{14} = s_{1 5} = 0.006$, $s_{24} = -s_{2 5} = -0.02$, $s_{34} = s_{3 5} = 0.036$, with all other phases set to zero, and take $m_4 = m_5 = 1$~TeV.
The top panel displays the distinct asymmetries (with the line and colour code as in Fig.~\ref{fig:AsymMass3211:m4m5}); the lower panels respectively illustrate the dependency of the form factors - absolute value (on the left) and argument (on the right) - on $\phi_4$.
}
\label{fig:l1l2l2AsymPhi}
\end{figure}

The asymmetries in the cLFV 3-body decays are also sensitive to the Majorana phases; we illustrate the dependency on Fig.~\ref{fig:l1l2l2AsymPhi}
in which (in full analogy to Fig.~\ref{fig:l1l2l2AsymDelta})
we display the asymmetries as a function of one the Majorana phases, $\varphi_4$.

\subsubsection{Indistinguishable leptons in the final state - (i) }

Continuing this simplified survey, we now address the case of 
two indistinguishable charged leptons in the final state of the 3-body decay. We begin with the well-known $\ell_\alpha \to 3 \ell_\beta$ decays, in which case only three asymmetries are non-vanishing ($P$, $P^\prime$ and $T$). 

\begin{figure}[h!]
\centering
\includegraphics[width=0.48 \textwidth] {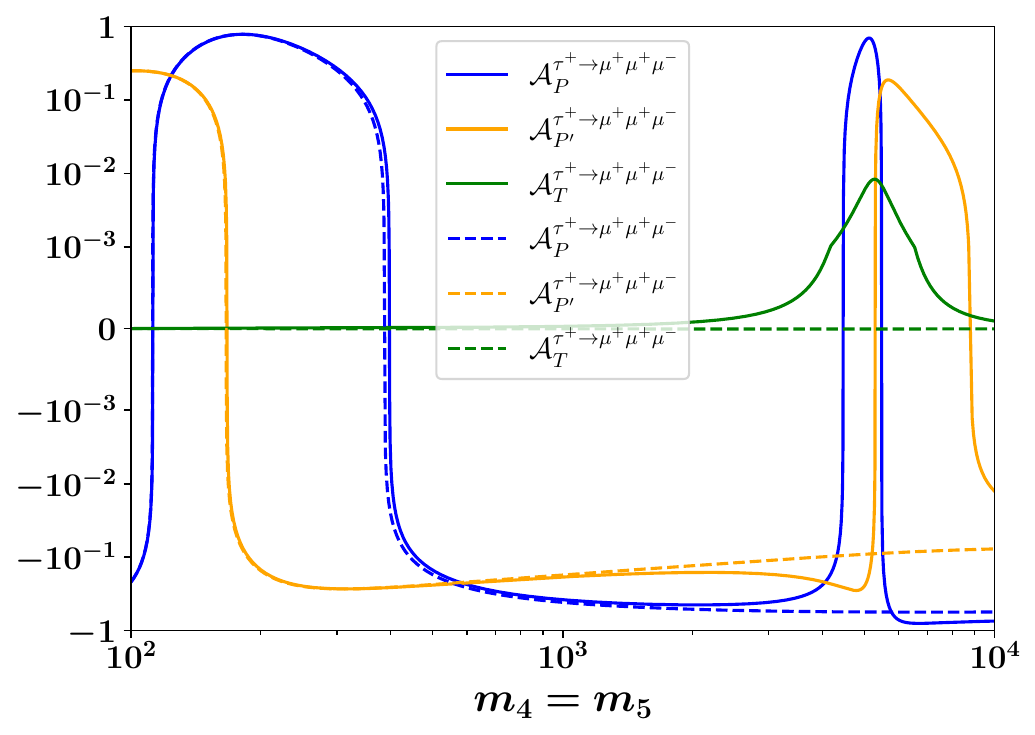}
\hspace*{3mm}
\includegraphics[width=0.48 \textwidth]{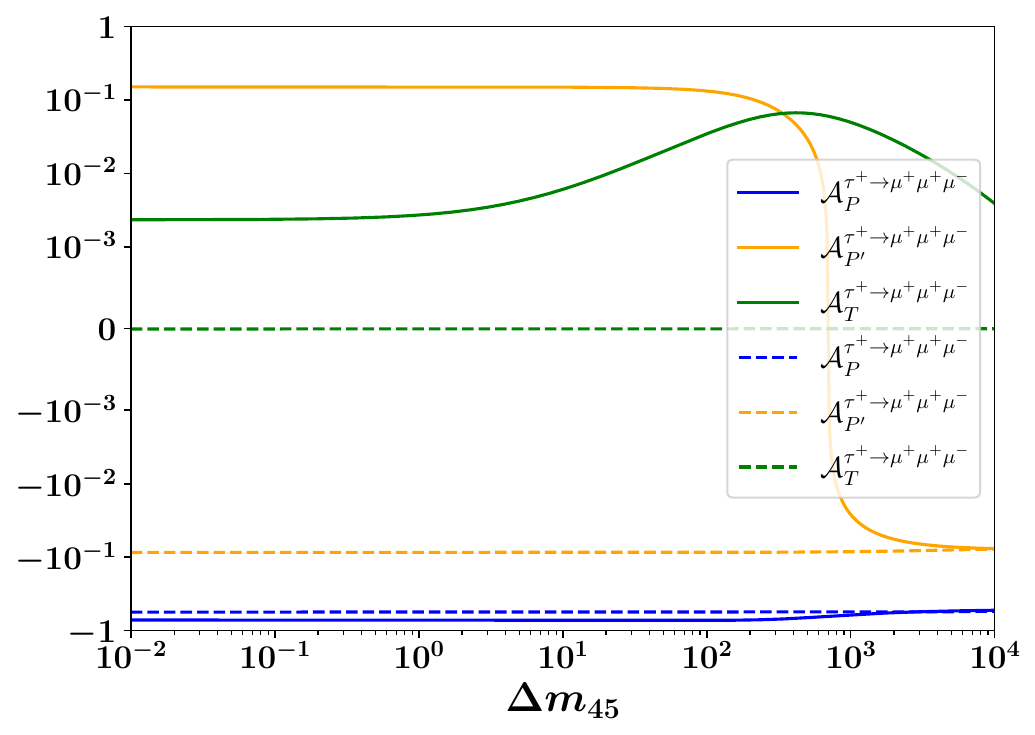}
\caption{On the left, asymmetries for the $\tau^+ \to \mu^+ \mu^+ \mu^-$ decay, as a function of the (degenerate) heavy sterile masses $m_{4,5}$ (in GeV); on the right, as a function of the mass difference between the heavy states, $\Delta m_{45}$, setting $m_4=6$~TeV. We choose $s_{14} = -s_{1 5} = 0.0006$, $s_{24} = s_{2 5} = 0.008$, $s_{34} = s_{3 5} = 0.038$; we have also considered  $\delta_{24} = 2 \pi/3$ (full lines) or $0$ (dashed lines), and $\varphi_4 = 2 \pi/3$, with all other phases set to 0. The $P$, $P^\prime$ and $T$ asymmetries are respectively denoted by blue, yellow and green lines.
}
\label{fig:l3lAsymMass-splitting}
\end{figure}

\begin{figure}[h!]
\centering
\includegraphics[width=0.48 \textwidth] {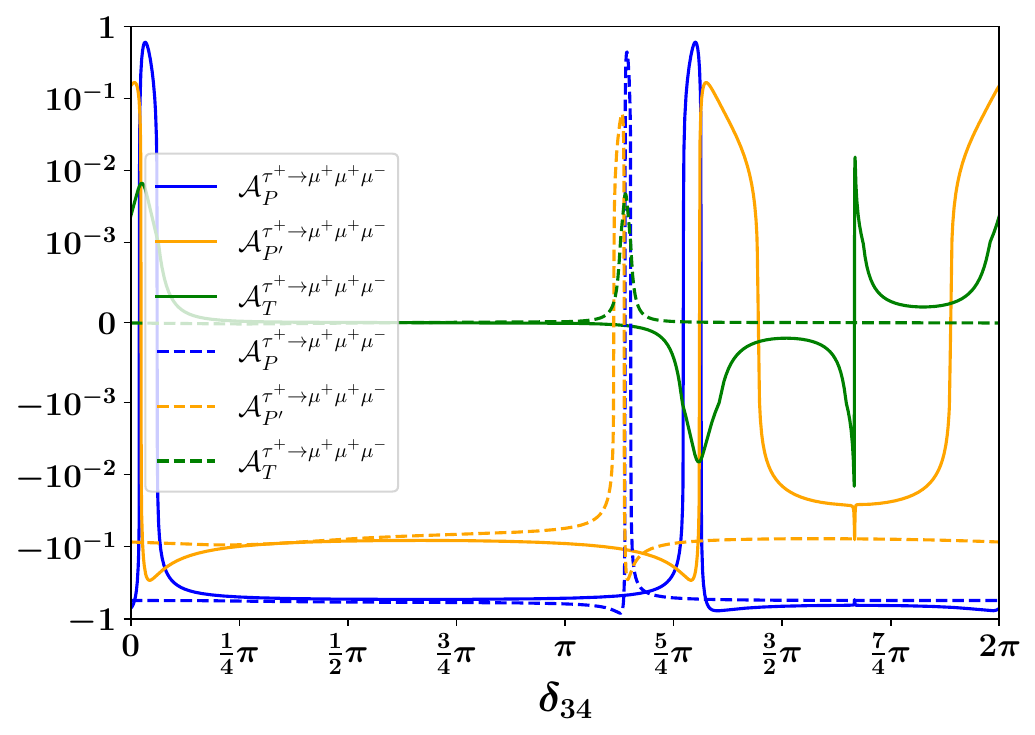}
\includegraphics[width=0.48 \textwidth] {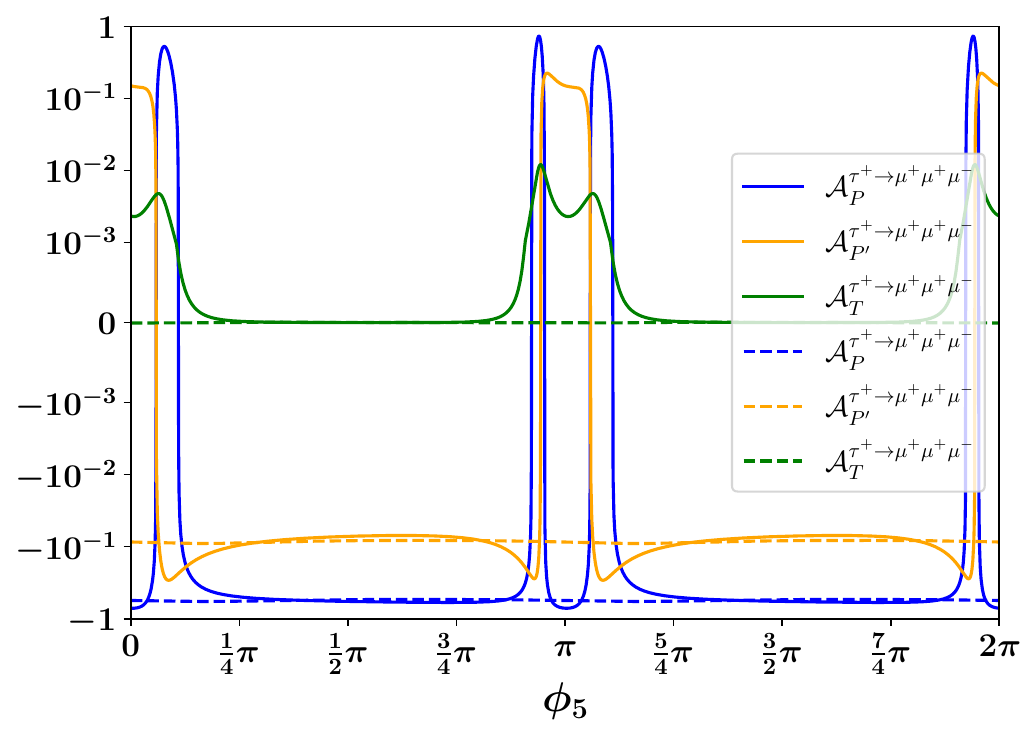}
\caption{On the left, asymmetries for the $\tau^+ \to \mu^+ \mu^+ \mu^-$ decay, as a function of the Dirac phase $\delta_{34}$ with $\delta_{24} = 2 \pi/3$ (or $0$ for dashed lines), with all other phases set to 0. On the right panel, asymmetries as a function of the Majorana phase $\varphi_5$ 
for $\delta_{24} = 2 \pi/3$ (or $0$ for dashed lines) and  $\varphi_4 = 2 \pi/3$ 
(with all other phases set to 0). 
We have set $m_4=m_5=6$~TeV, and $s_{14} = -s_{1 5} = 0.0006$, $s_{24} = s_{2 5} = 0.008$, $s_{34} = s_{3 5} = 0.038$. Colour code as in Fig.~\ref{fig:l3lAsymMass-splitting}.}
\label{fig:l3lAsymDelta-phi}
\end{figure}

For the cLFV $\tau \to 3\mu$ decay,
Fig.~\ref{fig:l3lAsymMass-splitting} displays the dependency of the relevant asymmetries on the mass scale and mass splittings of the heavy states, while Fig.~\ref{fig:l3lAsymDelta-phi} illustrates the impact of a single Dirac and Majorana phase.
While the overall behaviour is not unexpected from the study of the asymmetries in case (ii-a), it is worth considering the differences associated with the interference of the non-vanishing Dirac and Majorana phases (associated with the full lines): although we do not depict the associated form factors here, let us notice that around $6-7$~TeV there is a significant constructive interference of both phases in the $Z$-penguin contributions, which is at the source of the observable ``peaks'' for the asymmetries. 
The other form factors are also sensitive to these effects, with the exception of those arising from box diagrams. 

\begin{figure}[h!]
\centering
\includegraphics[width=0.48 \textwidth] {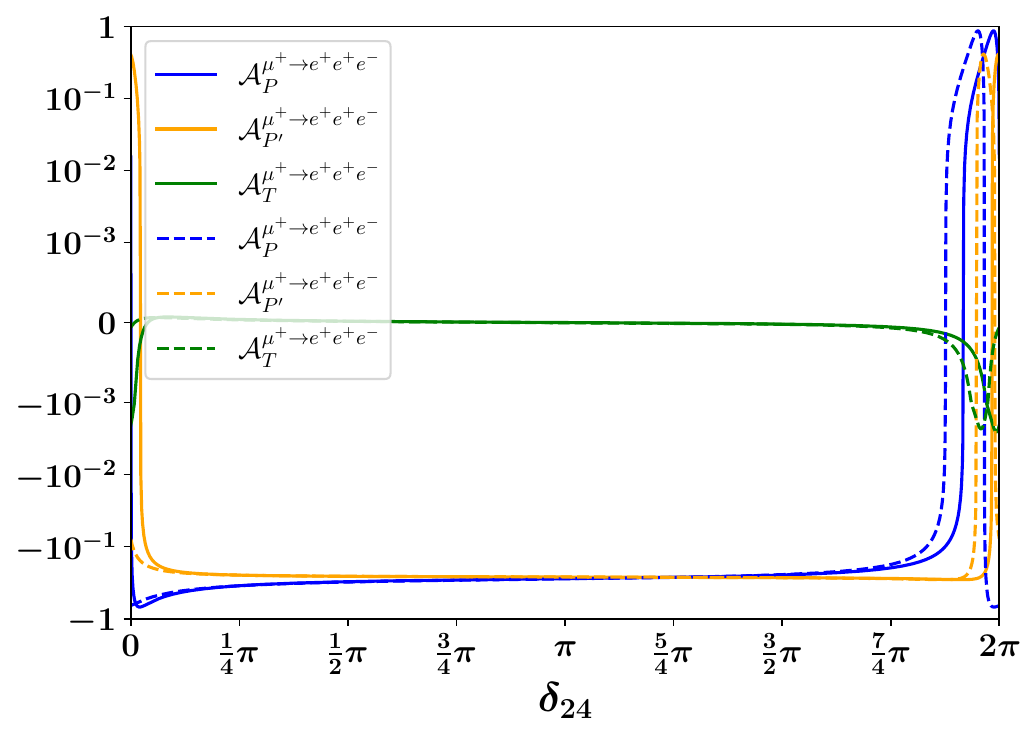}
\includegraphics[width=0.48 \textwidth] {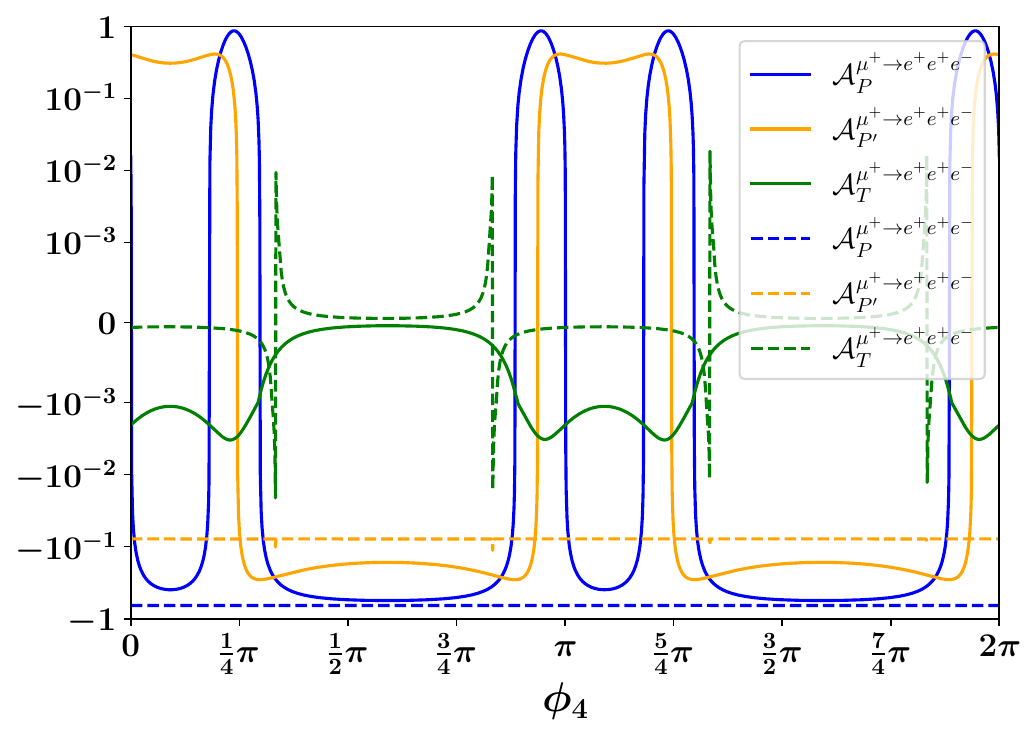}
\caption{On the left, asymmetries for the $\mu^+ \to e^+ e^+ e^-$ decay, as a function of the Dirac phase $\delta_{24}$ with $\delta_{14} = 2 \pi/3$ (or $0$ for dashed lines), with all other phases set to 0. On the right panel, asymmetries as a function of the Majorana phase $\varphi_4$ 
for $\delta_{14} = \pi/32$ (or $0$ for dashed lines) and  $\varphi_5 = 5 \pi/6$ 
(with all other phases set to 0). 
We have set $m_4=m_5=1$~TeV, and taken $s_{14} = -s_{1 5} = 0.0006$, $s_{24} = s_{2 5} = 0.008$, $s_{34} = s_{3 5} = 0.038$. Colour code as in Fig.~\ref{fig:l3lAsymMass-splitting}.}
\label{fig:l3lAsymDelta-phi:mu}
\end{figure}

Even for the apparently simple case of $\ell_\alpha \to 3\ell_\beta$ decays, and still working for an illustrative benchmark point (fixed heavy spectrum), the results displayed in both panels of Figs.~\ref{fig:l3lAsymDelta-phi} and~\ref{fig:l3lAsymDelta-phi:mu} are quite striking. 
The richness of the CPV sources present (even in this very minimal ad-hoc extension) opens the door to surprisingly complex scenarios. 
Each of the displayed asymmetries, $P$, $P^\prime$ and $T$ dramatically changes, from sizeable to near-zero values, further experiencing sign-flips, a consequence of interference between the effects of same-nature phases (Dirac-Dirac and Majorana-Majorana) or cross-natured (Dirac-Majorana). Again, we recall that at this point we are not conducting a phenomenological analysis, and that the results here presented can be in fact experimentally disfavoured or ruled out. 
Once - as we will subsequently do in the comprehensive phenomenological studies of Section~\ref{sec:full-results} - all the phases are taken into account, in association with a variety of active-sterile mixing possibilities, one expects even more complex situations.
The angular asymmetries are extremely sensitive to the phases, and including them as observables in the cLFV set of observables is crucial.

\subsubsection{Double cLFV transitions - (ii-b)}
The final considered case, in which the lepton flavour changes by two units, corresponds to 
$\ell_\alpha^+ \to \ell_\beta^+ \ell_\beta^+ \ell_\delta^-$ transitions, as for example
$\tau^+ \to \mu^+ \mu^+ e^-$. At one-loop, these transitions are exclusively dominated by box diagrams (no other contribution). 
Thus, and as expected, in view of the 
single operator contributions, the considered asymmetries are fixed:
\begin{equation}
    \mathcal{A}^{P}_{\tau^+ \to \ell_\beta^+ \ell_\beta^+ \ell_\gamma^-}\,=-\,1\,,
    \quad
    \mathcal{A}^{P^\prime}_{\tau^+ \to \ell_\beta^+ \ell_\beta^+ \ell_\gamma^-}\,=\, \mathcal{A}^{T}_{\tau^+ \to \ell_\beta^+ \ell_\beta^+ \ell_\gamma^-}\,=\,0\,.
\end{equation}

\FloatBarrier
\subsection{Prospective study of the asymmetries in cLFV three-body decays}\label{sec:full-results}

The previous section offered a preliminary understanding of the role of the different parameters in what 
concerns the studied asymmetries in the cLFV 3-body decays. In what follows, we now consider a full phenomenological study, investigating sizeable regions in parameter space, and allowing for all CPV phases to be simultaneously non-vanishing. Leading to the results presented in this section, the following ranges for the parameters were considered (and randomly scanned over):
\begin{align}\label{eq:scan}
&    \vert \sin (\theta_{1\ 4,5}) \vert \in [6.0 \times 10^{-5}, 6.0 \times 10^{-3}]\,,
    \nonumber\\
&    \vert \sin (\theta_{2\ 4,5}) \vert \in [1.9 \times 10^{-4},0.036]\,,\nonumber\\
&    \vert \sin (\theta_{3\ 4,5}) \vert \in [8.3 \times 10^{-4},0.13]\,,\nonumber\\
   & m_{4}  \in [10^3, 10^4] \ \text{GeV}\,, \quad \quad
    \Delta m_{4 5}  \in [10^{-2}, 5 \times 10^3] \ \text{GeV}\,,\nonumber\\
&    \delta_{\alpha i} \in [0,2 \pi]\,, \quad \quad 
    \varphi_{i} \in [0,2 \pi]\,.
\end{align}
We have also imposed all available constraints on such HNL extensions, including cLFV bounds, among others, as discussed in Appendix~\ref{app:HNL-extensions}. 
Moreover, we incorporate into our study the concept of observability: in other words,  and unless otherwise explicitly stated, one only evaluates or discusses asymmetries which are associated with cLFV rates which lie within future sensitivity ranges of the dedicated facilities. 
To render the discussion more straightforward, this information is presented in Table~\ref{table:BR3body}. 
\renewcommand{\arraystretch}{1.3}
\begin{table}[h!]
    \centering
    \hspace*{-2mm}{\small\begin{tabular}{|c|c|c|}
    \hline
    Observable & Current bound & Future sensitivity  \\
    \hline\hline
    $\text{BR}(\mu \to 3 e)$    &
     \quad $<1.0\times 10^{-12}$ \quad (SINDRUM~\cite{Bellgardt:1987du})    &
     \quad $10^{-15(-16)}$ \quad (Mu3e~\cite{Blondel:2013ia})   \\
    $\text{BR}(\tau \to 3 e)$   &
    \quad $<2.7\times 10^{-8}$ \quad (Belle~\cite{Hayasaka:2010np})&
    \quad $5\times10^{-10}$ \quad (Belle II~\cite{Kou:2018nap})     \\
    $\text{BR}(\tau \to 3 \mu )$    &
    \quad $<1.9\times 10^{-8}$ \quad (Belle II~\cite{Belle-II:2024sce})  &
    \quad $5\times10^{-10}$ \quad (Belle II~\cite{Kou:2018nap})     \\
    & & \quad $1.4 \times 10^{-10}$ \quad (STCF~\cite{Achasov:2023gey})\\
    & & \quad$5\times 10^{-11}$\quad (FCC-ee~\cite{Abada:2019lih})\\
     $\text{BR}(\tau^+ \to e^+ \mu^+ \mu^- )$    &
    \quad $<2.7\times 10^{-8}$ \quad (Belle~\cite{Hayasaka:2010np})  &
    \quad $5\times10^{-10}$ \quad (Belle II~\cite{Kou:2018nap})     \\
    $\text{BR}(\tau^+ \to \mu^+ e^+ e^- )$    &
    \quad $<1.8\times 10^{-8}$ \quad (Belle~\cite{Hayasaka:2010np})  &
    \quad $5\times10^{-10}$ \quad (Belle II~\cite{Kou:2018nap})     \\
    $\text{BR}(\tau^+ \to e^+ e^+ \mu^- )$    &
    \quad $<1.5\times 10^{-8}$ \quad (Belle~\cite{Hayasaka:2010np})  &
    \quad $3\times10^{-10}$ \quad (Belle II~\cite{Kou:2018nap})     \\
    $\text{BR}(\tau^+ \to \mu^+ \mu^+ e^- )$    &
    \quad $<1.7\times 10^{-8}$ \quad (Belle~\cite{Hayasaka:2010np})  &
    \quad $4\times10^{-10}$ \quad (Belle II~\cite{Kou:2018nap})     \\
    \hline
    \end{tabular}}
    \caption{Current experimental bounds and future sensitivities on relevant leptonic cLFV observables. Notice that limits are given at $90\%\:\mathrm{C.L.}$, and that Belle II projected sensitivities rely on an integrated luminosity of $50\:\mathrm{ab}^{-1}$.}
    \label{table:BR3body}
\end{table}
\renewcommand{\arraystretch}{1.}

\begin{figure}[h!]
\centering
\includegraphics[width=0.48 \textwidth] {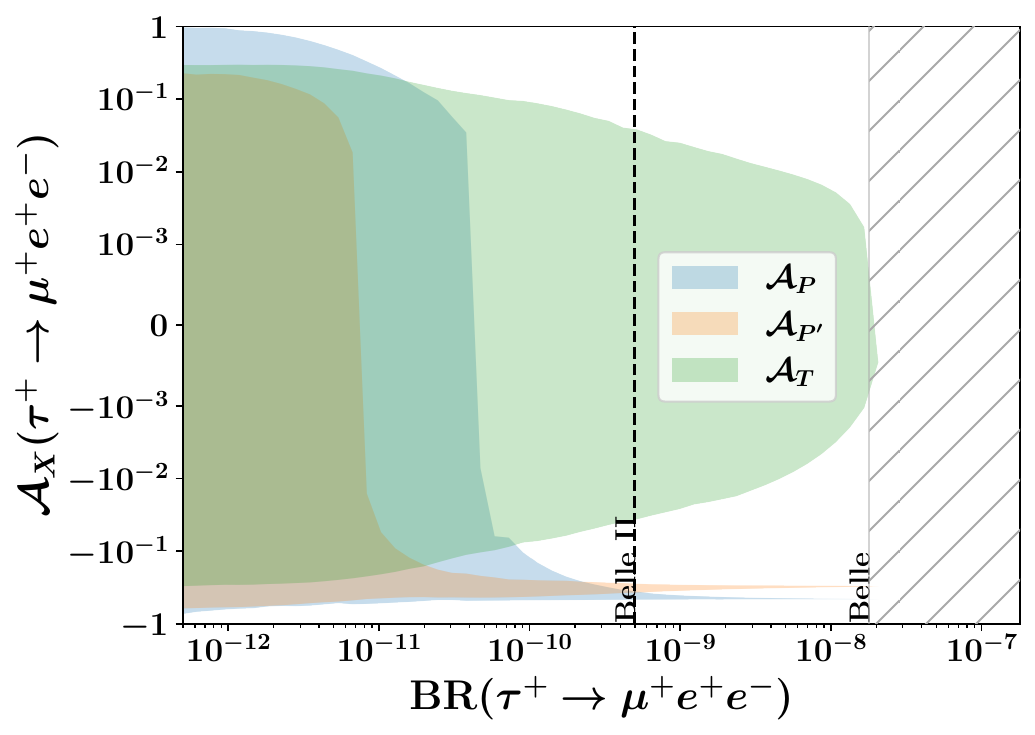}
\hspace*{3mm}
\includegraphics[width=0.48 \textwidth] {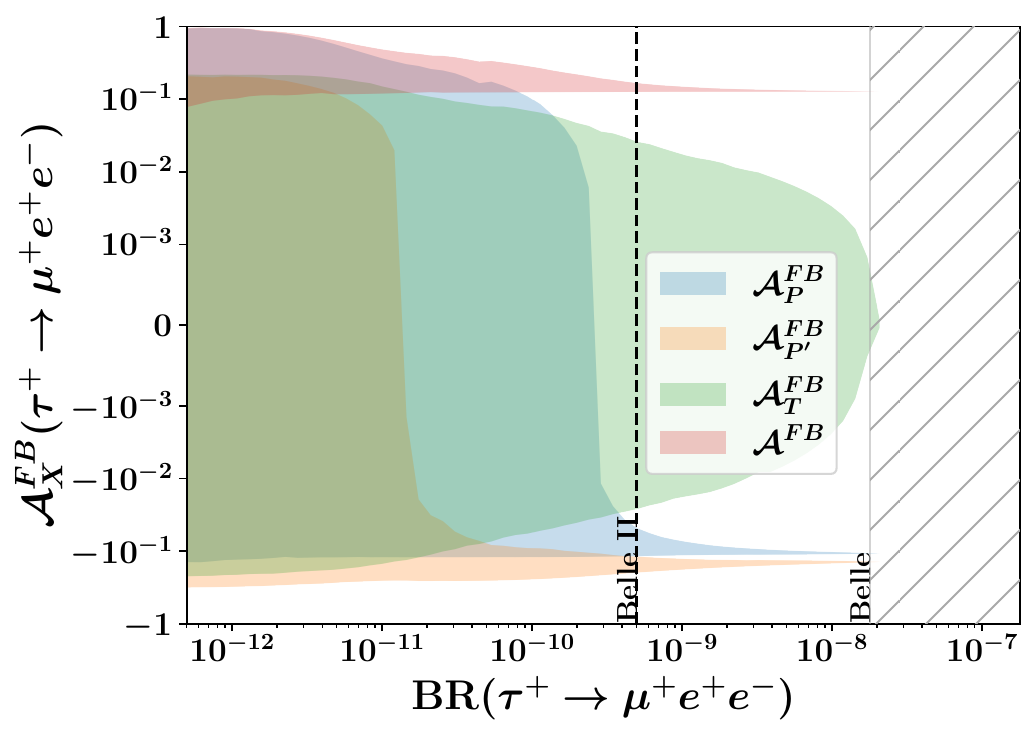}
\\
\includegraphics[width=0.48 \textwidth] {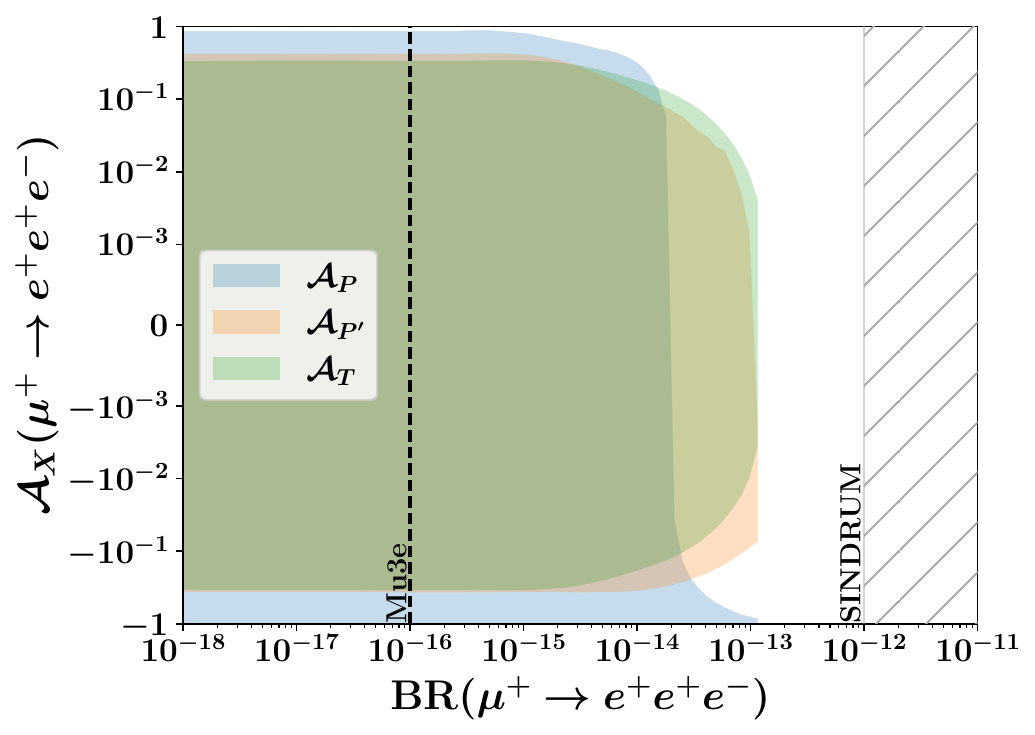}
\hspace*{3mm}
\includegraphics[width=0.48 \textwidth] {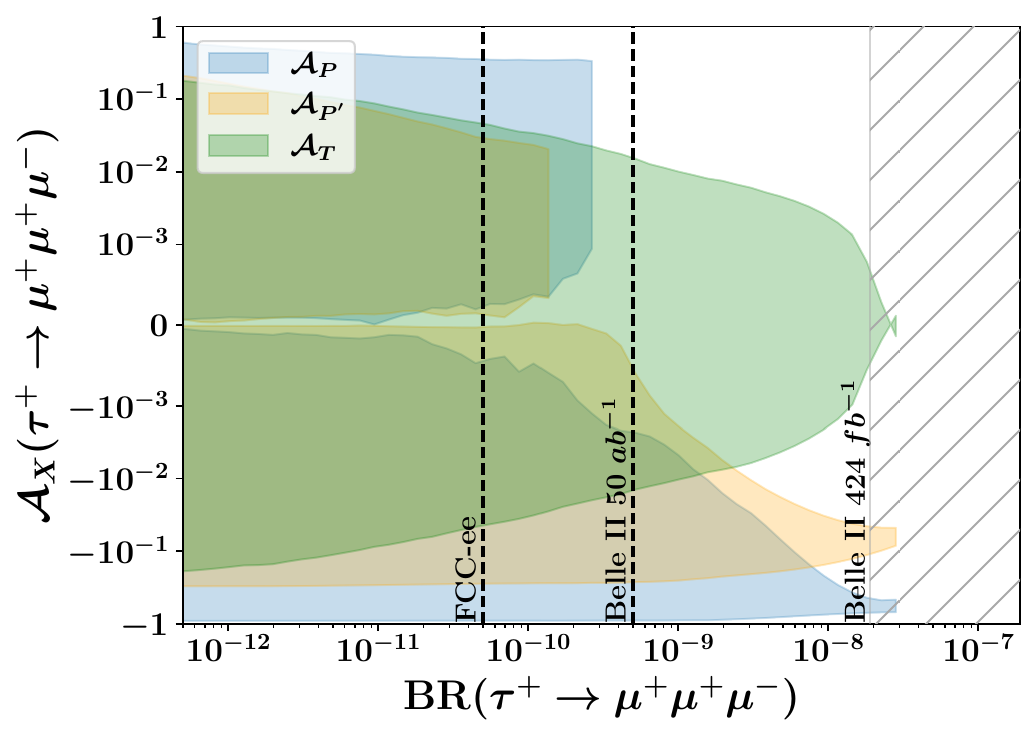}
\caption{Expected ranges of the asymmetries vs. the corresponding branching ratios. On the top row, 
$\tau^+ \to \mu^+ e^+ e^-$ decays (on the left 
$P$, $P^\prime$ and $T$ asymmetries, while on the right the corresponding FB ones). 
On the lower panels, $\mu^+ \to e^+ e^+ e^-$ (left) and $\tau^+ \to \mu^+ \mu^+ \mu^-$ (right). Only points complying with available constraints are displayed.
}
\label{fig:Asyms-BR}
\end{figure}

A first overview of the prospects to experimentally observe the asymmetries associated with the cLFV decay modes is given in Fig.~\ref{fig:Asyms-BR}. We display the $P$, $P^\prime$
and $T$ asymmetries (as well as the FB counterparts when relevant) versus the associated branching ratio of the cLFV decays. Leading to the plots, we have scanned over the ranges summarised in Eqs.~(\ref{eq:scan}); only phenomenologically viable points (in agreement with all imposed constraints)
are shown.

As can be seen for the upper row (corresponding to distinguishable final state leptons, i.e. (ii-a)), with the exception of the $T$-asymmetry, all other asymmetries converge to narrow bands of constant values (around the few percent level), a consequence of the $Z$-penguin dominant contributions. However, notice that for regimes in parameter space leading to rates beyond future sensitivity, several operators  contribute at comparative level, and a considerable spread then occurs for the predicted asymmetries. 
For the $T$ asymmetry of the $\tau^+ \to \mu^+ e^+ e^-$ decays, and although no significant difference can be seen for the upper and lower boundaries, the maximal extrema tend to be associated with values for the decay rate close to the future sensitivity of Belle II (see Table~\ref{table:BR3body}). 

Concerning the $\ell_\alpha \to 3 \ell_\beta$ decays, displayed in the lower row of Fig.~\ref{fig:Asyms-BR}, several comments are in order. The compact areas (or thick bands) associated with the 
$\mu \to 3 e$ asymmetries correspond to the formal extrema of the latter, with increasingly larger spans corresponding to regimes for which there is no clear operator dominance (typically the case for smaller associated branching ratios). 
This is all the more evident for the 
$\tau \to 3 \mu$ asymmetries.
Notice that conflict with current bounds from CR($\mu-e$, Au) preclude larger values for BR($\mu \to 3e$).

For completeness, and from the exploration 
of our data sets, we have numerically identified the extreme values for the distinct asymmetries (independently of experimental considerations). The upper and lower values are summarised in Table~\ref{tab:AsymExtrema}.  
\renewcommand{\arraystretch}{1.3}
\begin{table}[h!]
    \centering
    \begin{tabular}{cc}
    \begin{tabular}{c||c|c}
         & min & max\\ \hline
        $\mathcal{A}^{P}_{\mu^+ \to e^+ e^+ e^-}$ & $-1$ & $0.88$\\ \hline
        $\mathcal{A}^{P}_{\tau^+ \to e^+ e^+ e^-}$ & $-1$ & $0.93$\\ \hline
        $\mathcal{A}^{P}_{\tau^+ \to \mu^+ \mu^+ \mu^-}$ & $-0.89$ & $0.86$\\ \hline
        $\mathcal{A}^{P}_{\tau^+ \to \mu^+ e^+ e^-}$ & $-1$ & $0.97$\\ \hline
        $\mathcal{A}^{P}_{\tau^+ \to e^+ \mu^+ \mu^-}$ & $-0.91$ & $0.96$\\ \hline
        $\mathcal{A}^{P}_{\tau^+ \to \ell_\beta^+ \ell_\beta^+ \ell_\gamma^-}$ & $-1$ & $-1$\\ \hline
 \end{tabular} \hspace*{5mm}
 & \hspace*{5mm}
    \begin{tabular}{c||c|c}
         & min & max\\ \hline
        $\mathcal{A}^{P^\prime}_{\mu^+ \to e^+ e^+ e^-}$ & $-0.37$ & $0.43$\\ \hline
        $\mathcal{A}^{P^\prime}_{\tau^+ \to e^+ e^+ e^-}$ & $-0.35$ & $0.30$\\ \hline
        $\mathcal{A}^{P^\prime}_{\tau^+ \to \mu^+ \mu^+ \mu^-}$ & $-0.30$ & $0.53$\\ \hline
        $\mathcal{A}^{P^\prime}_{\tau^+ \to \mu^+ e^+ e^-}$ & $-0.73$ & $0.29$\\ \hline
        $\mathcal{A}^{P^\prime}_{\tau^+ \to e^+ \mu^+ \mu^-}$ & $-0.65$ & $0.58$\\ \hline
        $\mathcal{A}^{P^\prime}_{\tau^+ \to \ell_\beta^+ \ell_\beta^+ \ell_\gamma^-}$ & $0$ & $0$\\ \hline
  \end{tabular}
  \vspace*{8mm}
\\
   \begin{tabular}{c||c|c}
         & min & max\\ \hline
        $\mathcal{A}^{T}_{\mu^+ \to e^+ e^+ e^-}$ & $-0.35$ & $0.35$\\ \hline
        $\mathcal{A}^{T}_{\tau^+ \to e^+ e^+ e^-}$ & $-0.27$ & $0.27$\\ \hline
        $\mathcal{A}^{T}_{\tau^+ \to \mu^+ \mu^+ \mu^-}$ & $-0.34$ & $0.34$\\ \hline
        $\mathcal{A}^{T}_{\tau^+ \to \mu^+ e^+ e^-}$ & $-0.31$ & $0.31$\\ \hline
        $\mathcal{A}^{T}_{\tau^+ \to e^+ \mu^+ \mu^-}$ & $-0.48$ & $0.48$\\ \hline
        $\mathcal{A}^{T}_{\tau^+ \to \ell_\beta^+ \ell_\beta^+ \ell_\gamma^-}$ & $0$ & $0$\\ \hline
 \end{tabular}\hspace*{5mm}
 &\hspace*{5mm}
\raisebox{-5mm}{  \begin{tabular}{c||c|c}
         & min & max\\ \hline
    $\mathcal{A}^{FB}_{\tau^+ \to \mu^+ e^+ e^-}$ & $0.02$ & $0.96$\\ \hline
        $\mathcal{A}^{FB}_{\tau^+ \to e^+ \mu^+ \mu^-}$ & $-0.03$ & $0.91$\\ \hline
        $\mathcal{A}^{FB\, P}_{\tau^+ \to \mu^+ e^+ e^-}$ & $-0.25$ & $0.95$\\ \hline
        $\mathcal{A}^{FB\, P}_{\tau^+ \to e^+ \mu^+ \mu^-}$ & $-0.35$ & $0.89$\\ \hline
        $\mathcal{A}^{FB\, P^\prime}_{\tau^+ \to \mu^+ e^+ e^-}$ & $-0.36$ & $0.23$\\ \hline
        $\mathcal{A}^{FB\, P^\prime}_{\tau^+ \to e^+ \mu^+ \mu^-}$ & $-0.28$ & $0.39$\\ \hline
        $\mathcal{A}^{FB\, T}_{\tau^+ \to \mu^+ e^+ e^-}$ & $-0.23$ & $0.23$\\ \hline
        $\mathcal{A}^{FB\, T}_{\tau^+ \to e^+ \mu^+ \mu^-}$ & $-0.30$ & $0.30$\\ \hline
    \end{tabular}     } 
    \end{tabular}
    \caption{Values for the extrema of the asymmetries (obtained from the numerical scan).}
    \label{tab:AsymExtrema}
\end{table}
\renewcommand{\arraystretch}{1.}

In the near future, experiments dedicated to searches for cLFV muon transitions are expected to make significant progress, with either more stringent bounds, or hopefully the observation of these SM-forbidden processes. 
It is thus interesting to assess the potential for simultaneous observation of several cLFV quantities: decay (conversion) rates and asymmetries. 
Since - as shown in~\cite{Abada:2021zcm} - the CP violating phases can disrupt expected correlation patterns, increasing the number of explored observables is paramount to inferring whether or not cLFV stems from the presence of HNL. 

\begin{figure}[h!]
\centering
\includegraphics[width=0.48 \textwidth] {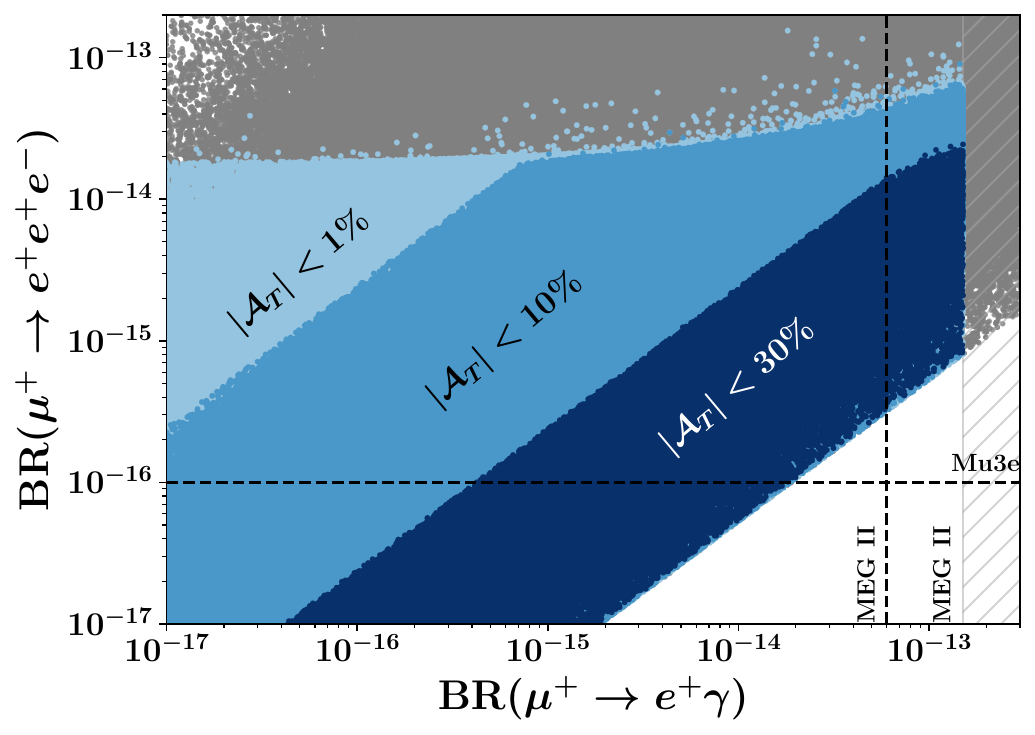}\hspace*{8mm}
\includegraphics[width=0.48 \textwidth] {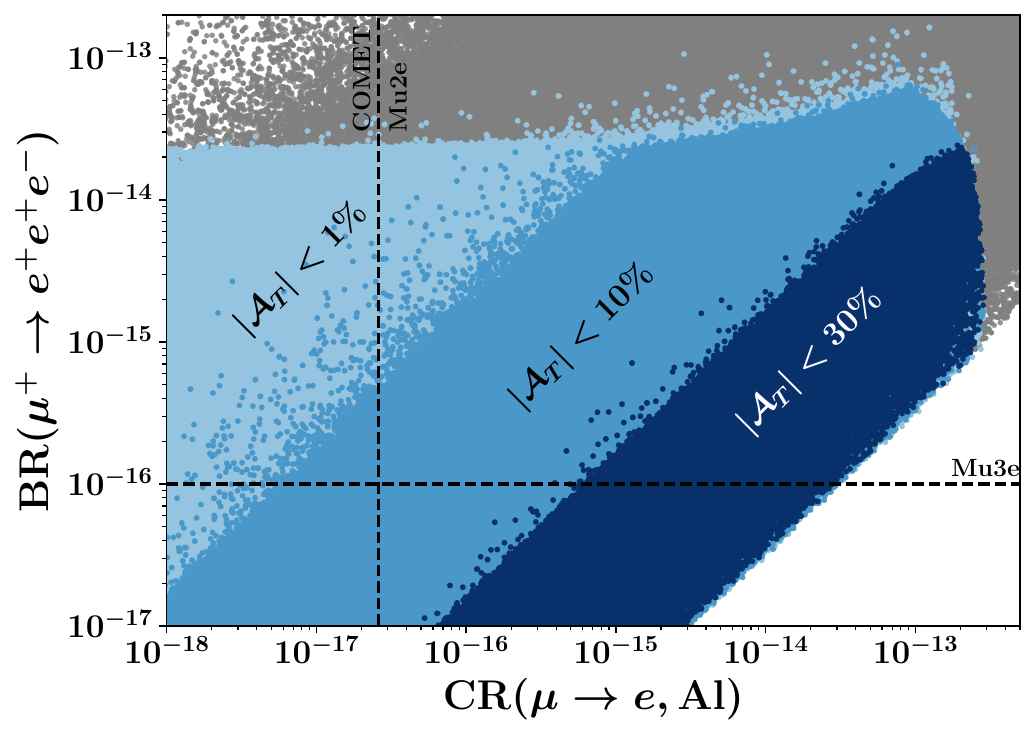}
\caption{Projected values for the $T$-asymmetry in association with $\mu \to 3 e$ decays. On the left, in association with BR($\mu \to 3e$) vs. BR($\mu \to e \gamma$); on the right, 
BR($\mu \to 3e$) vs. CR($\mu-e$, Al). The blue-coloured bands denote regimes for the maximal possible values of the absolute value of the $T$ asymmetry. Grey (and grey-dashed) regions correspond to exclusion due to conflict with experimental bounds. Horizontal and vertical lines denote future sensitivity and current experimental bounds.}
\label{fig:Mue_TAsym}
\end{figure}

In Fig.~\ref{fig:Mue_TAsym} we thus display the expected ranges for the $T$-asymmetry (in association with $\mu \to 3 e$ decays), in the plane spanned by the latter decay, and by either $\mu \to e \gamma$ and the neutrinoless muon-electron conversion in nuclei.
As visible from both panels of Fig.~\ref{fig:Mue_TAsym}, in the regions of future ``observability'' for the muon-electron cLFV processes, one can have values of the $T$-asymmetry as large as\footnote{Notice that the blue coloured bands of Fig.~\ref{fig:Mue_TAsym} correspond to a superposition of darker points on top of lighter regions; for instance, for BR($\mu \to e \gamma$)$\sim \mathcal{O}(10^{-13})$ and 
BR($\mu \to 3e$)$\sim \mathcal{O}(10^{-15})$, the asymmetry can range from 25\% to close to zero.
} 25\%.  
As expected, the diagonal nature of the asymmetry bands reflects the interference of the dipole contribution with more involved four-lepton vector operators (especially $Z$-penguins).

It is crucial to notice how the joint observation of the three cLFV rates and the associated $T$-asymmetry
are instrumental in probing this class of extensions:
as an example, notice that for a future measurement of BR($\mu \to 3e$)$\sim \mathcal{O}(10^{-15})$ associated with $\mathcal{A}^T \sim \mathcal{O}(20\%)$, $\mu-e$ conversion must necessarily be measured with a rate larger than $10^{-14}$.

\subsection{Angular distributions}
\begin{figure}[h!]
\centering
\includegraphics[width=0.48 \textwidth] {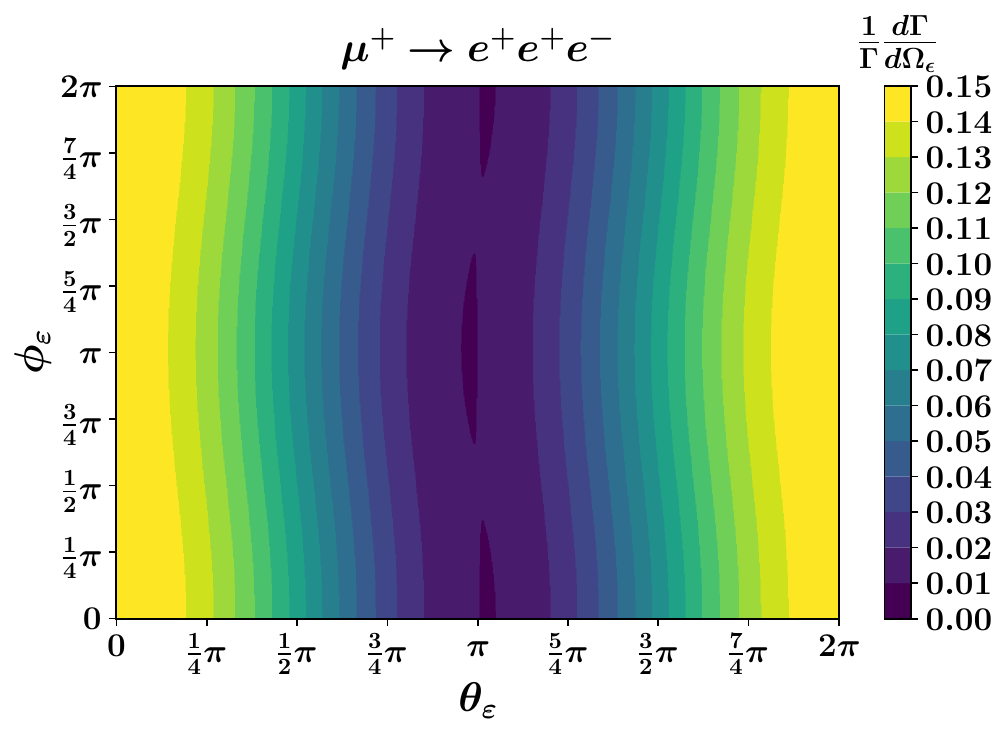}
\\
\includegraphics[width=0.48 \textwidth] {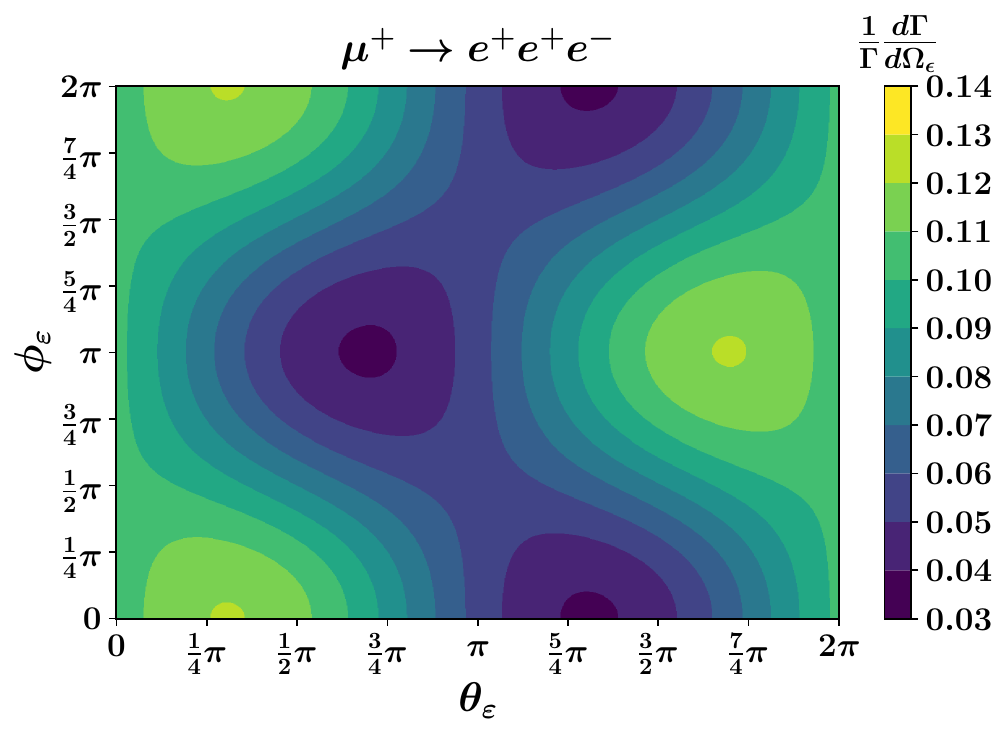}
\hspace*{3mm}
\includegraphics[width=0.48 \textwidth] {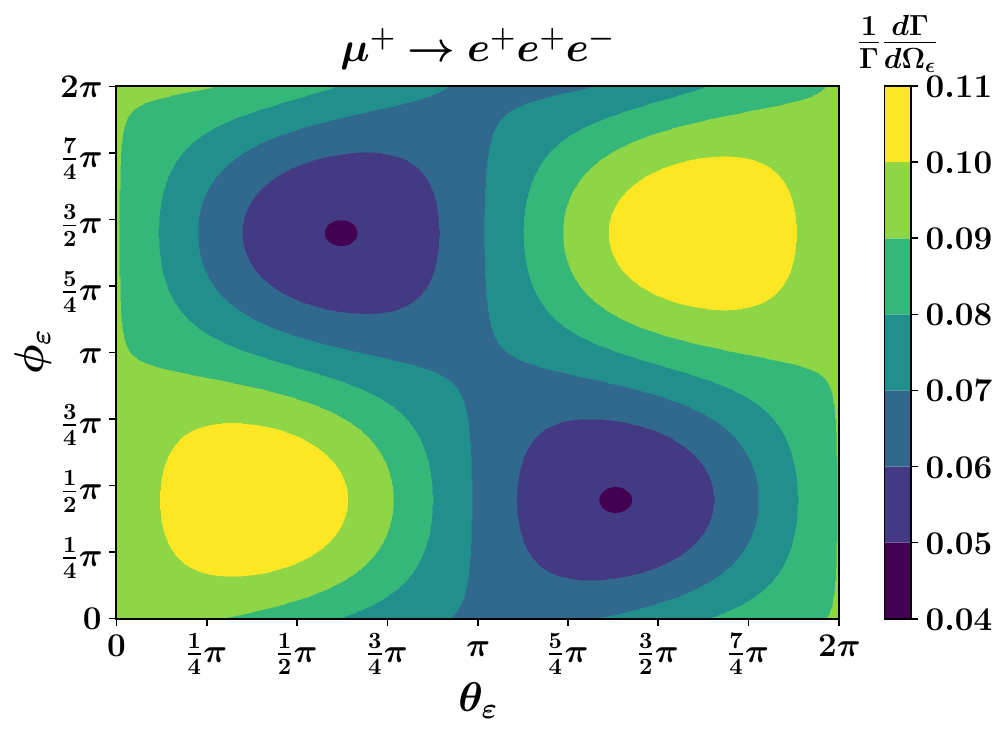}
\caption{Double differential polarised decays assuming an average polarisation of $100 \%$ for the decaying particle in $\mu \to 3 e$ decays. We vary the polarised angle $\theta_\varepsilon$ along the $x$-axis and $\phi_\varepsilon$ along the $y$-axis and represent the double differential polarised decays through coloured contours (see colour palette). 
Different benchmark points were chosen such that they maximise the $P$ (top panel), $P^\prime$ (bottom-left panel), and $T$ asymmetry (bottom-right panel); all lead to cLFV rates within future experimental reach.
}
\label{fig:PolDiffDec}
\end{figure}

In addition to the integrated observables (total cLFV rates and asymmetries), it is instructive to study the full angular distribution of the decays in $\Omega_\varepsilon$.
Depending on the dominant operator (or rather the dominant angular coefficient) at work, one generally expects very distinctive angular distributions.
In Fig.~\ref{fig:PolDiffDec} we present the angular distributions of the electron momentum with respect to the muon spin in $\mu^+\to e^+e^+e^-$ for three different parameter benchmark choices, which respectively maximise the angular coefficients $\mathcal A_{P,P',T}$. The relevant data for each point is given in Table~\ref{table:BP}. 
\renewcommand{\arraystretch}{1.3}
\begin{table}[h!]
    \centering
    \hspace*{-2mm}{\small\begin{tabular}{|c|ccccccccc|ccc|c|}
    \hline
     & $m_4$  & $\Delta m_{45}$ &
    $\theta_{1 i}$ & $\theta_{2 i}$ & $\theta_{3 i}$ & $\delta_{1 i}$ & $\delta_{2 i}$ & $\delta_{3 i}$ & $\varphi_i$  & $\mathcal{A}_P$ & $\mathcal{A}_P^\prime$ & $\mathcal{A}_T$ & {\footnotesize BR($\mu \to 3e$)}\\
    \hline\hline
    P1 & $8$ & 0& $0.0031$ &  $-0.0025$ &  $0.026$ &  $4.5$ &  $5.3$ &  $4.8$ & $0.4$ &\underline{0.87} &0.07 &0.03 & $4.7\times 10^{-16} $ \\
         &     &  & $-0.0031$ &  $0.0025$ &  $0.026$
      &  $4.5$ &  $5.3$ &  $4.8$ &  $1.1$ & & &   
       &  \\
    \hline
    P2 & $2.5$ & 0& $0.0009$ &  $-0.0023$ &  $0.034$ &  $4.5$ &  $3.3$ &  $0.6$ & $5.3$ & 0.29 &\underline{0.49} &0.05 
    & $2.4\times 10^{-16}$\\
         &     &  & $0.0009$ &  $0.0023$ &  $0.034$
      &  $4.5$ &  $-3.3$ &  $-0.6$ &  $4.0$ & & &   &
         \\
    \hline
    P3& $9$ & 0.5& $0.0017$ &  $0.00074$ &  $0.049$ &  $5.1$ &  $0.2$ &  $4.8$ & $2.4$ &0.14 &0.06 &\underline{0.34} & 
    $1.1\times 10^{-15}$\\
         &     &  & $0.0017$ &  $0.0067$ &  $-0.0014$
      &  $5$.9 &  $4.5$ &  $5.5$ &  $5.9$ & & &   
     &    \\
    \hline
\end{tabular}}
    \caption{Benchmark points maximising the different asymmetries (as depicted in Fig.~\ref{fig:PolDiffDec}). For a given point, the upper (lower) lines correspond to (rounded values of) $\theta_{a 4 (5)}$, $\delta_{a 4 (5)}$ and $\varphi_{4(5)}$, with $a=1-3$. Masses and mass splittings in TeV. We underline the maximal values for each of the asymmetries.}
    \label{table:BP}
\end{table}
\renewcommand{\arraystretch}{1.}

As can be seen, the distributions show very distinctive patterns, potentially allowing, should enough statistics be available, for further refined binned analyses beyond the integrated asymmetries.
This could help to eliminate flat directions in effective (Wilson) coefficient space and increase the model discriminating power of the $\ell_\alpha\to\ell_\beta\ell_\gamma\ell_\delta$ systems.

After this brief survey of the angular observables, we will henceforth rely on the integrated asymmetries, as these could be measured even with lower associated statistics.

\subsection{Future prospects for tau-lepton decays}

\begin{figure}[h!]
\centering
\includegraphics[width=0.48 \textwidth] {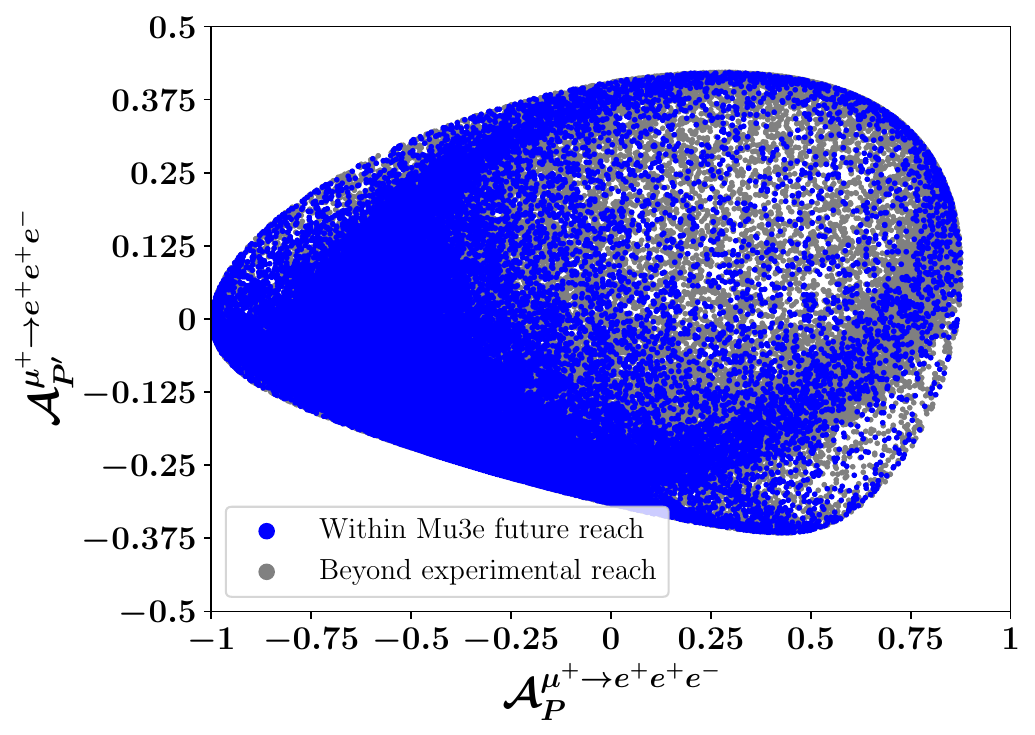}
\hspace*{3mm}
\includegraphics[width=0.48 \textwidth] {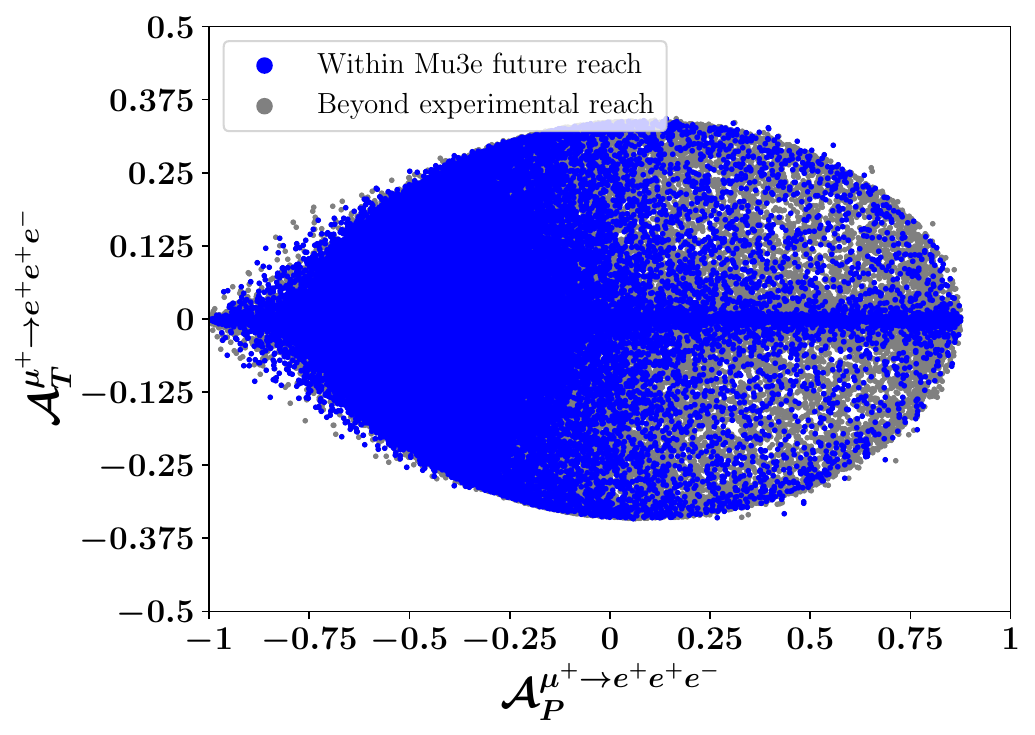}
\\
\includegraphics[width=0.48 \textwidth] {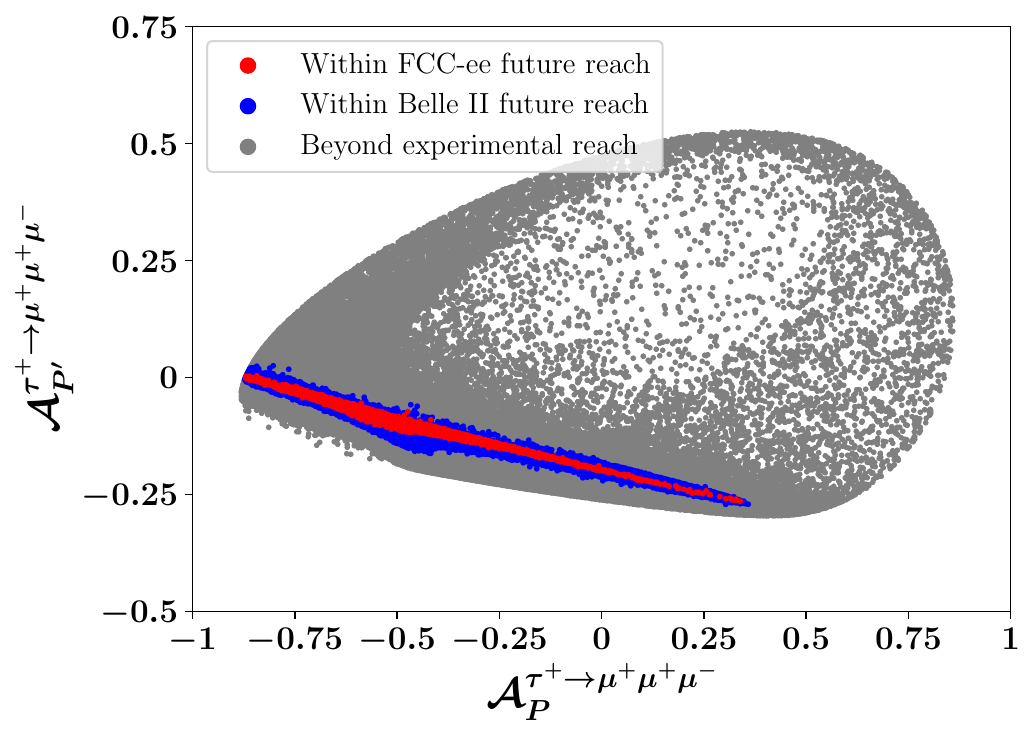}
\hspace*{3mm}
\includegraphics[width=0.48 \textwidth] {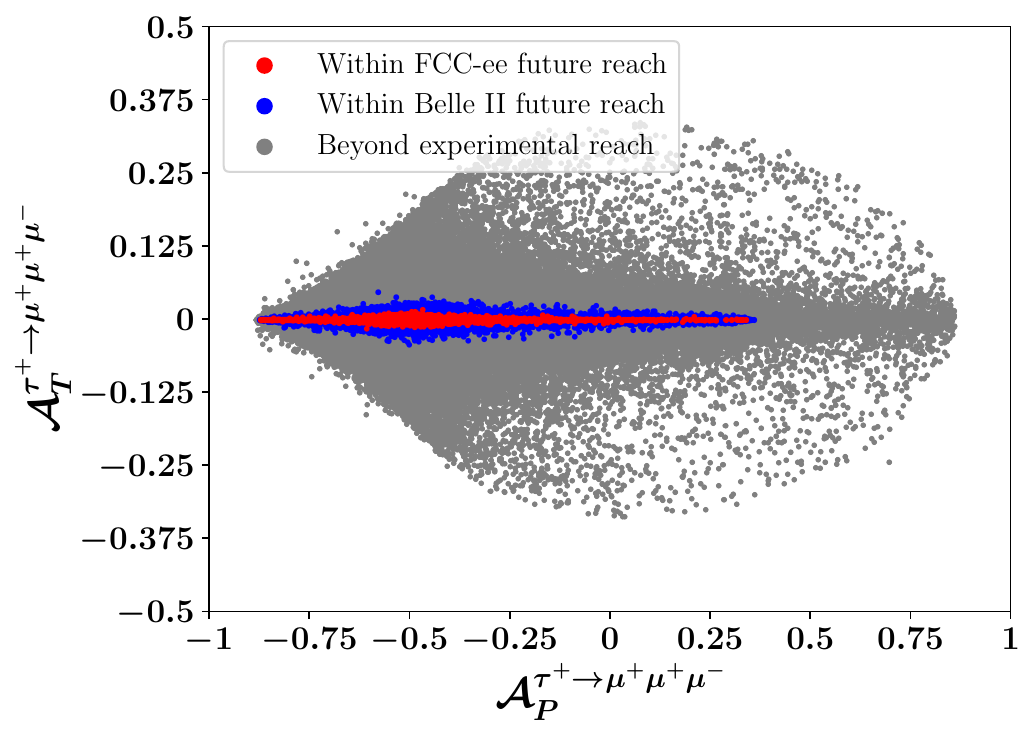}
\caption{Joint prospects for $\mu \to 3 e$ asymmetries (top panels) and for $\tau \to 3 \mu$ 
asymmetries (bottom panels). 
All points displayed are in agreement with current experimental bounds; the blue (grey) points are within (beyond) future sensitivity for the decays under consideration. 
For the case of $\tau$ decay asymmetries, blue points correspond to Belle II's future reach, and red to FCC-ee's expected sensitivity.}
\label{fig:l3l_Asym_vs_Asym}
\end{figure}

A final important aspect to investigate concerns the prospects for the experimental observation of different asymmetries (associated with a unique cLFV decay, or the comparison of asymmetries related to different flavour transitions). 
In order to provide a comprehensive view, in Fig.~\ref{fig:l3l_Asym_vs_Asym} we 
present the ranges of pairs of asymmetries for 
$\tau \to 3 \mu$ and $\mu \to 3 e$ decays. 
We display (phenomenologically allowed) points both within and beyond future sensitivity,  cf.~Table~\ref{table:BR3body}.

From Fig.~\ref{fig:l3l_Asym_vs_Asym}, and
for the case of $\mu \to 3 e$ decays, one readily verifies that the expected future sensitivity of Mu3e offers a good coverage of the full range of the presented asymmetries. (We notice that the grey region encompasses rates from the future sensitivity to values as tiny as $\mathcal{O}(10^{-40})$.) 

The situation is however 
very different when one considers $\tau \to 3\mu$ decays. 
The moderate expected improvement offered by Belle II (and FCC-ee) in what concerns the future sensitivity only allows accessing a very small portion of the asymmetry range\footnote{Recall that the large decay rates that are within future reach correspond to regions of $Z$-penguin dominance.}. 
Nevertheless, the comparatively smaller interval for experimental accessibility concerning the cLFV $\tau$ modes offers the possibility of further testing this minimal HNL ad-hoc extension of the SM, as the ranges for the $P$ and $P^\prime$ ($T$) asymmetries in $\tau \to 3e$ are strongly correlated and quite narrow. 

\begin{figure}[h!]
\centering
\includegraphics[width=0.48 \textwidth] {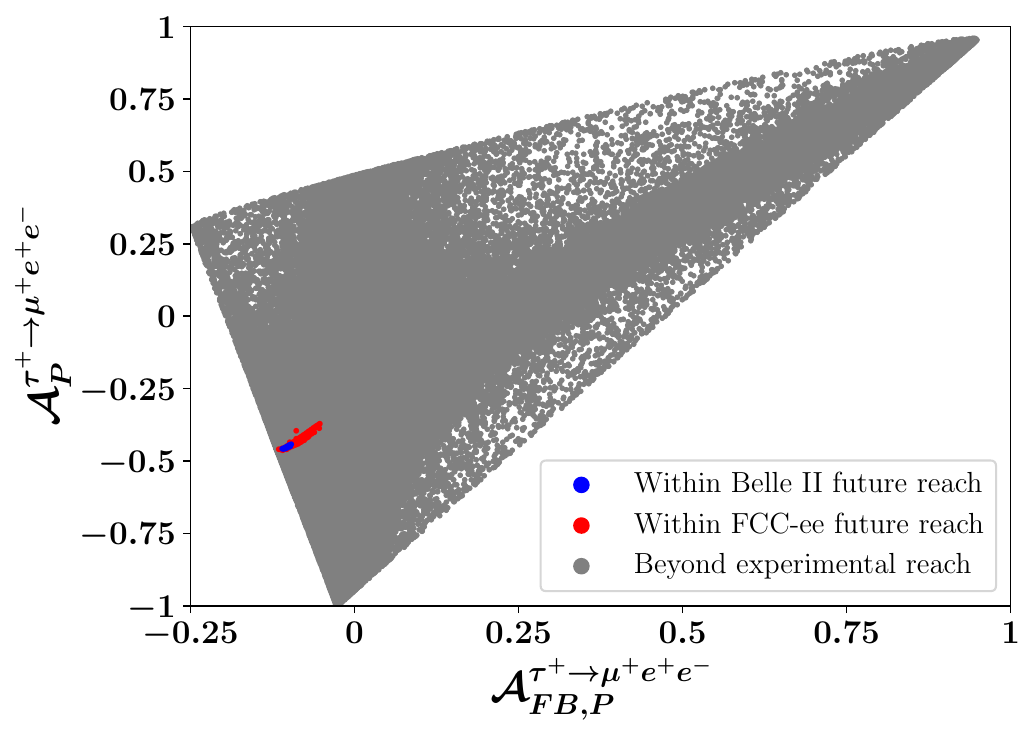}
\hspace*{3mm}
\includegraphics[width=0.48 \textwidth] {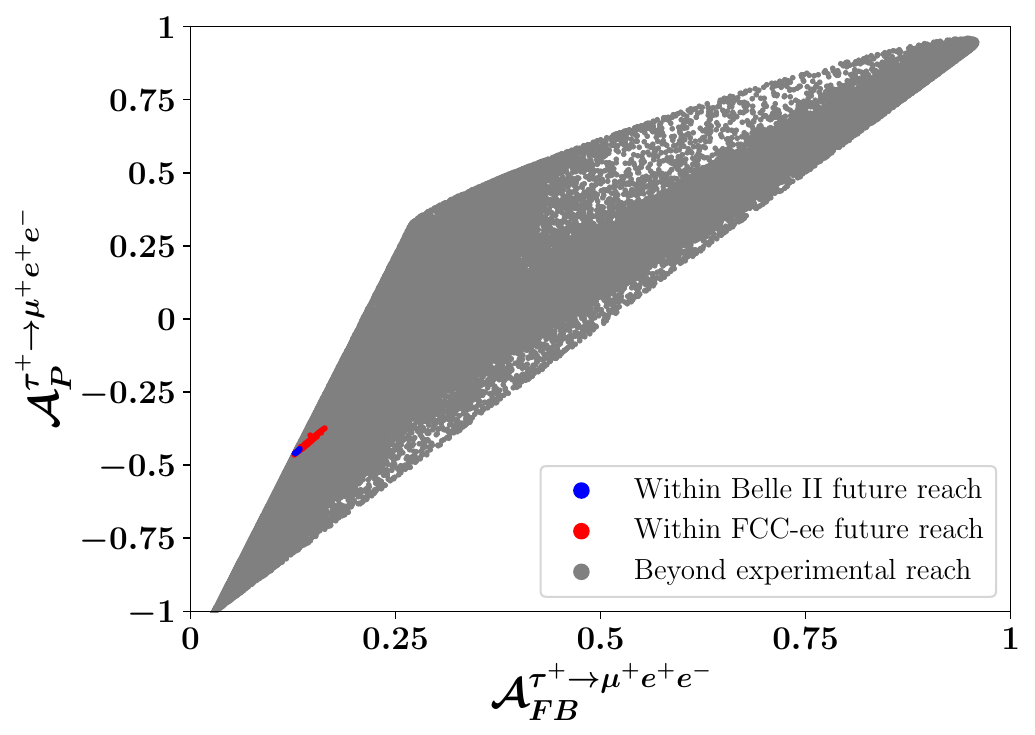}
\\
\includegraphics[width=0.48 \textwidth] {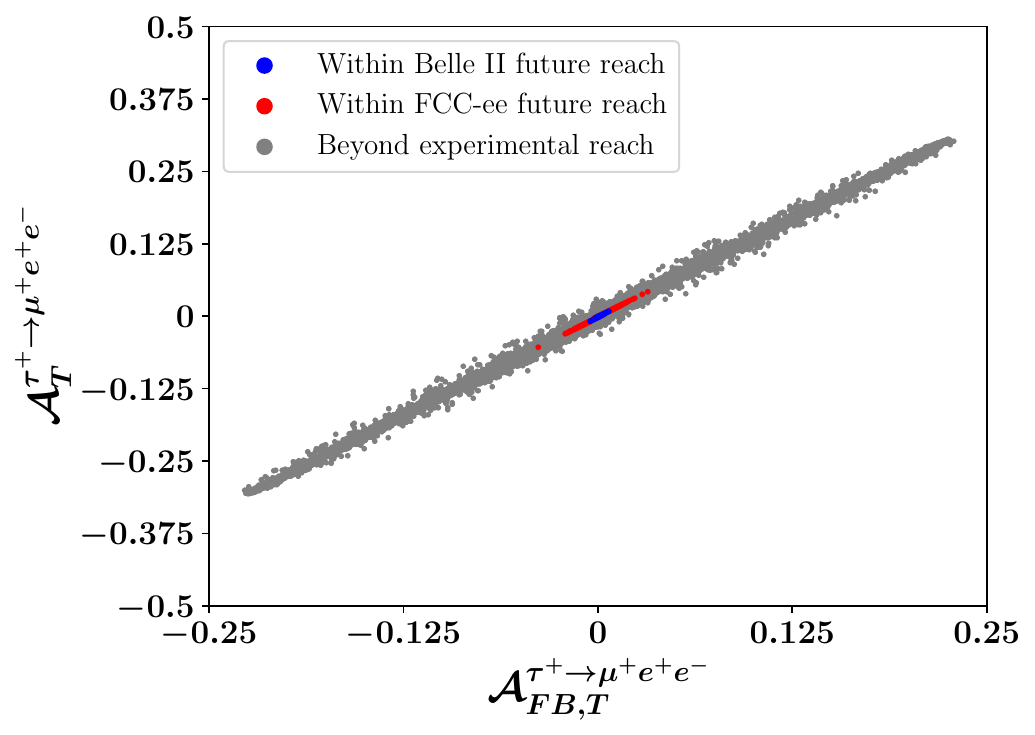}
\hspace*{3mm}
\includegraphics[width=0.48 \textwidth] {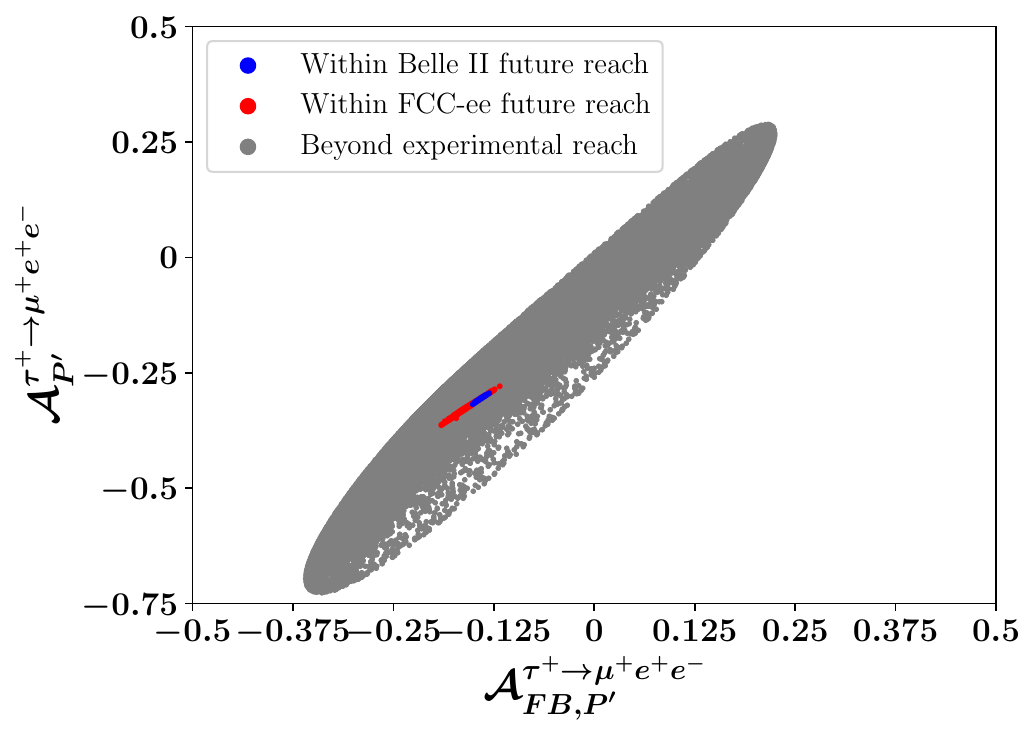}
\\
\includegraphics[width=0.48 \textwidth] {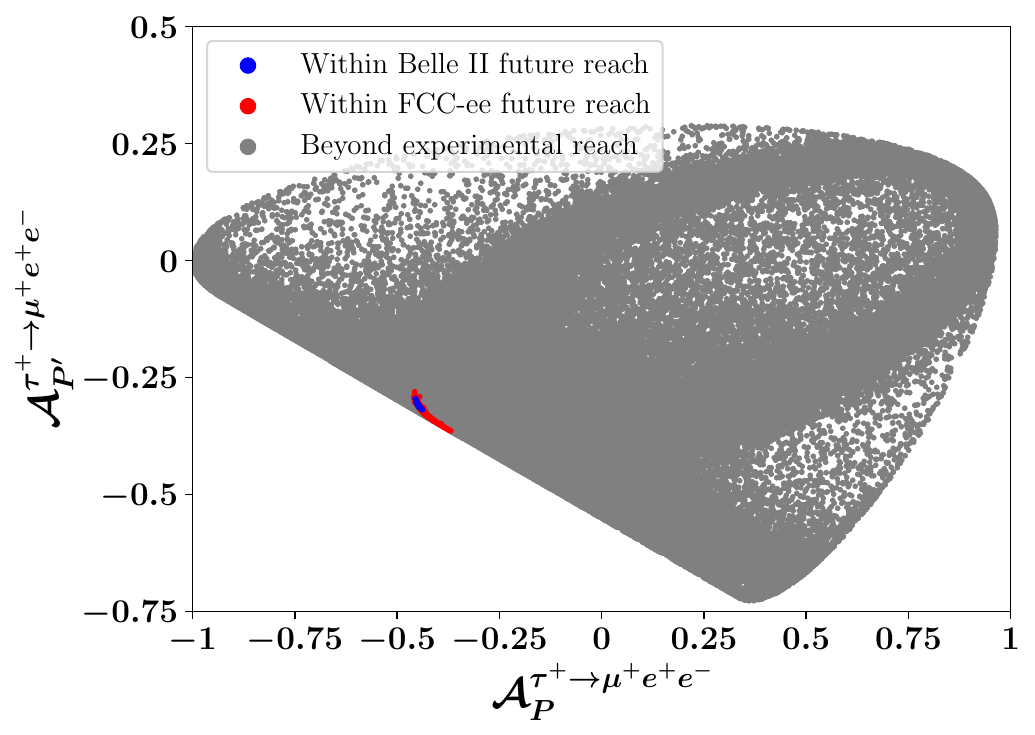}
\hspace*{3mm}
\includegraphics[width=0.48 \textwidth] {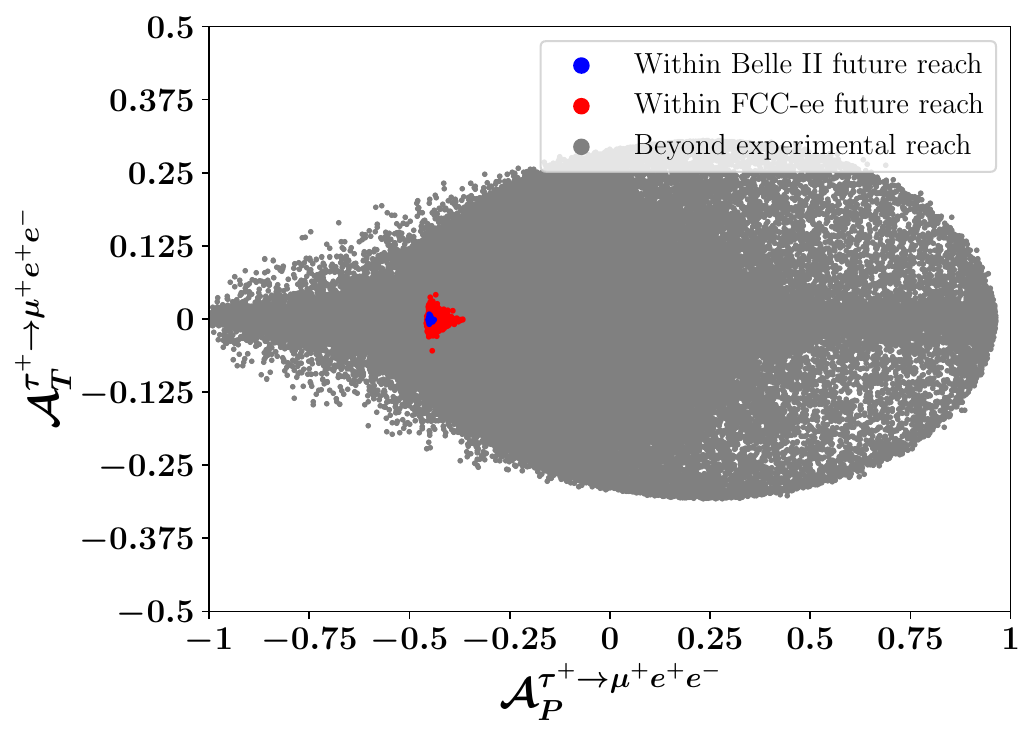}
\caption{Joint prospects for $\tau^+ \to \mu^+ e^+ e^-$ asymmetries.
Colour code as in Fig.~\ref{fig:l3l_Asym_vs_Asym}.
}
\label{fig:TaMuee_Asym_vs_Asym}
\end{figure}

Carrying out the same study, but for $\tau^+ \to \mu^+ e^+ e^-$ decays further exacerbates the conclusions drawn for $\tau \to 3\mu$ decays: this is shown in the panels of Fig.~\ref{fig:TaMuee_Asym_vs_Asym}. 

The asymmetries associated with the $\tau^+ \to \mu^+ e^+ e^-$ mode
(and in general 
$\tau^+ \to \ell_\beta^+ \ell_\gamma^+ \ell_\gamma^-$ decays) reinforce the previous findings: within the experimentally accessible regimes (cf. Table~\ref{table:BR3body}), the predictions cover a tiny fraction of the full range for the asymmetries\footnote{We also notice that from both Figs.~\ref{fig:l3l_Asym_vs_Asym} and~\ref{fig:TaMuee_Asym_vs_Asym}, one verifies that the asymmetries are confined to regions bounded by the extrema listed in Table~\ref{tab:AsymExtrema}.}, to the extent that in most cases this minimal model becomes predictive (and thus falsifiable). 
 
\begin{figure}[h!]
\centering
\includegraphics[width=0.48 \textwidth] {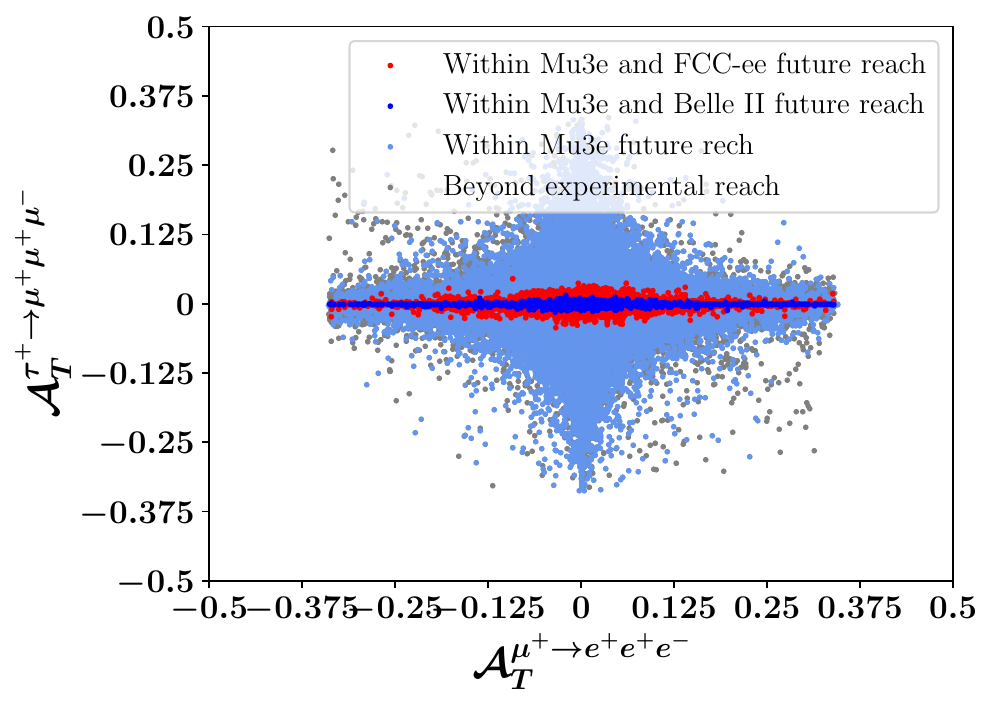}
\caption{Prospects for $T$-asymmetries in association with  
$\ell_\alpha \to 3 \ell_\beta$ decays. 
All points displayed are in agreement with current experimental bounds; light blue points are within future Mu3e experimental reach, while dark blue points correspond to having both decays within future expected reach (Mu3e and Belle II). Finally, red points correspond to expected sensitivity of FCC-ee to cLFV tau decays, still within Mu3e reach. Grey points lie beyond any future experimental sensitivity.
}
\label{fig:Tasym:mu3e-tau3mu}
\end{figure}

We conclude our discussion by directly comparing 
simple tau and muon cLFV 3-body decays: $\ell_\alpha \to 3 \ell_\beta$. 
While for the $P^{(\prime)}$ asymmetries there is little to be learnt, it is nevertheless interesting to consider the $T$-asymmetry, which we display in Fig.~\ref{fig:Tasym:mu3e-tau3mu}. It summarises clearly the previous discussion, and synthesises the potential of this minimal HNL extension to be 
studied via cLFV decays and associated asymmetries.

\section{Outlook}\label{sec:concs}
In a clear departure from SM prediction, the discovery of neutrino oscillations clearly signalled that individual flavours were not conserved in the neutral lepton sector, and opened a wide door for the violation of charged lepton flavours. Numerous NP constructions (among them those encompassing a mechanism of neutrino mass generation) can be at the source of observable contributions to cLFV processes. Interestingly, heavy sterile fermions are an integral part of several well-motivated models of neutrino mass generation and, if not excessively heavy, can also lead to sizeable contributions to cLFV observables. 

Generic NP models featuring heavy neutral leptons further open the possibility of additional sources of CP violation, in the form of new Dirac and Majorana phases. 
In recent years, the impact of the CPV phases on cLFV observables has been extensively addressed~\cite{Abada:2021zcm,Abada:2022asx,Kriewald:2024cnt} - strikingly, both Dirac and Majorana CPV phases can strongly affect na\"ive correlation patterns between observables (in fact leading to a loss of correlation), which in turn hinders one's capacity to test HNL extensions as the underlying source of the observed cLFV. 

In view of these findings, it becomes increasingly important to extend (and diversify) the number of observables under scrutiny, in order to isolate relative phases between the contributing operators. 
In this work we have thus proposed to investigate the prospects for a number of asymmetries in association with cLFV three-body decays, in the context of a minimal ``ad-hoc'' extension of the SM via 2 sterile states.  
In particular, we have considered $\mu \to 3 e$ decays - which will be explored at Mu3e, as well as all possible (purely leptonic) cLFV tau decay modes. 
For the relevant cases, we have considered 
$P$, $P^\prime$ and $T$ asymmetries, as well as ``Forward-Backward'' ones. 
In our study, we have carried out a full and independent 
computation of all above mentioned angular asymmetries, clarifying several points in what concerns phase space integration, and providing detailed expressions for all studied observables.  
In particular, and for the case of 
tau decays, we considered the various possible final state configurations (leading to (un)distinguishable charged leptons).  
The results of our study clearly demonstrate that - and as expected - CP violating phases can lead to significant contributions to the asymmetries, in association with decay rates within future experimental sensitivity. 
In certain scenarios, one can potentially have asymmetries as large as (or even above) 10\%.
Furthermore, the patterns in angular distributions can be very distinctive. 
Should a sizeable cLFV signal be observed and should enough statistics eventually become available, full binned analysis could further increase the model discriminating power of these observables.

We have also emphasised the crucial role of synergetic measurements of different observables - in particular for the $\mu-e$ sector. 
Even for the most general case of non-vanishing CP violating phases (which would prevent falsifying the model as $\mu \to 3e$ and $mu-e$ no longer are correlated, see~\cite{Abada:2021zcm}), this simple extension of the SM can be probed relying on the joint observation of cLFV rates and 
asymmetries: a future measurement of BR($\mu \to 3e$)$\sim \mathcal{O}(10^{-15})$ associated with $\mathcal{A}^T \sim \mathcal{O}(20\%)$, suggested that $\mu-e$ conversion would necessarily be measured with a rate larger than $10^{-14}$. 

In this manuscript we have also highlighted the probing power of the asymmetries associated with tau decays. 
Since the expected improvement in associated experimental sensitivity is less striking than for muon cLFV channels, 
only a fraction of the potentially allowed asymmetries realistically lie within observation reach; in turn this implies that this NP model becomes almost predictive regarding the possibly observed values of the asymmetries.
Beyond this simple study relying on an ad-hoc construction, we can envisage a full analysis of low-scale seesaw models, encompassing also further direct CP asymmetries as outlined in~\cite{Abada:2022asx}.
For instance in the case of the Inverse Seesaw, relying on the parametrisations developed in~\cite{Kriewald:2024rlg}, one can systematically differentiate regimes of operator dominances to potentially disentangle the source of leptonic CP violation: from the PMNS-sector, heavy-light mixings or heavy-heavy mixings.
Such a study, potentially allowing also for connections to low-scale leptogenesis scenarios, will be left for future work.

From the point of view of the distinct operators at work, notice that while phases disrupt the clear cLFV operator dominance (as occurred regarding $Z$-penguin dominance in the CP conserving case, which led to strong correlations for muon-electron conversion in nuclei and $\mu \to 3e$ decays), they now also lead to a non-negligible interference between operators - which can be investigated through measurements of the asymmetries. 
As also clear from our discussion, the angular observables together with CP-odd asymmetries play a crucial role in identifying the operator set responsible for cLFV. 
This will pave a bottom-up approach to construct more elaborate (well-motivated) models which can account for neutrino mass generation, and accommodate the experimentally measured observables. 

Clearly, if either $\mu$ or $\tau$ cLFV transitions are observed in the (near) future, they will bring forth a very rich experimental programme; in this sense,   
the exploration of the most comprehensive set of associated observables is of paramount importance. 
In particular, and as we have argued here, the exploitation of angular observables and asymmetries should be foreseen in advance, and carried out in association with any possible measurement.

\section*{Acknowledgements}
JK is supported by the Slovenian Research Agency under the research core funding 
No. P1-0035 and in part by the research grants J1-3013 and N1-0253. 
This project has received support from the IN2P3 (CNRS) Master Project, ``Hunting for Heavy Neutral Leptons'' (12-PH-0100).

\appendix
\section{Computation of the decay asymmetries}\label{app:asym-computation}

This appendix is dedicated to a detailed presentation of the computation of asymmetries in association with generic $1 \to 3$ decays.
Following the schematic decay depicted in Fig.~\ref{fig:ref-frame}, and for the definitions 
of momenta (and polarisation) given in Eq.~(\ref{eq:kinematics:convention}) and Eq.~(\ref{eq:reducedmomenta}), one obtains the following 
Lorentz invariant quantities:
\begin{align}\label{eq:lorentz-inv}
    p^2 & = m_\alpha^2, \quad k_1^2 = m_\beta^2, 
    \quad k_2^2 = m_\gamma^2, \quad k_3^2 = m_\delta^2\,,\nonumber \\
    p \cdot k_1 & = \frac{1}{2} (m_\alpha^2 + m_\beta^2 - s_1)\,,\nonumber \\
    p \cdot k_2 & = \frac{1}{2} (m_\alpha^2 + m_\gamma^2 - s_2)\,,\nonumber \\
    p \cdot k_3 & = \frac{1}{2} (m_\alpha^2 + m_\delta^2 - s_3)\,,\nonumber \\
    k_1 \cdot k_2 & = \frac{1}{2} (s_3 - m_\beta^2 - m_\gamma^2)\,,\nonumber \\
    k_1 \cdot k_3 & = \frac{1}{2} (s_2 - m_\beta^2 - m_\delta^2)\,,\nonumber \\
    k_2 \cdot k_3 & = \frac{1}{2} (s_1 - m_\gamma^2 - m_\delta^2)\,,
\end{align}
for which the mass and flavour indices have been defined 
according to Eq.~(\ref{eq:genericdecay}), which we recall below
\begin{equation}
    \ell_\alpha^+ (p) \to 
    \ell_\beta^+ (k_1)\, \ell_\gamma^+ (k_2)\, \ell_\delta^-(k_3)\,. \nonumber
\end{equation}
For the present studies, and as illustrated in Fig.~\ref{fig:ref-frame}, we will 
define the polar angle $\theta_\varepsilon$ to lie between the polarisation vector and the outgoing negatively charged particle;  the azimuthal angle $\phi_\varepsilon$ is taken to lie between the projected polarisation vector in the decay plane and the outgoing negatively charged particle.
(Notice that the above assignment relies on the definition of the angles between final states with opposite charges, which is experimentally preferable.)
One thus obtains the following invariants involving the polarisation vector:
\begin{align}
   k_1 \cdot \varepsilon & = \frac{1}{2 m_\alpha^3  \lambda \! \left( \frac{m_\delta^2}{m_\alpha^2}, \frac{s_3}{m_\alpha^2} \right)} \left[ (m_\alpha^2 - m_\delta^2 + s_3) (m_\alpha^2 + m_\beta^2 - s_1) - 2 m_\alpha^2 (s_3 - m_\gamma^2 + m_\beta^2) \right] P \cos \theta_\varepsilon \nonumber \\
    & + \frac{\sqrt{s_3}}{2} \lambda \! \left( \frac{m_\gamma^2}{s_3}, \frac{m_\beta^2}{s_3} \right) P \sin \theta \sin \theta_\varepsilon \cos \phi_\varepsilon\,, \nonumber\\
    k_2 \cdot \varepsilon & = \frac{1}{2 m_\alpha^3  \lambda \! \left( \frac{m_\delta^2}{m_\alpha^2}, \frac{s_3}{m_\alpha^2} \right)} \left[ (m_\alpha^2 - m_\delta^2 + s_3) (m_\alpha^2 + m_\gamma^2 - s_2) - 2 m_\alpha^2 (s_3 + m_\gamma^2 - m_\beta^2) \right] P \cos \theta_\varepsilon \nonumber \\
    & - \frac{\sqrt{s_3}}{2} \lambda \! \left( \frac{m_\gamma^2}{s_3}, \frac{m_\beta^2}{s_3} \right) P \sin \theta \sin \theta_\varepsilon \cos \phi_\varepsilon\,,\nonumber\\
    k_3 \cdot \varepsilon & = - \frac{m_\alpha}{2} \lambda \! \left( \frac{m_\delta^2}{m_\alpha^2}, \frac{s_3}{m_\alpha^2} \right) P \cos \theta_\varepsilon,\nonumber\\
    \varepsilon^{k_1 k_2 k_3 \varepsilon} & = \frac{m_\alpha^2}{4} \sqrt{s_3} \lambda \! \left( \frac{m_\delta^2}{m_\alpha^2}, \frac{s_3}{m_\alpha^2} \right) \lambda \! \left( \frac{m_\gamma^2}{s_3}, \frac{m_\beta^2}{s_3} \right) P \sin \theta \sin \theta_\varepsilon \cos \phi_\varepsilon\,, \nonumber\\
    \sin \theta & = \frac{1}{m_\alpha^2 \lambda \! \left( \frac{m_\beta^2}{m_\alpha^2}, \frac{s_3}{m_\alpha^2} \right) \lambda \! \left( \frac{m_\gamma^2}{s_3}, \frac{m_\beta}{s_3} \right)} \left[ 4 s_2 s_1 - 2 (m_\alpha^2 - m_\gamma^2) (m_\delta^2 - m_\beta^2) - 2 (m_\alpha^2 - m_\beta^2) (m_\delta^2 - m_\gamma^2) \right. \nonumber \\
    & - \left. \frac{2}{s_3} \left( (m_\alpha^2 - m_\delta^2)^2 (m_\gamma^2 + m_\beta^2) + (s_2 - s_1) (m_\alpha^2 - m_\delta^2) (m_\gamma^2 - m_\beta^2) + (m_\alpha^2 + m_\delta^2) (m_\gamma^2 - m_\beta^2)^2 \right) \right]^{1/2}\,,
\end{align}
in which $\lambda(x,y)$ is the reduced Källén function
\begin{equation}
    \lambda(x,y) = \sqrt{(1 - x - y)^2 - 4 xy}\,.
    \label{eqn:kallen}
\end{equation}
Leading to the definition of the phase space integral bounds, it readily follows from the above equations that there are only two independent variables; we thus choose 
$s_1$ and $s_2$. While the boundary for $s_1$ is straightforward to obtain, for $s_2$ one finds (once $s_1$ is fixed)
\begin{align}
    s_2^{\pm} (s_1) & = m_\beta^2 + m_\delta^2 + \frac{(m_\alpha^2 - m_\beta^2 - s_1) (s_1 - m_\gamma^2 + m_\delta^2)}{2 s_1} \pm \frac{m_\alpha^2}{2} \lambda \! \left( \frac{m_\beta^2}{m_\alpha^2}, \frac{s_1}{m_\alpha^2} \right) \lambda \! \left( \frac{m_\gamma^2}{s_1}, \frac{m_\delta^2}{s_1} \right)\,.
\end{align}
It is important to point out that the above bounds (i.e. $s_{1,2}^{\pm}$ are not always sufficient to fully evaluate the asymmetries, especially in the case of indistinguishable final states. 
While the $\theta_\varepsilon$ angle can be defined in all cases
(as the negatively charged particle in the final state can always be identified), the $\phi_\varepsilon$ angle is ill-defined for final states comprising identical charged leptons. 
This would lead to the vanishing of the asymmetries associated with $\phi_\varepsilon$ (i.e. $P^\prime$ and $T$). 
In order to overcome this, one must distinguish final states relying on their energies:
the integration region can be separated into ``Forward''
and ``Backward'' subregions, respectively for $E_1 > E_2$ and $E_1 < E_2$ (with $E$ the energy of the outgoing particle in the rest frame of the decaying one). 
From an experimental point of view this would corresponds to identifying the highest energy final state as $\ell_\beta^+$, and the lowest one $\ell_\gamma^+$ (in the case in which $\beta = \gamma$).

The integration domain can thus be defined as 
\begin{align}\label{eq:IntegrationRegion}
&
\begin{cases}
    (m_\beta + m_\gamma)^2  \leq  s_1 \leq (m_\alpha - m_\beta)^2\\
    s_2^- (s_1) \leq s_2 (s_1) \leq s_2^+ (s_1)\\
    E_1 > E_2
\end{cases}
\nonumber\\
\Leftrightarrow
&
\begin{cases}
    m_\beta^2 + m_\alpha m_\gamma \leq  s_1 \leq \frac{1}{2} (m_\alpha^2 - 2 m_\beta^2 + m_\gamma^2)\\
    s_1 \leq s_2 (s_1) \leq s_2^+ (s_1)
\end{cases}
\cup \quad 
\begin{cases}
    (m_\beta + m_\gamma)^2 \leq  s_1 \leq m_\beta^2 + m_\alpha m_\gamma\\
    s_2^-(s_1) \leq s_2 (s_1) \leq s_2^+ (s_1)
\end{cases}\,,
\end{align}
where the integration domain is now the union of two subdomains.
An extra symmetry factor ($2$) must be introduced in the definition of the asymmetries to take into account the ``Backward'' subregion. Moreover, it is important to notice that one now has an additional region corresponding to $s_1 \leq s_2^-$. 
It does not contribute to any asymmetry in the limit of vanishing final state masses (with the exception of the squared dipole term featuring a soft collinear divergence).

Interestingly, and in the case of distinguishable final states, we can still identify these subregions, and 
introduce a new set of ``Forward-Backward'' ($FB$) asymmetries. The $FB$ asymmetries are thus defined in a similar way to Eq.~(\ref{eq:asym:def}), but now 
\begin{align}
    \int_\Omega \to \left(\int_{E_1 > E_2} - \int_{E_1 < E_2}\right)\,,
\end{align}
allowing to define new sets of observables: the 
$FB$, $FB,P$, $FB,P^\prime$ and $FB,T$ asymmetries.

\section{HNL extensions of the SM}\label{app:HNL-extensions} 
In this appendix we summarise some phenomenological features of SM extensions via sterile fermions, from the new terms in the interaction Lagrangian to the most relevant phenomenological constraints.  

\subsection{Modified interaction Lagrangian}
The presence of sterile states which have non-negligible mixings with the active neutrinos will lead to modified charged and neutral currents, which can be cast in the physical basis as
\begin{align}\label{eq:lagrangian:WGHZ}
& \mathcal{L}_{W^\pm}\, =\, -\frac{g_w}{\sqrt{2}} \, W^-_\mu \,
\sum_{\alpha=1}^{3} \sum_{j=1}^{3 + n_S} \mathcal{U}_{\alpha j} \bar \ell_\alpha 
\gamma^\mu P_L \nu_j \, + \, \text{H.c.}\,, \nonumber \\
& \mathcal{L}_{Z^0}^{\nu}\, = \,-\frac{g_w}{2 \cos \theta_w} \, Z_\mu \,
\sum_{i,j=1}^{3 + n_S} \bar \nu_i \gamma ^\mu \left(
P_L {C}_{ij} - P_R {C}_{ij}^* \right) \nu_j\,, \nonumber \\
& \mathcal{L}_{Z^0}^{\ell}\, = \,-\frac{g_w}{4 \cos \theta_w} \, Z_\mu \,
\sum_{\alpha=1}^{3}  \bar \ell_\alpha \gamma ^\mu \left(
{\bf C}_{V} - {\bf C}_{A} \gamma_5 \right) \ell_\alpha\,, \nonumber \\
& \mathcal{L}_{H^0}\, = \, -\frac{g_w}{2 M_W} \, H  \,
\sum_{i\ne j= 1}^{3 + n_S}  {C}_{ij}  \bar \nu_i\left(
P_R m_i + P_L m_j \right) \nu_j + \, \text{H.c.}\ , \nonumber \\
& \mathcal{L}_{G^0}\, =\,\frac{i g_w}{2 M_W} \, G^0 \,
\sum_{i,j=1}^{3 + n_S} {C}_{ij}  \bar \nu_i  
\left(P_R m_j  - P_L m_i  \right) \nu_j\,+ \, \text{H.c.}, \nonumber  \\
& \mathcal{L}_{G^\pm}\, =\, -\frac{g_w}{\sqrt{2} M_W} \, G^- \,
\sum_{\alpha=1}^{3}\sum_{j=1}^{3 + n_S} \mathcal{U}_{\alpha j} 
\bar \ell_\alpha\left(
m_i P_L - m_j P_R \right) \nu_j\, + \, \text{H.c.}\,, 
\end{align}
with $n_S$ denoting the number of sterile fermions added to the SM content, and with   
\begin{equation}\label{eq:Cij:def}
    {C}_{ij} = \sum_{\rho = 1}^3
  \mathcal{U}_{i\rho}^\dagger \,\mathcal{U}_{\rho j}^{\phantom{\dagger}}\:. 
\end{equation}
In the above, the indices 
$\alpha, \rho = 1, \dots, 3$ denote the flavour of the charged leptons, while $i, j = 1, \dots, 3+n_S$ correspond to the physical (massive) 
neutrino states; $P_{L,R} = (1 \mp \gamma_5)/2$, 
$g_w$ denotes the weak coupling constant, and
$\cos^2 \theta_w =  M_W^2 /M_Z^2$.
The coefficients ${\bf C}_{V}$ and ${\bf C}_{A}$ 
parametrise the SM vector and axial-vector currents 
for the interaction of neutrinos with charged leptons, respectively given by 
${\bf C}_{V} = \frac{1}{2} + 2 \sin^2\theta_w$ and 
${\bf C}_{A} = \frac{1}{2}$.

\subsection{A simplified model: SM ad-hoc extensions via two sterile fermions }
In this work we rely on a minimal ``ad-hoc'' extension of the SM via two generations of (Majorana) heavy sterile states. Such a simplified model, the so-called ``3+2'', can be perceived as a first step in phenomenological analysis of HNL extensions of the SM, in particular of low-scale realisations of type I seesaw mechanisms of neutrino mass generation (such as the inverse seesaw, or the linear seesaw, among other possibilities).

The neutral spectrum is composed of 5 Majorana states: 3 light (mostly active) neutrinos, and two additional heavy ones, with masses $m_{4,5}$. It is important to clarify that no assumption is made on the underlying mechanism at the origin of neutrino masses; one hypothesises that leptonic mixings are effectively encoded in an enlarged $5\times5$ unitary mixing matrix, $\mathcal U$. 
The upper left $3\times3$ block of the latter thus corresponds to the left-handed leptonic mixing matrix, the would-be PMNS, $\tilde{U}_\text{PMNS}$.

Under such an ad-hoc approach,  $\mathcal{U}$
can be parametrised through five subsequent rotations $R_{ij}$ (with $i\neq j$), and a diagonal matrix which encodes the four physical Majorana phases, $\varphi_i$ (see for instance~\cite{Abada:2015trh})
\begin{eqnarray}
    \mathcal{U} \,= \,R_{45}\,R_{35}\,R_{25}\,R_{15}\,
    R_{34}\,R_{24}\,R_{14}\,R_{23}\,R_{13}\,R_{12}\times\mathrm{diag}(1, e^{i\varphi_2}, e^{i\varphi_3}, e^{i\varphi_4}, e^{i\varphi_5})\,.
    \label{eqn:allrot}
\end{eqnarray}
The above rotations are of the form (illustrated by $R_{45}$):
\begin{equation}\label{eq:R45}
    R_{45} = \begin{pmatrix}
                1 & 0 & 0 & 0 & 0\\
                0 & 1 & 0 & 0 & 0\\
                0 & 0 & 1 & 0 & 0\\
                0 & 0 & 0 & \cos\theta_{45} & \sin \theta_{45} e^{-i\delta_{45}}\\
                0 & 0 & 0 & -\sin\theta_{45} e^{i\delta_{45}} & \cos\theta_{45}
            \end{pmatrix}\,.
\end{equation}
In the presence of non-vanishing active-sterile mixings (parametrised via $\theta{\alpha 4}$ and 
$\theta{\alpha 5}$), the would-be PMNS mixing matrix will be non-unitary; it proves convenient to parametrise the deviations from unitarity~\cite{Schechter:1980gr,Gronau:1984ct} via a matrix 
$\eta$~\cite{FernandezMartinez:2007ms}
\begin{equation}
\label{eq:defPMNSeta}
U_\text{PMNS} \, \to \, \tilde U_\text{PMNS} \, = \,(\mathbb{1} - \eta)\, 
U_\text{PMNS}\,.
\end{equation}
The deviations from unitarity of the would-be PMNS mixing matrix are at the source of modified leptonic currents (neutral and charged), which in turn open the door to extensive contributions to a wide array of processes. The latter include cLFV transitions and decays, lepton number violating processes (neutrinoless double beta decays, as well as semileptonic tau and meson decays), contributions to observables parametrising deviations from lepton flavour universality, among others, as we will subsequently discuss.

\subsection{Constraints on HNL extensions of the SM}
Firstly, the extended mixing matrix must comply with oscillation data. In our analysis, we have fixed the (mostly) active mixing parameters - i.e. the entries of the $R_{12}$, $R_{13}$ and $R_{23}$ matrices - to lie in the $3\sigma$ interval of the NuFIT~6.0 most recent results (without atmospheric data)~\cite{Esteban:2024eli}; for simplicity we take a normal ordering for the light neutrino spectrum\footnote{Notice that the cLFV rates - including the 3-body decays under consideration - are mostly independent of the ordering of the spectrum and of the lightest neutrino mass.}, and fix the lightest neutrino mass scale to $10^{-4}$~eV. 

As a consequence of the above mentioned departure from unitarity of the would-be PMNS matrix, 
the presence of the new (heavy) sterile states will also impact a number of observables\footnote{For the considered mass regimes (around the TeV), constraints on the heavy sterile states from direct searches at colliders and cosmological limits are in general not competitive with the flavour and EW bounds.} (in addition to cLFV processes). The latter include $Z$-pole observables (direct and indirectly, from modifications of the Fermi constant $G_F$ and the weak mixing angle, $\theta_w$), 
observables sensitive to the violation of lepton flavour universality (LFU), and charged lepton flavour violating processes (which will be separately discussed in the following subsection). 

In our analysis, we thus consider the generic bounds on $\eta$ as derived in~\cite{Blennow:2023mqx}, in particular for the following observables,\\
\noindent - $Z$ pole observables (LEP): $\Gamma_Z$, $\sigma^0_\text{had}$, $R_{e,\mu,\tau}$ (respectively denoting the total $Z$ width, the hadronic cross-section), and the leptonic ratios ($R_\ell=\Gamma_Z^\ell/\Gamma_Z^\text{had}$). Of further importance are the contributions to the  invisible $Z$ width, $\Gamma_Z^\text{inv}$, which is typically reduced in the presence of significant deviations of the PMNS from unitarity.\\
\noindent - LFU sensitive ratios, built from the comparison of leptonic decays of light mesons to electrons and muons, including $R^K_{e\mu}$, $R^\pi_{e\mu}$~\cite{Abada:2013aba,Abada:2012mc},  and (semi-)leptonic decays of the $\tau$-lepton. 
(Notice that these are particularly sensitive to modified $W\ell\nu$ interactions.)\\
\noindent
- Also taken into account in the constraints of~\cite{Blennow:2023mqx} are modifications of $G_F$, $\sin^2\theta_w$, the mass of the $W$-boson among others.

\medskip
The presence of the new Majorana heavy states can also lead to modifications of the predictions regarding neutrinoless double beta decays, in particular of the effective mass $m_{ee}$~\cite{Blennow:2010th,Abada:2014nwa}
\begin{equation}
\label{eq:def:0nubb_nS}
m_{ee} \simeq \sum_{i=1}^{3+n_s} \, \mathcal U_{e i}^2 \, p^2 \, \frac{m_{ i}}{p^2-m_{i}^2} \simeq \sum_{i=1}^3 \, \mathcal U^2_{e i} \, m_i + \sum_{k=4}^{3+n_s} \mathcal U_{e k}^2 \, p^2 \, \frac{m_{k}}{p^2-m_{k}^2} \; ,
\end{equation}
with the virtual momentum $p^2 \simeq -(100 \, \mathrm{MeV})^2$. In the present study we have taken into account upper limits on
$m_{ee}^\mathrm{eff} \lesssim  (28 \div 122)\:\mathrm{MeV}$, from the KamLAND-ZEN collaboration~\cite{KamLAND-Zen:2024eml}. 

\medskip
In the presence of CP violating phase, one should also consider the implications for the electric dipole moments of charged leptons; a dedicated analysis for ``3+2'' simple extensions of the SM~\cite{Abada:2015trh} suggested that once all cLFV constraints had been imposed, one typically found contributions to $d_e$ below $10^{-30}$~$e$.cm  (and thus in agreement with the most recent bound
$|d_e| < 4.1 \times 10^{-30}~e$.cm~\cite{Roussy:2022cmp}, with the exception of regimes associated with very sizeable values of $\theta_{14}$, which are excluded in our current study. 
(For the muon and tau EDMs the predictions lay orders of magnitude below experimental sensitivity~\cite{Abada:2015trh}).

\medskip
Finally, let us emphasise perturbative unitarity constraints~\cite{Chanowitz:1978mv,Durand:1989zs,Bernabeu:1993up,Fajfer:1998px,Ilakovac:1999md}, leading to the following bounds on the HNL decay widths
\begin{equation}
    \frac{\Gamma(N_i)}{m_{N_i}}<\frac{1}{2},\quad \text{for}\quad i\geq4\,,
\end{equation}
also a consequence of modified charged current interactions.

\subsection{Charged lepton flavour violation constraints on HNL extensions}
The most important bounds to be considered in this study in general arise from current searches for rare cLFV transitions and decays, which lead to stringent constraints on HNL masses and mixing parameters. 
In addition to the cLFV three-body decays (which are at the core of the present study), the most relevant processes include radiative decays, $\mu - e$ conversion in atoms, and cLFV $Z$ boson decays. We collect the latter bounds in Table~\ref{tab:cLFV_lep} (which includes limits on 3-body decays already given in the main body of the manuscript). 

\renewcommand{\arraystretch}{1.3}
\begin{table}[h!]
    \centering
    \hspace*{-2mm}{\small\begin{tabular}{|c|c|c|}
    \hline
    Observable & Current bound & Future sensitivity  \\
    \hline\hline
    $\text{BR}(\mu\to e \gamma)$    &
    \quad $<4.2\times 10^{-13}$ \quad (MEG~\cite{TheMEG:2016wtm})   &
    \quad $6\times 10^{-14}$ \quad (MEG II~\cite{Baldini:2018nnn}) \\
    $\text{BR}(\tau \to e \gamma)$  &
    \quad $<3.3\times 10^{-8}$ \quad (BaBar~\cite{Aubert:2009ag})    &
    \quad $3\times10^{-9}$ \quad (Belle II~\cite{Kou:2018nap})      \\
    $\text{BR}(\tau \to \mu \gamma)$    &
     \quad $ <4.4\times 10^{-8}$ \quad (BaBar~\cite{Aubert:2009ag})  &
    \quad $10^{-9}$ \quad (Belle II~\cite{Kou:2018nap})     \\
    \hline
    $\text{BR}(\mu \to 3 e)$    &
     \quad $<1.0\times 10^{-12}$ \quad (SINDRUM~\cite{Bellgardt:1987du})    &
     \quad $10^{-15(-16)}$ \quad (Mu3e~\cite{Blondel:2013ia})   \\
    $\text{BR}(\tau \to 3 e)$   &
    \quad $<2.7\times 10^{-8}$ \quad (Belle~\cite{Hayasaka:2010np})&
    \quad $5\times10^{-10}$ \quad (Belle II~\cite{Kou:2018nap})     \\
    $\text{BR}(\tau \to 3 \mu )$    &
    \quad $<3.3\times 10^{-8}$ \quad (Belle~\cite{Hayasaka:2010np})  &
    \quad $5\times10^{-10}$ \quad (Belle II~\cite{Kou:2018nap})     \\
    & & \quad$5\times 10^{-11}$\quad (FCC-ee~\cite{Abada:2019lih})\\
        $\text{BR}(\tau^- \to e^-\mu^+\mu^-)$   &
    \quad $<2.7\times 10^{-8}$ \quad (Belle~\cite{Hayasaka:2010np})&
    \quad $5\times10^{-10}$ \quad (Belle II~\cite{Kou:2018nap})     \\
    $\text{BR}(\tau^- \to \mu^-e^+e^-)$ &
    \quad $<1.8\times 10^{-8}$ \quad (Belle~\cite{Hayasaka:2010np})&
    \quad $5\times10^{-10}$ \quad (Belle II~\cite{Kou:2018nap})     \\
    $\text{BR}(\tau^- \to e^-\mu^+e^-)$ &
    \quad $<1.5\times 10^{-8}$ \quad (Belle~\cite{Hayasaka:2010np})&
    \quad $3\times10^{-10}$ \quad (Belle II~\cite{Kou:2018nap})     \\
    $\text{BR}(\tau^- \to \mu^-e^+\mu^-)$   &
    \quad $<1.7\times 10^{-8}$ \quad (Belle~\cite{Hayasaka:2010np})&
    \quad $4\times10^{-10}$ \quad (Belle II~\cite{Kou:2018nap})     \\
    \hline
    $\text{CR}(\mu- e, \text{N})$ &
     \quad $<7 \times 10^{-13}$ \quad  (Au, SINDRUM~\cite{Bertl:2006up}) &
    \quad $10^{-14}$  \quad (SiC, DeeMe~\cite{Nguyen:2015vkk})    \\
    & &  \quad $2.6\times 10^{-17}$  \quad (Al, COMET~\cite{Krikler:2015msn,COMET:2018auw,Moritsu:2022lem})  \\
    & &  \quad $8 \times 10^{-17}$  \quad (Al, Mu2e~\cite{Bartoszek:2014mya})\\
    \hline
    $\mathrm{BR}(Z\to e^\pm\mu^\mp)$ & \quad$< 4.2\times 10^{-7}$\quad (ATLAS~\cite{Aad:2014bca}) & \quad$\mathcal O (10^{-10})$\quad (FCC-ee~\cite{Abada:2019lih})\\
    $\mathrm{BR}(Z\to e^\pm\tau^\mp)$ & \quad$< 4.1\times 10^{-6}$\quad (ATLAS~\cite{ATLAS:2021bdj}) & \quad$\mathcal O (10^{-10})$\quad (FCC-ee~\cite{Abada:2019lih})\\
    $\mathrm{BR}(Z\to \mu^\pm\tau^\mp)$ & \quad$< 5.3\times 10^{-6}$\quad (ATLAS~\cite{ATLAS:2021bdj}) & \quad $\mathcal O (10^{-10})$\quad (FCC-ee~\cite{Abada:2019lih})\\
    \hline
    \end{tabular}}
    \caption{Current experimental bounds and future sensitivities on relevant leptonic cLFV observables, with limits quoted at $90\%\:\mathrm{C.L.}$ (For the case of Belle II sensitivities, these correspond to an integrated luminosity of $50\:\mathrm{ab}^{-1}$.)}
    \label{tab:cLFV_lep}
\end{table}
\renewcommand{\arraystretch}{1.}

Although we do provide the analytical expressions for the 3-body cLFV decays in Appendix~\ref{app:3body:HNL}, for the radiative decays and neutrinoless conversion in nuclei in the context of HNL extensions of the SM, we refer to~\cite{Ilakovac:1994kj,Alonso:2012ji,Abada:2018nio,Abada:2022asx,Riemann:1982rq,Illana:1999ww,Mann:1983dv,Illana:2000ic,Ma:1979px,Gronau:1984ct,Deppisch:2004fa,Deppisch:2005zm,Dinh:2012bp,Abada:2014kba,Abada:2015oba,Abada:2015zea,Abada:2016vzu,Arganda:2014dta}, in which 
the expressions for the latter observables are provided. 

\section{cLFV 3-body decays}\label{app:3body:HNL}
Below we collect the formulae for the relevant cLFV decay rates (form factors and loop functions) 
which are discussed in the manuscript. Further expressions (including those for radiative decays, or neutrinoless muon-electron conversion in nuclei) can be found in~\cite{Ilakovac:1994kj,Alonso:2012ji,Abada:2018nio,Abada:2022asx,Riemann:1982rq,Illana:1999ww,Mann:1983dv,Illana:2000ic,Ma:1979px,Gronau:1984ct,Deppisch:2004fa,Deppisch:2005zm,Dinh:2012bp,Abada:2014kba,Abada:2015oba,Abada:2015zea,Abada:2016vzu,Arganda:2014dta, Abada:2021zcm}, among many others.

\subsection{Form factors}

Following~\cite{Ilakovac:1994kj, Alonso:2012ji},  in the context of SM extensions via a number $n_s$ of heavy neutral leptons, the three-body decay rates can be cast as 
\begin{eqnarray}
    \mathrm{BR}(\ell_\alpha\to 3\ell_\beta) &=&
    \frac{\alpha_w^4}{24576\,\pi^3}\,\frac{m_{\alpha}^5}{M_W^4}\,
\frac{1}{\Gamma_{\alpha}}\times\left\{2\left|\frac{1}{2}F_\text{box}^{\alpha
      3\beta} +F_Z^{\alpha\beta} - 2 s_w^2\,(F_Z^{\alpha\beta} -
    F_\gamma^{\alpha\beta})\right|^2 \right.  \nonumber\\ 
     &+& \left. 4 s_w^4\, |F_Z^{\alpha\beta} -
    F_\gamma^{\alpha\beta}|^2 + 16
    s_w^2\,\mathrm{Re}\left[(F_Z^{\alpha\beta} + \frac{1}{2}F_\text{box}^{\alpha
        3\beta})\,G_\gamma^{\alpha \beta
        \ast}\right]\right.\nonumber\\ 
     &-&\left. 48 s_w^4\,\mathrm{Re}\left[(F_Z^{\alpha\beta} -
      F_\gamma^{\alpha\beta})\,G_\gamma^{\alpha\beta \ast}\right] + 32
    s_w^4\,|G_\gamma^{\alpha\beta}|^2\left[\log\frac{m_{\beta}^2}{m_{\alpha}^2}
      - \frac{11}{4}\right] \right\}\,, 
\end{eqnarray}
for the case of same-flavoured final leptons. For the case of tau decays, several combinations can be present in the final state, so that one has to consider $\ell_\beta^+\to \ell_\gamma^+ \ell_\alpha^+ \ell_\alpha^-$, as well as $\ell_\beta^+\to \ell_\alpha^+ \ell_\gamma^+ \ell_\alpha^-$. For the former, the rates are given by
\begin{eqnarray}
\text{BR}(\ell_\alpha^+\to \ell_\gamma^+ \ell_\beta^+ \ell_\beta^-) &= &
     \frac{\alpha_w^4}{24576\,\pi^3}\,\frac{m_{\alpha}^5}{M_W^4}\,
\frac{1}{\Gamma_{\alpha}}\times\left\{\left|F_\text{box}^{\alpha
      \gamma \beta \beta} +F_Z^{\alpha\gamma} - 2 s_w^2\,(F_Z^{\alpha\gamma} -
    F_\gamma^{\alpha\gamma})\right|^2 \right.  \nonumber\\ 
     &+& \left. 4 s_w^4\, |F_Z^{\alpha\gamma} -
    F_\gamma^{\alpha\gamma}|^2 + 8
    s_w^2\,\mathrm{Re}\left[(F_Z^{\alpha\gamma} + F_\text{box}^{\alpha
      \gamma \beta \beta})\,G_\gamma^{\alpha \gamma
        \ast}\right]\right.\nonumber\\ 
     &-&\left. 32 s_w^4\,\mathrm{Re}\left[(F_Z^{\alpha\gamma} -
      F_\gamma^{\alpha\gamma})\,G_\gamma^{\alpha\gamma \ast}\right] + 32
    s_w^4\,|G_\gamma^{\alpha\gamma}|^2\left[\log\frac{m_{\beta}^2}{m_{\alpha}^2}
      - 3\right] \right\}\,,
\end{eqnarray}
while for the latter (which can only be induced through box contributions) one finds
\begin{eqnarray}
\text{BR}(\ell_\alpha^+\to  \ell_\gamma^+ \ell_\gamma^+ \ell_\beta^-) &= &
 \frac{\alpha_w^4}{49152\pi^3}\ \frac{m_\alpha^5}{M_W^4}\, 
 \frac{1}{\Gamma_\alpha}\, 
 \left| F_\text{box}^{\alpha
      \gamma \gamma \beta} \right|^2\,.
\end{eqnarray}
In the above, $M_W$ denotes the $W$ boson mass, $\alpha_w = g_w^2/4\pi$ the weak coupling, $s_w$ is the sine of the weak mixing angle; $m_{\beta}$ ($\Gamma_\beta$) is the mass (total width)
of the decaying charged lepton of flavour $\beta$.  
The form factors present in the above equations are given by~\cite{Alonso:2012ji, Ilakovac:1994kj} 
\begin{align}
    \label{eq:cLFV:FF}
    G_\gamma^{\alpha \beta} & = \sum_i \mathcal{U}_{\alpha i}^* \,\mathcal{U}_{\beta i} \,G_\gamma (x_i )\,, 
   \\
    F_\gamma^{\alpha \beta} & = \sum_i \mathcal{U}_{\alpha i}^* \,\mathcal{U}_{\beta i} \,F_\gamma(x_i )\,, 
    \\
    F_Z^{\alpha \beta} & = \sum_i \mathcal{U}_{\alpha i}^*   \left( \mathcal{U}_{\beta i} \,F_Z (x_i) + \sum_j \mathcal{U}_{\beta j} \left[ C_{i,j}^* \, G_Z(x_i, x_j) + C_{i,j} \, H_Z (x_i, x_j) \right]  \right)\,,
    \\
    F_{\text{Box}}^{\alpha \beta \gamma \delta} & = \sum_i \sum_j \left( \mathcal{U}_{\alpha i}^* \,\mathcal{U}_{\delta j}^* \left(\mathcal{U}_{\beta i} \,\mathcal{U}_{\gamma j} + \mathcal{U}_{\beta j} \,\mathcal{U}_{\gamma i} \right) F_{\text{XBox}} (x_i, x_j ) + \right.\nonumber \\
    & + \left.
        \mathcal{U}_{\alpha i}^*\, \mathcal{U}_{\delta i}^*\, \mathcal{U}_{\beta j} \mathcal{U}_{\gamma j} G_{\text{Box}} (x_i, x_j ) \right)\,,
\end{align} 
in which the sums run over the neutral mass eigenstates ($i,j=1,...,3+n_S$). 
The loop functions are given in the following subsection, with the corresponding arguments defined as $x_i ={m_{i}^2}/{M_W^2}$. We recall that in the above 
$\mathcal{U}$ is the $(3+n_s)\times (3+n_s)$ lepton mixing matrix (unitary) and $C_{ij}$ was defined 
in Eq.~(\ref{eq:Cij:def}).

As mentioned before, details on observables such as radiative decays and $\mu-e$ conversion in nuclei (not explicitly presented here) can be found in~\cite{Ilakovac:1994kj,Alonso:2012ji,Abada:2018nio,Abada:2022asx,Riemann:1982rq,Illana:1999ww,Mann:1983dv,Illana:2000ic,Ma:1979px,Gronau:1984ct,Deppisch:2004fa,Deppisch:2005zm,Dinh:2012bp,Abada:2014kba,Abada:2015oba,Abada:2015zea,Abada:2016vzu,Arganda:2014dta}.

\subsection{Loop functions}
The loop functions are defined as follows (in the Feynman gauge)
\begin{align}
    F_\gamma(x) & = \frac{7 x^3 - x^2 - 12 x}{12 (1 - x)^3} - \frac{x^4 - 10 x^3 + 12 x^2}{6 (1 - x)^4} \ln(x)\,, \\
    G_\gamma(x) & = -\frac{2 x^3 + 5 x^2 - x}{4 (1 - x)^3} - \frac{3 x^3}{2 (1 - x)^4} \ln(x)\,,\\
    F_Z(x) & = - \frac{5 x}{2 (1 - x)} - \frac{5 x^2}{2 (1 - x)^2} \ln(x)\,,\\
    G_Z(x,y) & = - \frac{1}{2 (x - y)} \left[ \frac{x^2 (1 - y)}{1 - x} \ln(x) - \frac{y^2 (1 - x)}{1 - y} \ln(y) \right]\,,\\
    H_Z(x,y) & = \frac{\sqrt{x y}}{4 (x - y)} \left[ \frac{x^2 - 4 x}{1 - x} \ln(x) - \frac{y^2 - 4 y}{1 - y} \ln(y) \right]\,,\\
    F_{\text{XBox}}(x,y) & = - \frac{1}{x - y} \left[ \left(1 + \frac{x y}{4}\right) \left(\frac{1}{1 - x} + \frac{x^2}{(1 - x)^2} \ln(x) - \frac{1}{1 - y} - \frac{y^2}{(1 - y)^2} \ln(y) \right) \right. \nonumber\\
    & - \left. 2 x y \left(\frac{1}{1 - x} + \frac{x}{(1 - x)^2} \ln(x) - \frac{1}{1 - y} - \frac{y}{(1 - y)^2} \ln(y) \right)\right]\,,\\
    G_{\text{Box}}(x,y) & = - \frac{\sqrt{x y}}{x - y} \left[ \left(4 + x y\right) \left(\frac{1}{1 - x} + \frac{x}{(1 - x)^2} \ln(x) - \frac{1}{1 - y} - \frac{y}{(1 - y)^2} \ln(y) \right) \right. \nonumber\\
    & - \left. 2 \left(\frac{1}{1 - x} + \frac{x^2}{(1 - x)^2} \ln(x) - \frac{1}{1 - y} - \frac{y^2}{(1 - y)^2} \ln(y) \right)\right]\,.
\end{align}

\section{Additional details regarding the computation of the amplitudes: phase space integration}\label{app:asy:details}
In this final appendix, we offer a detailed description of the steps leading to a near-analytical computation 
of the asymmetries, for the distinct cases of three-body cLFV decays. 

The differential polarised decay rate can be written as
\begin{align}\label{eq:diff-pol-rate}
    \frac{d \Gamma}{d s_1 \,d \,s_2 \,d \Omega_\varepsilon} &= \frac{\alpha_w^4}{65\, 536 \,\pi^5\, M_W^4 \,m_\alpha^5} \left[C_1(s_1, s_2) + C_2(s_1, s_2) \,X(s_1, s_2) \,P\, \cos \theta_\varepsilon \right.\nonumber \\
    & + \left. C_3(s_1, s_2)\, Y(s_1, s_2) \,P\, \sin \theta_\varepsilon \cos \phi_\varepsilon + C_4(s_1, s_2) \,Z(s_1, s_2)\, P \,\sin \theta_\varepsilon \sin \phi_\varepsilon \right]\,,
\end{align}
in which we recall that $P \equiv \vert \vec{P} \vert$, and $X,Y,Z$ are the prefactors associated with the scalar product between the polarisation vector and the momenta, defined as
\begin{align}
    X &= \lambda \! \left(\frac{m_\delta^2}{m_\alpha^2}, \frac{s_3}{m_\alpha^2} \right)^{-1},\\
    Y &= \frac{\sqrt{s_3}}{2} \lambda \! \left(\frac{m_\gamma^2}{s_3}, \frac{m_\beta^2}{s_3} \right) \sin \theta,\\
    Z &= \sqrt{s_3} \lambda \! \left( \frac{m_\delta^2}{m_\alpha^2}, \frac{s_3}{m_\alpha^2} \right) \lambda \! \left( \frac{m_\gamma^2}{s_3}, \frac{m_\beta^2}{s_3} \right) \sin \theta\,.
\end{align}
For the distinct cases, (i), (ii-a) and (ii-b),
the $C_i$ coefficients entering Eq.~(\ref{eq:diff-pol-rate}) are defined as
\begin{itemize}
    \item [(i)] $\ell_\alpha^+ \to \ell_\beta^+ \ell_\beta^+ \ell_\beta^-$:
    \begin{align}
        C_1 & =  \vert F_\gamma \vert^2 F_1(s_1,s_2) + \vert G_\gamma \vert^2 F_2(s_1,s_2) + \vert F_Z \vert^2 F_3(s_1,s_2) + \vert F_\text{Box} \vert^2 F_4(s_1,s_2)\nonumber \\
        & + \text{Re} \left(F_\text{Box} F_\gamma^* \right) F_5(s_1,s_2) + \text{Re} \left(F_\text{Box} G_\gamma^* \right) F_6(s_1,s_2) + \text{Re} \left(F_\text{Box} F_Z^* \right) F_7(s_1,s_2)\nonumber \\
        & + \text{Re} \left(F_Z F_\gamma^* \right) F_8(s_1,s_2) + \text{Re} \left(F_Z G_\gamma^* \right) F_9(s_1,s_2) + \text{Re} \left(G_\gamma F_\gamma^* \right) F_{10}(s_1,s_2)\,,
        \\
        C_2 & =  \vert F_\gamma \vert^2 G_1(s_1,s_2) + \vert G_\gamma \vert^2 G_2(s_1,s_2) + \vert F_Z \vert^2 G_3(s_1,s_2) + \vert F_\text{Box} \vert^2 G_4(s_1,s_2)\nonumber \\
        & + \text{Re} \left(F_\text{Box} F_\gamma^* \right) G_5(s_1,s_2) + \text{Re} \left(F_\text{Box} G_\gamma^* \right) G_6(s_1,s_2) + \text{Re} \left(F_\text{Box} F_Z^* \right) G_7(s_1,s_2)\nonumber \\
        & + \text{Re} \left(F_Z F_\gamma^* \right) G_8(s_1,s_2) + \text{Re} \left(F_Z G_\gamma^* \right) G_9(s_1,s_2) + \text{Re} \left(G_\gamma F_\gamma^* \right) G_{10}(s_1,s_2)\,,
        \\
        C_3 & = \left(\vert F_\gamma - F_Z \vert^2 + \text{Re} \left(F_Z F_\gamma^* \right)\right) H_1(s_1,s_2) + \vert G_\gamma \vert^2 H_2(s_1,s_2) + \text{Re} \left(F_\text{Box} G_\gamma^* \right) H_3(s_1,s_2)\nonumber \\
        & + \text{Re} \left(F_Z G_\gamma^* \right) H_4(s_1,s_2) + \text{Re} \left(G_\gamma F_\gamma^* \right) H_5(s_1,s_2)\,,
        \\
        C_4 & =\text{Im} \left(F_\text{Box} G_\gamma^* \right) H_6(s_1,s_2) + \text{Im} \left(F_\gamma G_\gamma^* \right) H_7(s_1,s_2) + \text{Im} \left(F_Z G_\gamma^* \right) H_8(s_1,s_2)\,;
    \end{align}
    \item [(ii-a)] $\ell_\alpha^+ \to \ell_\gamma^+ \ell_\beta^+ \ell_\beta^-$:
    \begin{align}
        C_1 & =  \vert F_\gamma \vert^2 F_1(s_1,s_2) + \vert G_\gamma \vert^2 F_2(s_1,s_2) + \vert F_Z \vert^2 F_3(s_1,s_2) + \vert F_\text{Box} \vert^2 F_4(s_1,s_2)\nonumber \\
        & + \text{Re} \left(F_\text{Box} F_\gamma^* \right) F_5(s_1,s_2) + \text{Re} \left(F_\text{Box} G_\gamma^* \right) F_6(s_1,s_2) + \text{Re} \left(F_\text{Box} F_Z^* \right) F_7(s_1,s_2)\nonumber \\
        & + \text{Re} \left(F_Z F_\gamma^* \right) F_8(s_1,s_2) + \text{Re} \left(F_Z G_\gamma^* \right) F_9(s_1,s_2) + \text{Re} \left(G_\gamma F_\gamma^* \right) F_{10}(s_1,s_2)\,,
        \\
        C_2 & = \vert F_\gamma \vert^2 G_1(s_1,s_2) + \vert G_\gamma \vert^2 G_2(s_1,s_2) + \vert F_Z \vert^2 G_3(s_1,s_2) + \vert F_\text{Box} \vert^2 G_4(s_1,s_2)\nonumber \\
        & + \text{Re} \left(F_\text{Box} F_\gamma^* \right) G_5(s_1,s_2) + \text{Re} \left(F_\text{Box} G_\gamma^* \right) G_6(s_1,s_2) + \text{Re} \left(F_\text{Box} F_Z^* \right) G_7(s_1,s_2)\nonumber \\
        & + \text{Re} \left(F_Z F_\gamma^* \right) G_8(s_1,s_2) + \text{Re} \left(F_Z G_\gamma^* \right) G_9(s_1,s_2) + \text{Re} \left(G_\gamma F_\gamma^* \right) G_{10}(s_1,s_2)\,,
        \\
        C_3 & = \vert F_\gamma \vert^2 H_1(s_1,s_2) + \vert G_\gamma \vert^2 H_2(s_1,s_2) + \vert F_Z \vert^2 H_3(s_1,s_2) + \text{Re} \left(F_\text{Box} (F_\gamma-F_Z)^* \right) H_4(s_1,s_2) \nonumber \\
        & + \text{Re} \left(F_\text{Box} G_\gamma^* \right) H_5(s_1,s_2) + \text{Re} \left(F_Z F_\gamma^* \right) H_6(s_1,s_2) + \text{Re} \left(F_Z G_\gamma^* \right) H_7(s_1,s_2)\nonumber \\
        & + \text{Re} \left(G_\gamma F_\gamma^* \right) H_8(s_1,s_2)\,,
        \\
        C_4 & =\text{Im} \left(F_\text{Box} G_\gamma^* \right) H_9(s_1,s_2) + \text{Im} \left(F_\gamma F_\gamma^* \right) H_{10}(s_1,s_2) + \text{Im} \left(F_Z G_\gamma^* \right) H_{11}(s_1,s_2)\,;
    \end{align}
    \item [(ii-b)] $\ell_\alpha^+ \to \ell_\gamma^+ \ell_\beta^+ \ell_\beta^-$:
    \begin{align}
        C_1 & = \vert F_\text{Box} \vert^2 F_4(s_1,s_2)\,,
        \\
        C_2 & =  \vert F_\text{Box} \vert^2 G_4(s_1,s_2)\,,
        \\
        C_3 & =  0\,,
        \\
        C_4 & =  0\,.
    \end{align}
\end{itemize}
In addition to the form factors already presented in Appendix~\ref{app:3body:HNL}, we introduce numerous abbreviations: $F_i$ ($i=1-10$), $G_j$ ($j=1-10$)
and $H_k$ ($k=1-11$), 
whose definitions vary depending on the considered flavour composition of the cLFV 3-body decays. 

The $F_i$ functions are defined as:
\begin{itemize}
    \item [(i)] $\ell_\alpha^+ \to \ell_\beta^+ \ell_\beta^+ \ell_\beta^-$:
    \begin{align}
    F_1 &= - 4 s_w^4 m_\alpha^2 \left[m_\alpha^2 \left( 2 m_\beta^2 - 5 s_1 \right) - 9 m_\beta^2 s_1 + 5 s_1^2 + 4 s_1 s_2 \right] + (s_1 \leftrightarrow s_2)\,,\\
    F_2 &= 4 s_w^4 m_\alpha^4 \left[\frac{2 m_\beta^2 \left(m_\alpha^2 - m_\beta^2 \right)^2}{s_1^2} + \frac{m_\beta^2 \left(m_\alpha^4 - 3 m_\beta^2 m_\alpha^2 + 2 m_\beta^4 \right)}{s_1 s_2} \right.\nonumber \\
    &+ \left. \frac{m_\alpha^4 + 3 m_\beta^4 - 2 s_2 \left( m_\alpha^2 + 3 m_\beta^2 \right) + 2 s_2^2}{s_1} - 3 m_\beta^2\right] + (s_1 \leftrightarrow s_2)\,,\\
    F_3 &= 4 m_\alpha^2 \left[m_\beta^2 \left(m_\beta^2 \left(2 s_w^2 - 1\right) - m_\alpha^2 \left(2 s_w^4 - 2 s_w^2 + 1\right)\right) \right.\nonumber \\
    &+ \left. s_1 \left(m_\beta^2 \left(9 s_w^4 - 10 s_w^2 + 3\right) - m_\alpha^2 \left(5 s_w^4 - 4 s_w^2 + 1\right)\right) \right.\nonumber \\
    &- \left. s_1^2 \left(5 s_w^4 - 4 s_w^2 + 1\right)- s_1 s_2 \left(2 s_w^2 - 1\right)^2\right] + (s_1 \leftrightarrow s_2)\,,\\
    F_4 &= m_\alpha^2 \left[- m_\beta^2 \left(m_\beta^2 + m_\alpha^2\right) + s_1 \left(3 m_\beta^2+ m_\alpha^2\right)- s_1^2 - s_1 s_2\right] + (s_1 \leftrightarrow s_2)\,,\\
    F_5 &= 2 m_\alpha^2 s_w^2 \left[- m_\beta^2 \left(m_\beta^2 + m_\alpha^2\right) + s_1 \left(5 m_\beta^2 + 2 m_\alpha^2\right)- s_1^2 - 2 s_1 s_2\right] + (s_1 \leftrightarrow s_2)\,,\\
    F_6 &= 2 m_\alpha^4 s_w^2 \left[ \frac{m_\beta^2 \left(m_\alpha^2 - m_\beta^2\right)}{s_1} + m_\alpha^2 - 2 s_1 \right] + (s_1 \leftrightarrow s_2)\,,\\
    F_7 &= 2 m_\alpha^2 \left[m_\beta^2 \left(s_w^2 - 1\right) \left(m_\beta^2 + m_\alpha^2\right) - s_1 \left( \left(2 s_w^2 - 1\right) m_\alpha^2 + \left(5 s_w^2 - 3\right) m_\beta^2\right) \right.\nonumber \\
    &+ \left. \left(2 s_w^2 - 1\right) s_1^2 + \left(2 s_w^2 - 1\right) s_1 s_2\right] + (s_1 \leftrightarrow s_2)\,,\\
    F_8 &= 4 m_\alpha^2 s_w^2 \left[-m_{\beta }^4+s_1 \left(\left(2-5 s_w^2\right) m_{\alpha }^2+\left(5-9 s_w^2\right) m_{\beta }^2\right)+\left(2 s_w^2-1\right) m_{\alpha }^2 m_{\beta }^2 \right. \nonumber \\
    &+ \left.s_1^2 \left(5 s_w^2-2\right)+2 s_1 s_2 \left(2 s_w^2-1\right)\right] + (s_1 \leftrightarrow s_2),\\
    F_9 &= 4 m_\alpha^4 s_w^2 \left[\frac{m_\beta^2 \left(3 s_w^2 - 1\right) \left(m_\beta^2 - m_\alpha^2\right)}{s_1} + \left(1 - 2 s_w^2\right) m_\alpha^2 + 3 s_w^2 m_\beta^2 \right.\nonumber \\
    &+ \left. \left(3 s_w^2 - 2\right) s_1\right] + (s_1 \leftrightarrow s_2)\,,\\
    F_{10} &= 4 m_\alpha^4 s_w^4 \left[\frac{3 m_\beta^2 \left(m_\alpha^2 - m_\beta^2\right)}{s_1} + 2 m_\alpha^2 - 3 m_\beta^2 - 3 s_1 \right] + (s_1 \leftrightarrow s_2)\,;
    \end{align}
    \item [(ii-a)] $\ell_\alpha^+ \to \ell_\gamma^+ \ell_\beta^+ \ell_\beta^-$:
    \begin{align}
    F_1 &= 8 s_w^4 m_\alpha^2 \left[-2 \left(m_{\alpha }^2 m_{\gamma }^2+m_{\beta }^4\right)+2 s_2 \left(m_{\alpha }^2+2 m_{\beta }^2+m_{\gamma }^2\right)+s_1 \left(m_{\alpha }^2+m_{\gamma }^2\right)\right.\nonumber \\
    &- \left.s_1^2-2 s_2 s_1-2 s_2^2\right]\,,\\
    F_2 &= 8 s_w^4 m_\alpha^4 \left[-\left(m_{\alpha }^2+2 m_{\beta }^2+m_{\gamma }^2\right)+\frac{m_{\alpha }^4+2 m_{\beta }^4+m_{\gamma }^4-2 s_2 \left(m_{\alpha }^2+2 m_{\beta }^2+m_{\gamma }^2\right)+2 s_2^2}{s_1}\right.\nonumber \\
    &+\left.\frac{2 m_{\beta }^2 \left(m_{\alpha }^2-m_{\gamma }^2\right)^2}{s_1^2}+2 s_2\right]\,,\\
    F_3 &= 2 m_\alpha^2 \left[-m_{\beta }^2 m_{\gamma }^2+s_1 \left(1-2 s_w^2\right) \left(2 m_{\beta }^2+\left(1-2 s_w^2\right) m_{\alpha }^2+\left(1-2 s_w^2\right) m_{\gamma }^2\right)\right.\nonumber \\
    &+ \left.s_2 \left(8 s_w^4-4 s_w^2+1\right) \left(m_{\alpha }^2+2 m_{\beta }^2+m_{\gamma }^2\right)-m_{\alpha }^2 \left(m_{\beta }^2+\left(8 s_w^4-4 s_w^2+1\right) m_{\gamma }^2\right)\right.\nonumber \\
    &+ \left.\left(-8 s_w^4+4 s_w^2-1\right) m_{\beta }^4-s_1^2 \left(1-2 s_w^2\right)^2-2 s_1 s_2 \left(1-2 s_w^2\right)^2+s_2^2 \left(-8 s_w^4+4 s_w^2-1\right)\right]\,,\\
    F_4 &= 2 m_\alpha^2 \left[-\left(m_{\alpha }^2+m_{\beta }^2\right) \left(m_{\beta }^2+m_{\gamma }^2\right)+s_1 \left(m_{\alpha }^2+2 m_{\beta }^2+m_{\gamma }^2\right)\right.\nonumber \\
    &+ \left.s_2 \left(m_{\alpha }^2+2 m_{\beta }^2+m_{\gamma }^2\right)-s_1^2-2 s_2 s_1-s_2^2\right]\,,\\
    F_5 &= 4 m_\alpha^2 s_w^2 \left[-m_{\alpha }^2 m_{\gamma }^2-m_{\beta }^4+s_1 \left(m_{\alpha }^2+m_{\beta }^2+m_{\gamma }^2\right)+s_2 \left(m_{\alpha }^2+2 m_{\beta }^2+m_{\gamma }^2\right)\right.\nonumber \\
    &- \left.s_1^2-s_2^2-2 s_1 s_2\right]\,,\\
    F_6 &= 4 m_\alpha^4 s_w^2 \left[m_{\alpha }^2+\frac{m_{\beta }^2 \left(m_{\alpha }^2-m_{\gamma }^2\right)}{s_1}-s_1-s_2 \right]\,,\\
    F_7 &= 2 m_\alpha^2 \left[-m_{\beta }^2 m_{\gamma }^2-s_2 \left(2 s_w^2-1\right) \left(m_{\alpha }^2+2 m_{\beta }^2+m_{\gamma }^2\right)+s_1 \left(\left(1-2 s_w^2\right) m_{\alpha }^2\right.\right.\nonumber \\
    &- \left. \left.2 \left(s_w^2-1\right) m_{\beta }^2+\left(1-2 s_w^2\right) m_{\gamma }^2\right)-m_{\alpha }^2 \left(m_{\beta }^2+\left(1-2 s_w^2\right) m_{\gamma }^2\right)+\left(2 s_w^2-1\right) m_{\beta }^4\right.\nonumber \\
    &+ \left.s_1^2 \left(2 s_w^2-1\right)+s_2^2 \left(2 s_w^2-1\right)+s_1 s_2 \left(4 s_w^2-2\right)\right]\,,\\
    F_8 &= 4 s_w^2 m_\alpha^2 \left[s_1 \left(m_{\beta }^2+\left(1-2 s_w^2\right) m_{\alpha }^2+\left(1-2 s_w^2\right) m_{\gamma }^2\right)-s_2 \left(4 s_w^2-1\right) \left(m_{\alpha }^2+2 m_{\beta }^2+m_{\gamma }^2\right)\right.\nonumber \\
    &+ \left.\left(4 s_w^2-1\right) \left(m_{\alpha }^2 m_{\gamma }^2+m_{\beta }^4\right)+s_1^2 \left(2 s_w^2-1\right)+s_2 s_1 \left(4 s_w^2-2\right)+s_2^2 \left(4 s_w^2-1\right) \right]\,,\\
    F_9 &= 4 m_\alpha^4 s_w^2 \left[-\frac{\left(4 s_w^2-1\right) m_{\beta }^2 \left(m_{\alpha }^2-m_{\gamma }^2\right)}{s_1}+\left(1-2 s_w^2\right) m_{\alpha }^2\right.\nonumber \\
    &+ \left.2 s_w^2 \left(2 m_{\beta }^2+m_{\gamma }^2\right)+s_1 \left(2 s_w^2-1\right)-s_2\right]\,,\\
    F_{10} &= 4 m_\alpha^4 s_w^4 \left[2 \left(m_{\alpha }^2-2 m_{\beta }^2-m_{\gamma }^2\right)+\frac{4 m_{\beta }^2 \left(m_{\alpha }^2-m_{\gamma }^2\right)}{s_1}-2 s_1\right]\,;
    \end{align}
    \item [(ii-b)] $\ell_\alpha^+ \to \ell_\gamma^+ \ell_\beta^+ \ell_\beta^-$:
    \begin{align}
    F_4 &= - m_\gamma^2 \left( m_\alpha^2 + m_\beta^2 \right) + s_1 \left( m_\alpha^2 + m_\beta^2 + 2 m_\gamma^2 \right) - s_1^2 - s_1 s_2 + (s_1 \leftrightarrow s_2)\,.
    \end{align}
\end{itemize}
For the distinct cases, the $G_j$ functions can be written as
\begin{itemize}
    \item [(i)] $\ell_\alpha^+ \to \ell_\beta^+ \ell_\beta^+ \ell_\beta^-$:
    \begin{align}
    G_1 &= 4 s_w^4 \left[8 m_{\alpha }^2 m_{\beta }^2 \left(m_{\alpha }^2-m_{\beta }^2\right)-s_1^2 \left(5 m_{\alpha }^2+19 m_{\beta }^2\right)-s_2 s_1 \left(3 m_{\alpha }^2+17 m_{\beta }^2\right)\right.\nonumber \\
    &- \left.2 s_1 m_{\beta }^2 \left(m_{\alpha }^2-9 m_{\beta }^2\right)+5 s_1^3+13 s_2 s_1^2\right] + (s_1 \leftrightarrow s_2)\,,\\
    G_2 &= 4 s_w^4 m_\alpha^2 \left[-2 m_{\alpha }^2 m_{\beta }^2-3 m_{\alpha }^4+9 m_{\beta }^4+6 s_1 \left(m_{\alpha }^2-2 m_{\beta }^2\right)\right.\nonumber \\
    &+ \left.\frac{-2 m_{\alpha }^4 m_{\beta }^2+2 m_{\alpha }^2 m_{\beta }^4-2 s_1^2 \left(m_{\alpha }^2+5 m_{\beta }^2\right)+s_1 \left(m_{\alpha }^4+15 m_{\beta }^4\right)+2 s_1^3}{s_2}\right.\nonumber \\
    &+ \left.\frac{2 \left(m_{\beta }^3-m_{\alpha }^2 m_{\beta }\right)^2 \left(m_{\alpha }^2-2 m_{\beta }^2\right)}{s_1 s_2}+\frac{2 s_1 \left(m_{\beta }^3-m_{\alpha }^2 m_{\beta }\right)^2-4 m_{\beta }^4 \left(m_{\alpha }^2-m_{\beta }^2\right)^2}{s_2^2}\right.\nonumber \\
    &+ \left.2 s_1^2\right] + (s_1 \leftrightarrow s_2)\,,\\
    G_3 &= 4 \left[s_1^2 \left(\left(-5 s_w^4+4 s_w^2-1\right) m_{\alpha }^2+\left(-19 s_w^4+18 s_w^2-5\right) m_{\beta }^2\right)-s_2 s_1 \left(\left(3 s_w^4-4 s_w^2+1\right) m_{\alpha }^2\right. \right.\nonumber \\
    &+ \left. \left.\left(17 s_w^4-18 s_w^2+5\right) m_{\beta }^2\right)+s_1 \left(2 \left(2-3 s_w^2\right)^2 m_{\beta }^4-2 s_w^4 m_{\alpha }^2 m_{\beta }^2\right)\right.\nonumber \\
    &+ \left.2 \left(2 s_w^2-1\right) m_{\beta }^2 \left(m_{\alpha }^2-m_{\beta }^2\right) \left(\left(2 s_w^2-1\right) m_{\alpha }^2-m_{\beta }^2\right)+s_1^3 \left(5 s_w^4-4 s_w^2+1\right)\right.\nonumber \\
    &+ \left.s_2 s_1^2 \left(13 s_w^4-12 s_w^2+3\right)\right] + (s_1 \leftrightarrow s_2)\,,\\
    G_4 &= 2 m_{\beta }^2 \left(m_{\alpha }^4-m_{\beta }^4\right)-s_1^2 \left(m_{\alpha }^2+5 m_{\beta }^2\right)-s_1 s_2 \left(m_{\alpha }^2+5 m_{\beta }^2\right)+8 s_1 m_{\beta }^4+s_1^3+3 s_1^2 s_2 \nonumber \\
    & + (s_1 \leftrightarrow s_2)\,,\\
    G_5 &= 2 s_w^2 \left[-2 m_{\alpha }^2 m_{\beta }^4+4 m_{\alpha }^4 m_{\beta }^2-2 m_{\beta }^6-s_1^2 \left(2 m_{\alpha }^2+9 m_{\beta }^2\right)-s_1 s_2 \left(2 m_{\alpha }^2+9 m_{\beta }^2\right)\right.\nonumber \\
    &+ \left.12 s_1 m_{\beta }^4+2 s_1^3+6 s_1^2 s_2\right] + (s_1 \leftrightarrow s_2)\,,\\
    G_6 &= 2 m_\alpha^2 s_w^2 \left[m_{\beta }^2 \left(m_{\alpha }^2+3 m_{\beta }^2\right)-2 s_1 \left(m_{\alpha }^2+4 m_{\beta }^2\right)\right.\nonumber \\
    &+ \left.\frac{2 m_{\alpha }^2 m_{\beta }^2 \left(m_{\alpha }^2-m_{\beta }^2\right)+s_1 \left(m_{\beta }^4-m_{\alpha }^2 m_{\beta }^2\right)}{s_2}+2 s_1^2+2 s_1 s_2\right] + (s_1 \leftrightarrow s_2)\,,\\
    G_7 &= 2 \left[s_1^2 \left(\left(2 s_w^2-1\right) m_{\alpha }^2+\left(9 s_w^2-5\right) m_{\beta }^2\right)+s_1 s_2 \left(\left(2 s_w^2-1\right) m_{\alpha }^2+\left(9 s_w^2-5\right) m_{\beta }^2\right)\right.\nonumber \\
    &+ \left.4 s_1 \left(2-3 s_w^2\right) m_{\beta }^4+2 m_{\beta }^2 \left(s_w^2 m_{\alpha }^2 m_{\beta }^2+\left(1-2 s_w^2\right) m_{\alpha }^4+\left(s_w^2-1\right) m_{\beta }^4\right)\right.\nonumber \\
    &+ \left.s_1^3 \left(1-2 s_w^2\right)+s_1^2 s_2 \left(3-6 s_w^2\right)\right] + (s_1 \leftrightarrow s_2)\,,\\
    G_8 &= 2 s_w^2 \left[2 s_1^2 \left(\left(5 s_w^2-2\right) m_{\alpha }^2+\left(19 s_w^2-9\right) m_{\beta }^2\right)+4 s_1 \left(s_w^2 m_{\alpha }^2 m_{\beta }^2+3 \left(2-3 s_w^2\right) m_{\beta }^4\right)\right.\nonumber \\
    &+ \left.2 s_2 s_1 \left(\left(3 s_w^2-2\right) m_{\alpha }^2+\left(17 s_w^2-9\right) m_{\beta }^2\right)-4 m_{\beta }^2 \left(m_{\beta }^2-m_{\alpha }^2\right) \left(m_{\beta }^2+\left(2-4 s_w^2\right) m_{\alpha }^2\right)\right.\nonumber \\
    &+ \left.s_1^3 \left(4-10 s_w^2\right)+s_2 s_1^2 \left(12-26 s_w^2\right)\right] + (s_1 \leftrightarrow s_2)\,,\\
    G_9 &= 4 m_\alpha^2 s_w^2 \left[s_1 \left(\left(4 s_w^2-2\right) m_{\alpha }^2+2 \left(9 s_w^2-4\right) m_{\beta }^2\right)\right.\nonumber \\
    &+ \left.\frac{s_1 \left(3 s_w^2-1\right) m_{\beta }^2 \left(m_{\alpha }^2-m_{\beta }^2\right)-2 m_{\beta }^2 \left(m_{\beta }^2-m_{\alpha }^2\right) \left(\left(1-2 s_w^2\right) m_{\alpha }^2+s_w^2 m_{\beta }^2\right)}{s_2}\right.\nonumber \\
    &- \left.\left(\left(s_w^2-1\right) m_{\beta }^2 \left(m_{\alpha }^2+3 m_{\beta }^2\right)\right)+s_1^2 \left(2-5 s_w^2\right)+s_1 s_2 \left(2-5 s_w^2\right)\right] + (s_1 \leftrightarrow s_2)\,\\
    G_{10} &= 4 m_\alpha^2 s_w^4 \left[m_{\beta }^2 \left(m_{\alpha }^2+3 m_{\beta }^2\right)-2 s_1 \left(2 m_{\alpha }^2+9 m_{\beta }^2\right)\right.\nonumber \\
    &+ \left.\frac{2 m_{\beta }^2 \left(-3 m_{\alpha }^2 m_{\beta }^2+2 m_{\alpha }^4+m_{\beta }^4\right)+3 s_1 m_{\beta }^2 \left(m_{\beta }^2-m_{\alpha }^2\right)}{s_2}+5 s_1^2+5 s_1 s_2\right] + (s_1 \leftrightarrow s_2)\,;
    \end{align}
    \item [(ii-a)] $\ell_\alpha^+ \to \ell_\gamma^+ \ell_\beta^+ \ell_\beta^-$:
    \begin{align}
    G_1 &= 8 s_w^4 \left[m_{\alpha }^2 \left(-2 m_{\beta }^2 m_{\gamma }^2+4 m_{\beta }^4-2 m_{\gamma }^4\right)+4 m_{\alpha }^4 m_{\beta }^2-2 m_{\beta }^4 \left(m_{\beta }^2+m_{\gamma }^2\right)\right.\nonumber \\
    &- \left.s_1^2 \left(m_{\alpha }^2+m_{\beta }^2+2 m_{\gamma }^2\right)-2 s_2^2 \left(m_{\alpha }^2+3 m_{\beta }^2+2 m_{\gamma }^2\right)-s_1 s_2 \left(m_{\alpha }^2+6 m_{\beta }^2+5 m_{\gamma }^2\right)\right.\nonumber \\
    &+ \left.s_1 \left(m_{\alpha }^2 \left(m_{\gamma }^2-5 m_{\beta }^2\right)+m_{\beta }^2 m_{\gamma }^2+2 m_{\beta }^4+m_{\gamma }^4\right)\right.\nonumber \\
    &+ \left.2 s_2 \left(-m_{\alpha }^2 \left(m_{\beta }^2-2 m_{\gamma }^2\right)+3 m_{\beta }^2 m_{\gamma }^2+3 m_{\beta }^4+m_{\gamma }^4\right)+s_1^3+2 s_2^3+4 s_1 s_2^2+3 s_1^2 s_2\right]\,,
    \\
    G_2 &= 8 s_w^4 m_\alpha^2 \left[m_{\alpha }^2 \left(m_{\gamma }^2-m_{\beta }^2\right)-m_{\alpha }^4+3 m_{\beta }^2 m_{\gamma }^2+4 m_{\beta }^4+2 m_{\gamma }^4-s_2 \left(m_{\alpha }^2+8 m_{\beta }^2+5 m_{\gamma }^2\right)\right.\nonumber \\
    &+ \left.s_1 \left(m_{\alpha }^2-2 m_{\beta }^2-m_{\gamma }^2\right)\right.\nonumber \\
    &+ \left.\frac{1}{s_1} \left(m_{\alpha }^4 \left(m_{\beta }^2-m_{\gamma }^2\right)+4 m_{\alpha }^2 m_{\beta }^4-2 m_{\beta }^4 m_{\gamma }^2+m_{\beta }^2 m_{\gamma }^4-2 m_{\beta }^6-m_{\gamma }^6\right. \right.\\ \nonumber
    &- \left. \left.2 s_2^2 \left(m_{\alpha }^2+3 m_{\beta }^2+2 m_{\gamma }^2\right)+s_2 \left(2 m_{\gamma }^2 \left(m_{\alpha }^2+3 m_{\beta }^2\right)-2 m_{\alpha }^2 m_{\beta }^2+m_{\alpha }^4+6 m_{\beta }^4+3 m_{\gamma }^4\right)+2 s_2^3 \right)\right.\nonumber \\
    &+ \left.\frac{2 s_2 m_{\beta }^2 \left(m_{\alpha }^2-m_{\gamma }^2\right)-2 m_{\beta }^2 \left(m_{\alpha }^2-m_{\gamma }^2\right) \left(m_{\beta }^2+m_{\gamma }^2\right)}{s_1^2}+4 s_2^2+2 s_1 s_2\right]\,,
    \\
    G_3 &= 2 \left[-s_1^2 \left(2 s_w^2-1\right) \left(\left(2 s_w^2-1\right) m_{\alpha }^2+\left(2 s_w^2-3\right) m_{\beta }^2+2 \left(2 s_w^2-1\right) m_{\gamma }^2\right)\right.\nonumber \\
    &- \left.s_2^2 \left(8 s_w^4-4 s_w^2+1\right) \left(m_{\alpha }^2+3 m_{\beta }^2+2 m_{\gamma }^2\right)-2 s_1 s_2 \left(\left(2 s_w^4-4 s_w^2+1\right) m_{\alpha }^2\right.\right.\nonumber \\
    &+ \left.\left.\left(12 s_w^4-10 s_w^2+3\right) m_{\beta }^2+2 \left(5 s_w^4-4 s_w^2+1\right) m_{\gamma }^2\right)\right.\nonumber \\
    &+ \left.s_1 \left(4 \left(s_w^2-1\right)^2 m_{\beta }^2 m_{\gamma }^2+\left(1-2 s_w^2\right)^2 m_{\gamma }^4+2 m_{\alpha }^2 \left(\left(-10 s_w^4+6 s_w^2-1\right) m_{\beta }^2\right. \right.\right.\nonumber\\
    &+ \left.\left.\left.\left(2 s_w^4-4 s_w^2+1\right) m_{\gamma }^2\right)+\left(8 s_w^4-8 s_w^2+3\right) m_{\beta }^4\right)+s_2 \left(2 m_{\alpha }^2 \left(\left(-4 s_w^4+2 s_w^2-1\right) m_{\beta }^2\right.\right.\right.\nonumber \\
    &+ \left.\left.\left.\left(8 s_w^4-4 s_w^2+1\right) m_{\gamma }^2\right)+4 \left(6 s_w^4-3 s_w^2+1\right) m_{\beta }^2 m_{\gamma }^2+3 \left(8 s_w^4-4 s_w^2+1\right) m_{\beta }^4\right.\right.\nonumber \\
    &+ \left.\left.\left(8 s_w^4-4 s_w^2+1\right) m_{\gamma }^4\right)+4 \left(1-2 s_w^2\right)^2 m_{\alpha }^4 m_{\beta }^2+m_{\alpha }^2 \left(-2 \left(4 s_w^4-6 s_w^2+1\right) m_{\beta }^2 m_{\gamma }^2\right.\right.\nonumber \\
    &+ \left.\left.\left(16 s_w^4-8 s_w^2+3\right) m_{\beta }^4+\left(-8 s_w^4+4 s_w^2-1\right) m_{\gamma }^4\right)-m_{\beta }^2 \left(m_{\beta }^2+m_{\gamma }^2\right) \left(m_{\gamma }^2+\left(8 s_w^4\right.\right.\right.\nonumber \\
    &- \left.\left.\left.4 s_w^2+1\right) m_{\beta }^2\right)+s_1^3 \left(1-2 s_w^2\right)^2+3 s_1^2 s_2 \left(1-2 s_w^2\right)^2+s_2^3 \left(8 s_w^4-4 s_w^2+1\right)\right.\nonumber \\
    &+ \left.s_1 s_2^2 \left(16 s_w^4-12 s_w^2+3\right)\right]\,,
    \\
    G_4 &= 2 \left[\left(m_{\alpha }^2+m_{\beta }^2\right) \left(4 m_{\alpha }^2 m_{\beta }^2-\left(m_{\beta }^2+m_{\gamma }^2\right)^2\right)-s_1^2 \left(m_{\alpha }^2+3 m_{\beta }^2+2 m_{\gamma }^2\right)\right.\nonumber \\
    &- \left.2 s_2 s_1 \left(m_{\alpha }^2+3 m_{\beta }^2+2 m_{\gamma }^2\right)+s_1 \left(2 m_{\alpha }^2 \left(m_{\gamma }^2-m_{\beta }^2\right)+4 m_{\beta }^2 m_{\gamma }^2+3 m_{\beta }^4+m_{\gamma }^4\right)\right.\nonumber \\
    &- \left.s_2^2 \left(m_{\alpha }^2+3 m_{\beta }^2+2 m_{\gamma }^2\right)+s_2 \left(2 m_{\alpha }^2 \left(m_{\gamma }^2-m_{\beta }^2\right)+4 m_{\beta }^2 m_{\gamma }^2+3 m_{\beta }^4+m_{\gamma }^4\right)\right.\nonumber \\
    &+ \left.s_1^3+3 s_2 s_1^2+3 s_2^2 s_1+s_2^3 \right]\,,
    \\
    G_5 &= 4 s_w^2 \left[m_{\alpha }^2 \left(-3 m_{\beta }^2 m_{\gamma }^2+2 m_{\beta }^4-m_{\gamma }^4\right)+4 m_{\alpha }^4 m_{\beta }^2-m_{\beta }^4 \left(m_{\beta }^2+m_{\gamma }^2\right)\right.\nonumber \\
    &- \left.s_2^2 \left(m_{\alpha }^2+3 m_{\beta }^2+2 m_{\gamma }^2\right)-s_1 s_2 \left(2 m_{\alpha }^2+5 m_{\beta }^2+4 m_{\gamma }^2\right)-s_1^2 \left(m_{\alpha }^2+2 \left(m_{\beta }^2+m_{\gamma }^2\right)\right)\right.\nonumber \\
    &+ \left.s_1 \left(2 m_{\gamma }^2 \left(m_{\alpha }^2+m_{\beta }^2\right)-3 m_{\alpha }^2 m_{\beta }^2+2 m_{\beta }^4+m_{\gamma }^4\right)\right.\nonumber \\
    &+ \left.s_2 \left(-m_{\alpha }^2 \left(m_{\beta }^2-2 m_{\gamma }^2\right)+3 m_{\beta }^2 m_{\gamma }^2+3 m_{\beta }^4+m_{\gamma }^4\right)+s_1^3+s_2^3+3 s_1 s_2^2+3 s_1^2 s_2\right]\,,
    \\
    G_6 &= 4 m_\alpha^2 s_w^2 [m_{\gamma }^2 \left(m_{\alpha }^2+m_{\beta }^2\right)+2 m_{\beta }^4-s_1 \left(m_{\alpha }^2+3 m_{\beta }^2+m_{\gamma }^2\right)-s_2 \left(m_{\alpha }^2+3 m_{\beta }^2+m_{\gamma }^2\right)\nonumber \\
    &+ \frac{m_{\beta }^2 \left(m_{\alpha }^2-m_{\gamma }^2\right) \left(2 m_{\alpha }^2+m_{\beta }^2-m_{\gamma }^2\right)+s_2 m_{\beta }^2 \left(m_{\gamma }^2-m_{\alpha }^2\right)}{s_1}+s_1^2+s_2^2+2 s_1 s_2]\,,
    \\
    G_7 &= 2 \left[s_1^2 \left(\left(2 s_w^2-1\right) m_{\alpha }^2+\left(4 s_w^2-3\right) m_{\beta }^2+2 \left(2 s_w^2-1\right) m_{\gamma }^2\right)\right.\nonumber \\
    &+ \left.s_2^2 \left(2 s_w^2-1\right) \left(m_{\alpha }^2+3 m_{\beta }^2+2 m_{\gamma }^2\right)+s_1 \left(2 m_{\alpha }^2 \left(\left(3 s_w^2-1\right) m_{\beta }^2+\left(1-2 s_w^2\right) m_{\gamma }^2\right)\right.\right.\nonumber \\
    &- \left.\left.4 \left(s_w^2-1\right) m_{\beta }^2 m_{\gamma }^2+\left(3-4 s_w^2\right) m_{\beta }^4+\left(1-2 s_w^2\right) m_{\gamma }^4\right)\right.\nonumber \\
    &+ \left.s_2 \left(2 m_{\alpha }^2 \left(\left(s_w^2-1\right) m_{\beta }^2+\left(1-2 s_w^2\right) m_{\gamma }^2\right)+2 \left(2-3 s_w^2\right) m_{\beta }^2 m_{\gamma }^2+\left(3-6 s_w^2\right) m_{\beta }^4\right.\right.\nonumber \\
    &+ \left.\left.\left(1-2 s_w^2\right) m_{\gamma }^4\right)+s_1 s_2 \left(\left(4 s_w^2-2\right) m_{\alpha }^2+2 \left(5 s_w^2-3\right) m_{\beta }^2+4 \left(2 s_w^2-1\right) m_{\gamma }^2\right)\right.\nonumber \\
    &+ \left.m_{\alpha }^2 \left(2 \left(3 s_w^2-1\right) m_{\beta }^2 m_{\gamma }^2+\left(3-4 s_w^2\right) m_{\beta }^4+\left(2 s_w^2-1\right) m_{\gamma }^4\right)+4 \left(1-2 s_w^2\right) m_{\alpha }^4 m_{\beta }^2\right.\nonumber \\
    &+ \left.m_{\beta }^2 \left(m_{\beta }^2+m_{\gamma }^2\right) \left(\left(2 s_w^2-1\right) m_{\beta }^2-m_{\gamma }^2\right)+s_1^3 \left(1-2 s_w^2\right)\right.\nonumber \\
    &+ \left.s_2^3 \left(1-2 s_w^2\right)+s_1 s_2^2 \left(3-6 s_w^2\right)+s_1^2 s_2 \left(3-6 s_w^2\right)\right]\,,
    \\
    G_8 &= 4 s_w^2 \left[s_1^2 \left(\left(2 s_w^2-1\right) m_{\alpha }^2+2 \left(s_w^2-1\right) m_{\beta }^2+2 \left(2 s_w^2-1\right) m_{\gamma }^2\right)\right.\nonumber \\
    &+ \left.s_2^2 \left(4 s_w^2-1\right) \left(m_{\alpha }^2+3 m_{\beta }^2+2 m_{\gamma }^2\right)+s_1 \left(m_{\alpha }^2 \left(\left(10 s_w^2-3\right) m_{\beta }^2-2 \left(s_w^2-1\right) m_{\gamma }^2\right)\right.\right.\nonumber \\
    &- \left.\left.2 \left(s_w^2-1\right) m_{\beta }^2 m_{\gamma }^2+\left(2-4 s_w^2\right) m_{\beta }^4+\left(1-2 s_w^2\right) m_{\gamma }^4\right)\right.\nonumber \\
    &+ \left.s_2 \left(4 s_w^2-1\right) \left(m_{\alpha }^2 \left(m_{\beta }^2-2 m_{\gamma }^2\right)-3 m_{\beta }^2 m_{\gamma }^2-3 m_{\beta }^4-m_{\gamma }^4\right)\right.\nonumber \\
    &+ \left.s_1 s_2 \left(2 \left(s_w^2-1\right) m_{\alpha }^2+\left(12 s_w^2-5\right) m_{\beta }^2+2 \left(5 s_w^2-2\right) m_{\gamma }^2\right)+m_{\alpha }^2 \left(\left(4 s_w^2-3\right) m_{\beta }^2 m_{\gamma }^2\right.\right.\nonumber \\
    &+ \left.\left.\left(2-8 s_w^2\right) m_{\beta }^4+\left(4 s_w^2-1\right) m_{\gamma }^4\right)+4 \left(1-2 s_w^2\right) m_{\alpha }^4 m_{\beta }^2+\left(4 s_w^2-1\right) m_{\beta }^4 \left(m_{\beta }^2+m_{\gamma }^2\right)\right.\nonumber \\
    &+ \left.s_1^3 \left(1-2 s_w^2\right)+s_2^3 \left(1-4 s_w^2\right)+s_1 s_2^2 \left(3-8 s_w^2\right)+s_1^2 s_2 \left(3-6 s_w^2\right)\right]\,,
    \\
    G_9 &= 4 m_\alpha^2 s_w^2 \left[s_1 \left(\left(2 s_w^2-1\right) \left(m_{\alpha }^2+3 m_{\beta }^2\right)+\left(4 s_w^2-1\right) m_{\gamma }^2\right)+s_2 \left(\left(2 s_w^2-1\right) m_{\alpha }^2\right.\right.\nonumber \\
    &+ \left.\left.3 \left(4 s_w^2-1\right) m_{\beta }^2+\left(6 s_w^2-1\right) m_{\gamma }^2\right)\right.\nonumber \\
    &+ \left.\frac{s_2 \left(4 s_w^2-1\right) m_{\beta }^2 \left(m_{\alpha }^2-m_{\gamma }^2\right)-m_{\beta }^2 \left(m_{\alpha }^2-m_{\gamma }^2\right) \left(m_{\gamma }^2+\left(4 s_w^2-2\right) m_{\alpha }^2+\left(4 s_w^2-1\right) m_{\beta }^2\right)}{s_1}\right.\nonumber \\
    &+ \left.m_{\alpha }^2 \left(2 s_w^2 m_{\beta }^2+\left(1-2 s_w^2\right) m_{\gamma }^2\right)+\left(1-6 s_w^2\right) m_{\beta }^2 m_{\gamma }^2+\left(2-8 s_w^2\right) m_{\beta }^4-2 s_w^2 m_{\gamma }^4\right.\nonumber \\
    &+ \left.s_1^2 \left(1-2 s_w^2\right)+s_2^2 \left(1-4 s_w^2\right)+s_1 s_2 \left(2-6 s_w^2\right)\right]\,,
    \\
    G_{10} &= 8 m_\alpha^2 s_w^4 \left[m_{\alpha }^2 \left(m_{\gamma }^2-m_{\beta }^2\right)+3 m_{\beta }^2 m_{\gamma }^2+4 m_{\beta }^4+m_{\gamma }^4-s_1 \left(m_{\alpha }^2+3 m_{\beta }^2+2 m_{\gamma }^2\right)\right.\nonumber \\
    &- \left.s_2 \left(m_{\alpha }^2+6 m_{\beta }^2+3 m_{\gamma }^2\right)+\frac{2 m_{\beta }^2 \left(m_{\alpha }^2+m_{\beta }^2\right) \left(m_{\alpha }^2-m_{\gamma }^2\right)+2 s_2 m_{\beta }^2 \left(m_{\gamma }^2-m_{\alpha }^2\right)}{s_1}\right.\nonumber \\
    &+ \left.s_1^2+2 s_2^2+3 s_1 s_2\right]\,;
    \end{align}
    \item [(ii-b)] $\ell_\alpha^+ \to \ell_\gamma^+ \ell_\beta^+ \ell_\beta^-$:
    \begin{align}
    G_4 &= 2 \left(m_{\alpha }^2+m_{\beta }^2\right) \left(m_{\alpha }^2 m_{\beta }^2-m_{\gamma }^4\right)-s_1^2 \left(m_{\alpha }^2+m_{\beta }^2+4 m_{\gamma }^2\right)-s_2 s_1 \left(m_{\alpha }^2+m_{\beta }^2+4 m_{\gamma }^2\right)\nonumber \\
    &+ 4 s_1 \left(m_{\gamma }^2 \left(m_{\alpha }^2+m_{\beta }^2\right)-m_{\alpha }^2 m_{\beta }^2+m_{\gamma }^4\right)+s_1^3+3 s_2 s_1^2 + (s_1 \leftrightarrow s_2)\,.
    \end{align}
\end{itemize}
Finally the $H_k$ functions are defined as:
\begin{itemize}
    \item [(i)] $\ell_\alpha^+ \to \ell_\beta^+ \ell_\beta^+ \ell_\beta^-$:
    \begin{align}
    H_1 &= 8 m_\alpha^3 s_w^4 s_1 - (s_1 \leftrightarrow s_2)\,,\\
    H_2 &= 8 s_w^4 m_\alpha^5 \left[\frac{2 m_{\beta }^4-2 m_{\alpha }^2 m_{\beta }^2}{s_2^2}+\frac{-2 m_{\alpha }^2-m_{\beta }^2+2 s_1}{s_2}\right] - (s_1 \leftrightarrow s_2)\,,\\
    H_3 &= 2 s_w^2 m_\alpha^3 \left[\frac{2 m_{\beta }^4-s_1 \left(m_{\alpha }^2+3 m_{\beta }^2\right)+s_1^2}{s_2}+s_1\right] - (s_1 \leftrightarrow s_2)\,,\\
    H_4 &= 4 m_\alpha^3 s_w^2 \left[\frac{s_1 \left(\left(2 s_w^2-1\right) m_{\alpha }^2+\left(8 s_w^2-3\right) m_{\beta }^2\right)+2 \left(s_w^2 m_{\alpha }^2 m_{\beta }^2+\left(1-2 s_w^2\right) m_{\beta }^4\right)+s_1^2 \left(1-3 s_w^2\right)}{s_2}\right.\nonumber \\
    &+ \left.s_1 \left(1-3 s_w^2\right)\right] - (s_1 \leftrightarrow s_2)\,,\\
    H_5 &= 4 m_\alpha^3 s_w^4 \left[\frac{-2 m_{\alpha }^2 m_{\beta }^2+4 m_{\beta }^4-2 s_1 \left(m_{\alpha }^2+4 m_{\beta }^2\right)+3 s_1^2}{s_2}+3 s_1\right] - (s_1 \leftrightarrow s_2)\,,\\
    H_6 &= - m_\alpha^5 s_w^2 \left[\frac{m_{\alpha }^2+m_{\beta }^2-s_1}{s_2}\right] - (s_1 \leftrightarrow s_2)\,,\\
    H_7 &= - 2 m_\alpha^5 s_w^4 \left[\frac{2 \left(m_{\alpha }^2+m_{\beta }^2\right)-3 s_1}{s_2}\right] - (s_1 \leftrightarrow s_2)\,,\\
    H_8 &= - 2 m_\alpha^5 s_w^2 \left[\frac{\left(1-2 s_w^2\right) \left(m_{\alpha }^2+m_{\beta }^2\right)+s_1 \left(3 s_w^2-1\right)}{s_2}\right] - (s_1 \leftrightarrow s_2)\,;
    \end{align}
    \item [(ii-a)] $\ell_\alpha^+ \to \ell_\gamma^+ \ell_\beta^+ \ell_\beta^-$:
    \begin{align}
    H_1 &= 16 s_w^4 m_\alpha^3 \left[\left(3 m_{\beta }^2+m_{\gamma }^2\right)-s_2\right]\,,
    \\
    H_2 &= 16 s_w^4 m_\alpha^5 \left[\frac{2 m_{\beta }^2 \left(m_{\alpha }^2-m_{\gamma }^2\right)}{s_1^2}+\frac{m_{\alpha }^2+m_{\beta }^2-s_2}{s_1}-1\right]\,,
    \\
    H_3 &= 16 s_w^2 m_\alpha^3 \left[\left(3 s_w^2-1\right) m_{\beta }^2+s_w^2 m_{\gamma }^2-s_2 s_w^2\right]\,,
    \\
    H_4 &= 8 s_w^2 m_\alpha^3 m_\beta^2\,,
    \\
    H_5 &= 4 s_w^2 m_\alpha^3 \left[m_{\alpha }^2+2 m_{\beta }^2+m_{\gamma }^2 \right. \nonumber\\
    &+\left.\frac{m_{\alpha }^2 \left(m_{\beta }^2-m_{\gamma }^2\right)-m_{\beta }^2 \left(m_{\beta }^2+m_{\gamma }^2\right)+s_2 \left(m_{\alpha }^2+2 m_{\beta }^2+m_{\gamma }^2\right)-s_2^2}{s_1}-s_1-2 s_2\right]\,,
    \\
    H_6 &= 8 s_w^2 m_\alpha^3 \left[\left(1-6 s_w^2\right) m_{\beta }^2-2 s_w^2 m_{\gamma }^2+2 s_2 s_w^2\right]\,,
    \\
    H_7 &= 4 m_\alpha^3 s_w^2 \left[\frac{1}{s_1} \left(s_2 \left(\left(1-2 s_w^2\right) m_{\alpha }^2+\left(2-8 s_w^2\right) m_{\beta }^2+\left(1-6 s_w^2\right) m_{\gamma }^2\right)+m_{\alpha }^2 \left(\left(1-6 s_w^2\right) m_{\beta }^2\right. \right. \right. \nonumber\\
    &+\left.\left.\left.\left(2 s_w^2-1\right) m_{\gamma }^2\right)+\left(m_{\beta }^2+m_{\gamma }^2\right) \left(\left(4 s_w^2-1\right) m_{\beta }^2+2 s_w^2 m_{\gamma }^2\right)+s_2^2 \left(4 s_w^2-1\right)\right)+\left(1-2 s_w^2\right) m_{\alpha }^2\right. \nonumber\\
    &+\left.\left(2-6 s_w^2\right) m_{\beta }^2+\left(1-4 s_w^2\right) m_{\gamma }^2+s_1 \left(2 s_w^2-1\right)+s_2 \left(6 s_w^2-2\right)\right]\,,
    \\
    H_8 &= 2 \left[m_{\alpha }^2+3 m_{\beta }^2+2 m_{\gamma }^2\right. \nonumber\\
    &+\left.\frac{m_{\alpha }^2 \left(3 m_{\beta }^2-m_{\gamma }^2\right)-\left(m_{\beta }^2+m_{\gamma }^2\right) \left(2 m_{\beta }^2+m_{\gamma }^2\right)+s_2 \left(m_{\alpha }^2+4 m_{\beta }^2+3 m_{\gamma }^2\right)-2 s_2^2}{s_1}\right.\nonumber \\
    &- \left.s_1-3 s_2\right]\,,
    \\
    H_9 &= 2 s_w^2 m_\alpha^5 \left[\frac{m_\alpha^2 + m_\beta^2 - s_2}{s_1} - 1\right]\,,
    \\
    H_{10} &= 2 s_w^4 m_\alpha^5 \left[\frac{2 \left(m_{\alpha }^2+2 m_{\beta }^2+m_{\gamma }^2\right)-4 s_2}{s_1}-2\right]\,,
    \\
    H_{11} &= 2 s_w^2 m_\alpha^5 \left[\frac{m_{\beta }^2+\left(1-2 s_w^2\right) m_{\alpha }^2-2 s_w^2 \left(2 m_{\beta }^2+m_{\gamma }^2\right)+s_2 \left(4 s_w^2-1\right)}{s_1}+2 s_w^2-1\right]\,.
    \end{align}
\end{itemize}
\medskip
In order to obtain the final expressions for the different asymmetries, the integration over the phase space must be carried out. 

A final point - already mentioned in the main body of the text - concerns approximations usually adopted leading to analytical expressions for 3-body decays, in particular 
working in the limit of vanishing final state lepton masses. While taking $m_{e,\mu} \to 0$ upon the derivation of the decay width indeed allows to correctly approximate the full (numerical) computation, the same does not occur regarding the asymmetries, especially in the case of $\tau$ decays into a final state containing muons. 
As an illustrative example, consider the 
integrated values for $G_2$ upon the study of the asymmetries in  $\tau \to 3 \mu$ decays, corresponding to the phase space prefactor of the squared dipole contribution to the $P$ asymmetry.
One finds
\begin{align}
    &2 \left(\int_{m_\mu (m_\tau + m_\mu)}^{\frac{1}{2} (m_\tau^2 - m_\mu^2)} \int_{s_1}^{s_2^+(s_1)} d s_1 d s_2 + \int_{4 m_\mu^2}^{m_\mu (m_\tau + m_\beta)} \int_{s_2^-(s_1)}^{s_2^+(s_1)} d s_1 d s_2\right) \lambda \! \left(\frac{m_\mu^2}{m_\tau^2}, \frac{s_3}{m_\tau^2} \right)^{-1} G_2^{\tau \to 3 \mu}(s1,s2) \nonumber \\
    &\sim \frac{4}{3} s_w^4 m_{\tau }^8 \left(m_{\tau }^2 \left(4 \log \left(\frac{m_{\tau }^2}{m_{\mu }^2}\right)-23\right) + 24 (2 C+3) m_{\tau } m_{\mu } +m_{\mu }^2 \left(-3 \log \left(\frac{m_{\tau }^2}{m_{\mu }^2}\right) \left(3 \log \left(\frac{m_{\tau }^2}{m_{\mu }^2}\right)+16\right)\right.\right. \nonumber \\
    &+ \left.\left.6 \pi^2 - 15 C - 145\right)\right) + \mathcal{O}\!\left( m_\mu^3 \right)\,,
\end{align}
in which $C \approx 0.916$ is the Catalan's constant.
An order-by-order comparison of the numerical value of this asymptotic expansion with the numerical value of the exact integral results allows to readily identify the following issues: considering only terms up to $\mathcal{O}\!\left( m_\mu^0 \right)$ in the expansion would lead to errors of over $1000 \%$ between the full numerical result and the analytical approximation. This is due to an accidental cancellation involving the logarithms of the lepton mass ratios, and one must in fact consider terms up to second order in the lepton mass ratio to find an acceptable analytical approximation.
This problem is even worse for the $P^\prime$ and $T$ asymmetries, as the associated asymptotic expansions converge even more slowly towards the full numerical result.

Whenever feasible, we provide simple expressions for the asymmetries relying on asymptotic expansions, but we emphasise that the phase space integrals should be computed numerically.

{
\small
\bibliographystyle{JHEP}
\bibliography{Asym-cLFV-DKT}
}

\end{document}